\documentclass[journal,12pt,onecolumn]{IEEEtran}
\usepackage{lineno,hyperref}
\modulolinenumbers[5]

\usepackage[T1]{fontenc}%Selecao de codigos de fonte.
\usepackage[utf8]{inputenc}%Codificacao do documento
\usepackage{comment}
\usepackage{xcolor}
\usepackage{graphicx}
\usepackage{caption}
\usepackage{subcaption}
\usepackage{tcolorbox}
\usepackage{multirow}
\usepackage{rotating}
\usepackage{lscape}
\usepackage{adjustbox}
\usepackage{array, booktabs, ragged2e}
\usepackage{makecell}
\usepackage{tabulary}
\usepackage{longtable}%quebrar a tabela pra outra página
\usepackage{amsmath}
\usepackage{multirow}
\usepackage{graphicx}
\usepackage{booktabs}   
\usepackage{ltablex}
\usepackage{xltabular}
\usepackage{rotating}
\usepackage{ragged2e}
\usepackage{float}
\usepackage[normalem]{ulem}%texto cancelado

\usepackage{soul,xcolor}
\begin{document}

\title{Metropolitan Optical Networks: A Survey on New Architectures and Future Trends}

\author{Léia~Sousa~de~Sousa,
        André~C.~Drummond,~\IEEEmembership{Member,~IEEE}% <-this % stops a space
\thanks{L. Sousa and A. Drummond are with Department of Computer Science, University of Brasília, Brazil.}}

\maketitle

\begin{abstract}
Metropolitan optical networks are undergoing major transformations to continue being able to provide services that meet the requirements of the applications of the future. The arrival of the $5G$ will expand the possibilities for offering IoT applications, autonomous vehicles, and smart cities services while imposing strong pressure on the physical infrastructure currently implemented, as well as on static traffic engineering techniques that do not respond in an agile way to the dynamic and heterogeneous nature of the upcoming traffic patterns.
In order to guarantee the strictest quality of service and quality of experience requirements for users, as well as meeting the providers' objectives of maintaining an acceptable trade-off between cost and performance, new architectures for metropolitan optical networks have been proposed in the literature, with a growing interest starting from $2017$. However, due to the proliferation of a dozen of new architectures in recent years, many questions need to be investigated regarding the planning, implementation, and management of these architectures, before they could be considered for practical application.
This work presents a comprehensive survey of the new proposed architectures for metropolitan optical networks. Firstly, the main data transmission systems, equipment involved, and the structural organization of the new metro ecosystems are discussed. The already established and the novel architectures are presented, highlighting its characteristics and application, and comparative analysis among these architectures is carried out identifying the future technological trends. Finally, outstanding research questions are drawn to help direct future research on the field. Among the conclusions of this survey, the application of flexible spectrum allocation technologies seems to be the right, and necessary, evolution path for metropolitan optical network architectures.
\end{abstract}


\newpage
\setcounter{tocdepth}{3}
\tableofcontents

\section{Introduction}
Today we experience the ubiquitous usage of cloud services provided by worldwide geographically spread data centers. However, due to the long distances between the users and the data storage/computing infrastructure, high latency is experienced by transported traffic, so the usage experience is relevantly impacted for some applications. More recently, driven by advances in the IoT with $5G$ as an enabler technology, distributing data storage/processing by deploying computing and storage resources in proximity of the end users has become a major requirement for latency-sensitive services, leading to a new computation paradigm know as fog/edge computing \cite{paolucci2020disaggregated}. %\cite{paolucci2020disaggregated}.

The infrastructure of the Internet can be divided into three levels, the highest level, which comprises long distances nation wide infrastructure and interconnect large data centers is defined as the core transport networks, providing ultra high transport capacity. The lowest level is constituted by access networks, interconnecting end users to their local internet service providers. The bridge between the core networks and the access networks are the metropolitan transport networks \cite{muciaccia2019proposal}.

Nowadays, this ongoing change on the location of storage/computing systems towards the edge has already impacted the traffic profile in transport networks, shifting the load from the core level to metropolitan level \cite{9042241}. Moreover, the lowest degree of traffic aggregation in metro, due to the closer proximity with the traffic sources, naturally leads to a more dynamic and heterogeneous traffic profile, which imposes harder requirements for transporting it. All these changes have established the Metropolitan (or Metro) Optical Networks (MON) as the arena where both the academy and industry will focus their effort in search for innovation. 

The metro network environment is one of the network segments with considerable variety in terms of the presence of data traffic granularities, varying from below $1$ $Gb/s$ \cite{shen2018ultra} up to $800$ $Gb/s$ \cite{infinera800G}, with different modes of communication \cite{cugini2016receiver} and traffic profiles \cite{uzunidis2018dufinet}, different patterns of traffic distribution, both in time and space \cite{troia2019dynamic}, coexisting in the same network segment.
Also, design decisions such as the chosen network topology and architecture or the transmission system, result in a significant impact on the traffic generated in other network segments such as core networks (Internet backbone) and access networks \cite{thyagaturu2016software, rottondi2013routing}.
%\cite{rottondi2013routing}.

In this multifaceted scenario, network providers and service operators need infrastructure with lower Operational Expenditure (OPEX) and Capital Expenditure (CAPEX). While OPEX represents the financial amount earmarked for infrastructure maintenance, CAPEX refers to financial expenses directed towards the acquisition of new goods and equipment. Filterless Optical networks (FON) are a type of network that does not have filters installed on the switching nodes, thus they can't separate (or block) parts of the transmitted signal. As filters are one of the elements that most increase the cost of the network, this type of solution has been in evidence for the metro scenario, as highlighted in \cite{Ayoub:22}. But the authors point out that removing filters can result, over time, in higher absolute expenses on transponders.

The disaggregation of the network is driving unprecedented changes. With such an approach, providers have more freedom to deploy independent optical equipment from several different manufacturers (connectable optics, transponders, and Optical Cross-Connect (OXCs) / Reconfigurable Optical Add-Drop Multiplexer (ROADMs)), while an operating system for the network provides control, management, and interaction between these elements \cite{9132992,hernandez2020techno}. With this in mind, open-source and collaborative systems are under
development to exploit, in a more efficient way, network resources that are adaptable to each type of service offered, for example, the instantiation of numerous functions and virtual networks \cite{casellas2018enabling}.

The recent literature tutorial and surveys proposes various solutions for traffic engineering problem in optical networks \cite{Ayoub:22, thyagaturu2016software, rejiba2019survey,MILADICTESIC2019464}. While \cite{thyagaturu2016software, rejiba2019survey,MILADICTESIC2019464} are more generic and do not focus exclusively on MON, \cite{Ayoub:22} presents a technical-economic analysis and discusses the impacts on FON considering the instantiation of virtual networks, protection and survivability. %Handbook_of_ON%

 With prospects for adhering to the deployment of the Elastic Optical Network (EON) transport technology, and due to the feasibility of this technology to better meet the characteristics of metro networks, specific traffic engineering solutions for this network segment have been proposed concerning the tidal-traffic phenomenon characteristic \cite{yan2018tidal, yan2020area},
adaptive resource allocation \cite{8734478,kokkinos2019pattern, 8853968}, resource balancing \cite{yan2020area,8596108} and quality of service (QoS) guarantee \cite{liu2018joint}, but without taking into account the characteristics of the underlying network architecture. EON is a new standard for data transmission at high adaptive rates capable of maximizing the use of spectral resources.

On the other hand, projects for the implementation of new edge nodes with capacity for computing, processing, and data storage \cite{larrabeiti2020upcoming}, fog computing in many levels, as well micro data centers (mDC) \cite{liu2018joint,popescu2017impact} and edge-data center \cite{le2020survivable}, have caused changes in the physical infrastructure of the network. As these content delivery centers will be increasingly distributed throughout the metro network, although it is possible to take advantage of the equipment already installed \cite{infinera2020cost}, it is not feasible only to scale the physical resources in operation to meet the new demands~\cite{papanikolaou2018optimization}. New equipment has been designed to ensure capacity and dynamism with efficiency and low cost, such as transceivers~\cite{boffi2020multi,masood2020smart,li2017digital,moreolo2018modular} and optical switches~\cite{8204502WSS} based on space division multiplexing (SDM) and multi-core fibers \cite{liu2019space}, mainly following a modular \cite{calabretta2019photonic} and disaggregated \cite{hernandez2020techno} approach. 
Equipment that exploits SDM uses the spatial dimension to simultaneously deliver different data streams while creating parallel spatial channels. The concept of disaggregation can occur vertically or horizontally. With vertical disaggregation, even optical elements from different components can be separated, without loss of interoperability, between software and hardware. With horizontal disaggregation, functions are separated from operations into smaller subsystems, maintaining interoperability between components from different suppliers.

The main differential of this new generation of photonic hardware is its coherent multi-rate, multi-band, and multi-modulation transmission and detection properties \cite{paolucci2018filterless}, use of low-cost transmission lasers with software reconfiguration capabilities (SDN) \cite{nadal2019sdn}, optical integration with miniaturization and low energy consumption. Amid these innovations, legacy network architecture solutions such as pure OTN remain a considerable alternative for metropolitan network planning, including being redesigned in the form of Mobile-optimized OTN (M-OTN) technology. In pure OTN, the processing of data streams takes place in the electronic domain, while only the transport of these data takes place in the optical domain in a point-to-point manner, which ends up being very expensive for environments with large numbers of nodes, such as metropolitan networks. However, mainly for the fronthaul of 5G technology, M-OTN features a flexible time slot structure, with encrypted optical connections and simplified overhead, which reduces latency~\cite{infinera2020cost,8696394}.
%}}

With this wide range of possibilities, many metropolitan network architectures have been proposed resulting from the combinations of these new elements.  
Thus, network providers have been challenged to identify the ideal solution for the sustainability of their business, having to answer questions about ideal location/position for data within the network topology, and which network architecture/technology best benefits each type of service chain, as it is a high-cost investment in terms of maintaining operations, with some other challenges in terms of energy efficiency, latency, and scalability. As such innovations are inevitable, in addition to dealing with these issues, providers also need to identify new business opportunities that disruptive technologies can bring with the technological maturity of architectural proposals.

It is not possible to think of the $5G$ and $6G$ mobile communication systems without reflecting the concern with the optical network infrastructure that will support it on increasing the interconnection network while satisfying radio requirements, like low latency, for example~\cite{gangopadhyay20195g}.
This scenario of opportunities is favorable for the implementation of new architectures, gathered in this work, to accommodate and incorporate the operations and services necessary for the mobile network. The same feat cannot be achieved only with previous generations of the network due to the potential technical challenges linked to the adaptation of interfaces and the inability to coordinate high performance with low latency.

This work focuses on bringing together these architectural proposals for metropolitan data transport networks, highlighted in the recent state-of-the-art literature, to discuss its main characteristics, advantages and disadvantages. The architectural solutions are brought together in two large groups, being single and multi-layer architectures, and then an alignment of the proposals in the logical and nomenclature perspectives is presented, as well as a comparative analysis between them.
Single-layer architectures have the advantage of being more agile and fast, but achieving the expected elasticity for the future is a major technological challenge. Multi-layer networks, with a more facilitated approach for the implementation of programmable functions and for the aggregation of traffic, incur increased latency and effort in coordinating the layers involved. 

The main contribution of the work is the survey of the main architectures candidates to constitute the MONs of the next generation, with a mapping of their respective layers or hierarchical levels. Besides, research trends that will attract the attention of the research and development community in the coming years are also identified, as well as the main open questions identified regarding each architecture. As a main limitation, the difficulty of carrying out comparing all the architectures among themselves (individually, one by one) is pointed out, which was not possible to be done completely due to the nonexistence (the lack of identification) from sufficient sources in the literature, since, so far, for some of the architectures only seminal works have been identified. To the best of our knowledge, a survey has not been identified in the literature that gathers all architectural proposals with a focus on metropolitan optical networks. Possibly related solutions, but already in other fields of study, can be identified in an in-depth approach that includes areas such as technologies for access networks, mobile networks or even Crosshaul (Xhaul) architectures, which are not in the scope of this work.

\subsection*{Document Organization} \label{subsec:DocOrganization}

To address the topic of metropolitan optical networks, considering the themes highlighted in the current literature, this work was organized in six sections described as follows.

Section \ref{sec:background} contains the background of the work and discusses the fundamental properties of metro networks as transmission systems (Subsection \ref{subsec:techMetro}), structural (Subsection \ref{subsubsec:perspLogica}) and physical perspectives (Subsection \ref{subsub:PhysicalPerspective}) of optical nodes, that lay the foundation for understanding the rest of this work. The main metropolitan network topologies are also addressed because they are related to equipment choices in the construction of an architecture.
%This section concludes with the presentation of some ongoing projects aimed at optimizing the next generations of MON.
Section \ref{sec:arquitetura} discusses architectural proposals for the metro core and metro access segments. The access segment is included, although not the focus of this work, due to the unified view in the light of which some architectural proposals have recently emerged. The highlighted architectures are separated into two broad categories as identified in the literature: multi and single-layer. Single-layer architectures are organized into three types, which are the architectures with filter, filterless, and semi-filterless, that will be discussed individually.
Section \ref{sec:comparingArch} presents a comparison and implications on the architecture of metropolitan optical networks brought together. After comparison, the main themes discussed by each architecture are highlighted, showing the main fields of study in progress. Section \ref{sec:Trends} highlights research opportunities and literature gaps. Finally, Section \ref{sec:conclusao} presents conclusions and the final considerations. A list of acronyms is presented in Table \ref{tab:Acronym} after Section \ref{sec:conclusao}.

%\section{Arquitetura de nós ópticos}

%%%%%%%%%%%%%%%%%%%%%%%%%%%%%%%%%%%%%%%%%%%%%%%%%%%%%%%%%%%%%%
%%%%%%%%%%%%%%%%%%%%%%%%%%%%%%%%%%%%%%%%%%%%%%%%%%%%%%%%%%%%%%
%%%%%%%%%%%%%%%%%%%%%%%%%%%%%%%%%%%%%%%%%%%%%%%%%%%%%%%%%%%%%%
\section{Metropolitan Optical Networks}\label{sec:background}

This section discusses the main data transmission systems in Metropolitan Optical Networks (MON), that is, the main configurations of the optical spectrum defined for traffic transport, as well as the main equipment involved.
Metropolitan networks are formed by an electronic layer and an underlying optical layer. This organization makes it possible to divide the discharge complexity of designing networks at various levels and interconnecting them so that data can be packaged and aggregated in the electronic layer, and consequently, transported end-to-end through the optical layer \cite{Handbook_of_ON,katsalis2018towards}.
Typically, at the electronic layer, data packet switching is performed. In this process, the data to be sent is divided into small units and does not require the prior establishment of a path for sending these packets. Circuit switching predominates in the optical layer, in which an end-to-end optical path or circuit is established completely by reserving a portion of the optical spectrum bandwidth. Data in the form of pulses of light are sent through this circuit~\cite{kokkinos2019pattern,larrabeiti2020upcoming,kosmatos2019building}.
The object of study of this work are the optical metropolitan networks, and therefore, the discussions that follow will be concentrated in this network domain. In general, these optical networks can be classified as opaque, transparent or translucent. Transparent networks switches and transmits the signal exclusively on the optical medium and therefore Optical-Eletrical-Optical (OEO) devices are not required. If the signal conversion is required on all network nodes, this architecture is said to be opaque, however, there may be architectures that combine opaque and transparent nodes, and these are called translucent. This conversion process raises the cost of the network and further adds delay to network operations \cite{thyagaturu2016software}.

MONs are designed with short and medium distance links, less than $200$ $km$, usually within the limit of an optical single-span without amplification (around $80$ $km$)~\cite{li2017digital,ITUmonDefinition}. However, longer distances and smaller spans (multi-spans) are also possible, requiring some signal regeneration while driving the "pay as you grow" trend~\cite{paolucci2020disaggregated,boffi2020multi}. For these networks, there are a variety of transmission systems that can be used, as well as different types of hardware elements such as building blocks, which can be organized in various ways to compose the topology of the network. The set of transmission methods in optical networks stands out (Subsection \ref{subsec:techMetro}), with special emphasis on EON, since most of the works that will be discussed focus on this transport technology. 
It is also relevant to present the great diversity of equipment (Subsection \ref{subsec:equipaments}) that make up the ecosystem of MON, as well as the roles of various types of nodes in the infrastructure, since several possible hierarchical organizations are found in the literature, as will be shown in the course of this work.
Then the main topologies of MON will be emphasized (Subsection \ref{subsec:topologies}), since this is a preponderant factor for infrastructure planning, as well as a strong determinant for the choice of network elements to be implemented.

\subsection{Transmission Systems in Metropolitan Optical Networks (MONs)}\label{subsec:techMetro}

As with core networks, the optical transport technologies for metropolitan networks highlighted in the literature are subdivided into two main categories: fixed grid and flexible grid. In terms of fixed grid technology, the Wavelength Division Multiplexing (WDM)~\cite{calabretta2019photonic} is the main alternative in use today. Two WDM system modalities are available: Coarse Wavelength Division Multiplexing (CWDM) standardized by ITU $G.694.2$ \cite{lin2019three} and Dense Wavelength Division Multiplexing (DWDM) standardized by ITU $G.694.1$~\cite{8204502WSS}. Both are used together concerning metropolitan and access networks, with DWDM usually offering a greater number of channels, using fewer links and is recommended for distances greater than CWDM, but with a higher cost. However, to change from CWDM to DWDM, it is enough that the provider changes some equipment to support many channels with less granularity~\cite{8204502WSS,singleCarrier2016}. Besides, DWDM is especially used in filterless networks \cite{Ayoub:22, paolucci2018filterless, filer2019low}. In WDM networks, in order to meet a connection demand, it is necessary to solve the problem of routing and wavelength assignment (RWA)~\cite{wu2019analysis}.

The International Telecommunication Union (ITU) has proposed a set of standards to define the use of the optical spectrum. In one of these standards, that is, fixed grid, adopted by WDM, the optical spectrum is divided into partitions with a width of $50$ $GHz$ or $25$ $GHz$~\cite{boffi2020multi}. Each partition serves an optical channel, regardless of the bandwidth requirements of the transmitted signals. However, there is a limitation to accommodate signals due to the space between the channels, and on top of that there is a potential inefficiency in using these channels to accommodate very low rates, for example, as the allocated spectrum is $50$ $GHz$, if demand requires only $10$ $GHz$, means that $40$ $GHz$ of the optical spectrum is wasted.

Networks that implement WDM transmission and still bet on the concept of unbundling are built based on open-source software (OSS) concepts. The idea is to separate hardware and software elements from the WDM optical network in the analog (A-WDM) and digital (D-WDM) domains \cite{hernandez2020techno}. In the D-WDM domain, functions related to the transport of these analog channels are implemented in equipment such as multi-degree ROADMs (MD-ROADMs), line terminals (LTs), multiplexers (MUXs), and inline optical amplifiers (ILAs).
 In the D-WDM domain, the client's digital signals are adapted to the analog channels of the A-WDM domain, as is the case with the transponders (TPs) used in point-to-point (P2P) connections, muxponders (MPs) used to combine multiple sub-rates, and switchponders (SPs) that are cards that integrate an OTN switch with DWDM transponder \cite{hernandez2020techno}. Such equipment will be highlighted in the Subsection \ref{subsec:equipaments}.

To make flexible grid technology possible, parallel signal processing techniques, like OOFDM~\cite{MILADICTESIC2019464} and Nyquist-WDM, are implemented to achieve transmission rates beyond the limits of electronics. Due to the multiplexing of optical subcarriers as a single entity, it is possible to reduce the spacing between these subcarriers and improve spectral efficiency. Some candidate flexible transmission technologies are highlighted:
Nyquist-WDM (NWDM)~\cite{ujjwal2018review}, quasi-Nyquist WDM (qNWDM)~\cite{shiraki2019design}, Elastic Optical Networks (EONs) \cite{yan2020area} and Dense Elastic Optical Networks (DEONs).
%\footnote{Muita pretensão, mas a ideia é usar um nome para distinguir o sistema de transmissão EON de grade mais fina (usado na UDWS) da arquitetura UDWSN.}.

As EON is an evolution of the DWDM system \cite{ujjwal2018review}, the term DEON is being proposed in this work as a specialized version of EON that includes the use of spectrum spaces, that is, Frequency Slot Units (FSUs) narrower than the usual $12.5$ $GHz$, considered until now as the smallest possible spacing according to ITU-T recommendation $G.694.1$ \cite{glkabowski2020simulation}. In the literature it is possible to observe experiments in an attempt to reduce the size of the FSU \cite{rottondi2013routing, 7792281UDWSN,8025162OTN,zhang2016ultra,Zhang2018ExploitingEO}, especially in metropolitan optical networks, to provide better adaptation of the data rates practiced in this segment with what is made available of resources, which considerably increases the number of channels available for transmission.
This movement has led to the proposal of new network architectures with greater spectral granularity and that will implement specific equipment \cite{moreolo2018modular} to support finer filtering adjustment, as is expected with the Ultra-Dense Wavelength Switched Network (UDWSN) architecture \cite{shen2018ultra,7792281UDWSN,Zhang2018ExploitingEO},to be presented later in Subsection \ref{subsubsec:UDWSN}.

Nyquist wavelength-division multiplexing (NWDM) is a more flexible form of WDM that limits channel spacing to transmission rate by generating almost rectangular spectra, with negligible crosstalk and inter-symbol interference. This almost rectangular shape is due to pulses interspersed in the form of sync (cardinal sine) through Orthogonal Time-Division Multiplexing (OTDM). By allowing subcarriers so close in the composition of super-channels, NWDM improves spectral efficiency (SE) with minimal spacing~\cite{ujjwal2018review}. NWDM variations, called quasi-Nyquist Wavelength Division Multiplexing (qNWDM), can be achieved due to the difficulty in obtaining a perfect rectangular spectrum since its bandwidth is infinite and the band delimitation is due to deficient channel characteristics or by allocating the desired portion for an application \cite{shiraki2019design}. These Nyquist variations arise due to several roll-off factors that cause the spectrum to narrow or widen within the grid.

\subsection{Architecture Components}\label{subsec:equipaments}

%The basic structure of metropolitan optical transport networks are the nodes in the optical path layer. 

Metropolitan optical transport networks are composed of nodes in the optical path layer. The nodes perform various functions from a structural perspective, in the role of traffic aggregation hub (which does not generate traffic) or even edge nodes (on the border between two different network levels/layers). From a physical perspective, each node has a set of equipment that is necessary for the establishment of end-to-end optical paths, including bypass connections. Both perspectives are going to be presented in the following sections.

\subsubsection{Structural Perspective}\label{subsubsec:perspLogica}

In the metropolitan network segment, network nodes can also be defined from a structural perspective, with terminologies that refer to the roles that these nodes represent in each network segment, as well as the functionality of these segments.
This point deserves to be highlighted for aligning different expressions used to name the same region of interest within the structure of optical transport networks, both for areas of industry and research activity\cite{9042241,thyagaturu2016software}.
As it is one of the most heterogeneous network segments, both in terms of architectural and traffic characteristics, these definitions are useful to identify outbreaks where certain implementations need to be carried out more immediately or over a certain time interval, or even, to define the localization for the allocation of the data, where tasks and data on the network need to be close, mainly due to the emergence of new computing paradigms \cite{sivaraman2020network}.

\begin{figure}
	\centering
	\includegraphics[width=0.99\linewidth]{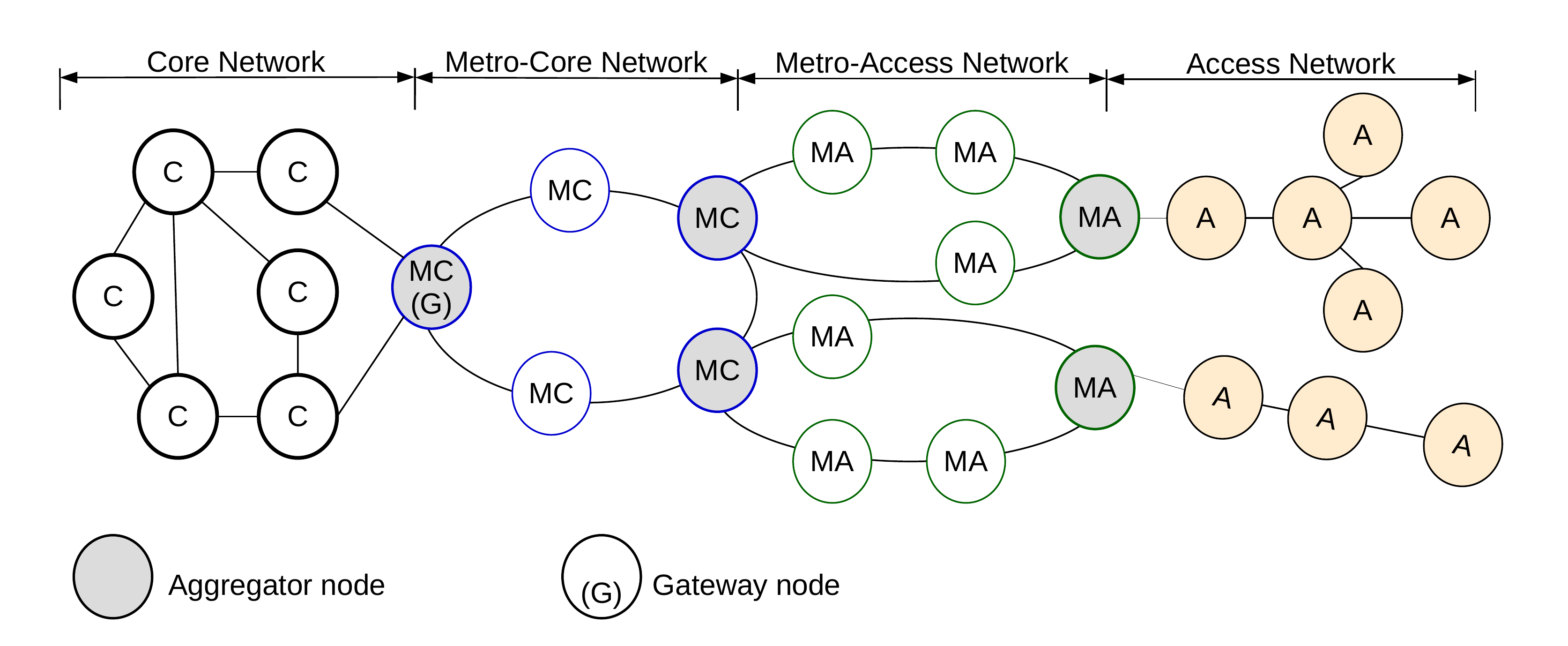}
	\caption{Structure of Optical Transport Networks}
	\label{fig:hierarquia}
\end{figure}

In the Figure \ref{fig:hierarquia} core network nodes with letter C are represented as black circles, the metropolitan segment is divided into two levels, being metropolitan-core network nodes with letters MC represented as blue circles, and metropolitan-access network nodes with letters MA represented by green circles. Further, the segment of nodes in the access network, with the letter A, is represented by circles filled in beige.  
The letter G in parenthesis identify the node that is the gateway to the core network. The fully colored gray nodes play the role of traffic aggregator for a given network level immediately below.

The switching nodes in the metropolitan networks are also called Metro-Core (MC) and are located in Points-of-Presence (POPs). The nodes in the Metro-Access network are also called Metro-Aggregation (MA) and are usually where the Central Offices (CO) are located \cite{shen2018ultra,Handbook_of_ON}. Another segment (not shown in the figure), which is positioned to the right of the access network, can be mentioned for a panoramic view of these interconnections. This segment concerns the device layer used to aggregate end customers and/or application of the Internet-of-Things (IoT) paradigm \cite{rejiba2019survey}.

The representation of the Figure \ref{fig:hierarquia} is adopted in \cite{shen2018ultra}, that considers metropolitan and access networks in its architecture. In \cite{lashgari2021end}, MA nodes from Figure \ref{fig:hierarquia} are called by metro nodes, and MC nodes are metro-core edge. Only on these nodes does traffic aggregation occur. 

Often the metropolitan network and access network architecture is presented as a single architecture, in an end-to-end perspective from the transmission point of view, as both the technology and the topology chosen for the access network determine the decision-making points in other segments of the network \cite{9042241, shen2018ultra}. 
In such environments, edge servers, normally facilitated by these mini-DCs or a small cluster of servers, usually connect to small base stations to share computing loads from mobile and IoT devices and provide a high QoS for end-users. Therefore, edge servers are the critical components in the edge computing environment. The new perspectives of organization of the network infrastructure have, more and more, overturned the segmented view and follows a unified conception of the network, since the edge nodes in the access network tend to become as robust and complex as the nodes in the metropolitan network \cite{muciaccia2019proposal}. 

The identification of these main points in MON infrastructures is also essential for the structural planning of $5G$ networks to offer new services. At a high level, the $5G$ network is formed by three main entities denominated  Remote Radio Units (RRU), Distributed Unit (DU) and Centralized Unit (CU). 
All these entities are interconnected by a fiber optic network.
On the margins of the nodes of the access network are the RRUs to collect the mobile traffic. Traffic is transported to the DU to be processed.  Due to the density of RRUs, several DUs are required as a way to improve service latency. CU is responsible for concentrating the traffic from multiple DUs in a region, as well as forwarding traffic to edge cloud. Edge cloud connects directly with core cloud \cite{lashgari2021end,paul2019traffic}. As an illustrative and simple comparison, this infrastructure required for 5G networks could be mapped with metropolitan and access optical networks as follow: CU and DU are located in MA nodes, with some CU point and edge cloud in MC nodes.
%RRU interconnects to DU through the fronthaul transport. DU connects to CU through midhaul transport. The interconnection of the cloud takes place through the backhaul transport. All this x-haul transport are based on optical network and they will need to keep up with the advance of different population and geographic densities. } 

For $6G$ networks, with more stringent latency and processing requirements, the idea is to further dilute the distribution levels of computing nodes, leading to the implementation of cloud and fog computing at many distributed levels with optical transport at higher data rates~\cite{yosuf2020cloud}.

For the segment of interest in this work, the network edge refers, at a higher level, to the first-hop network devices that applications use to connect to the network (such as access routers, network interface cards, virtual switches, and base stations on the mobile network) and devices. 
The same administrative entity can manage different parts of the metropolitan network, or can focus on offering its services in only part of that segment, such as: (i) a data center operator can control the edge servers and switches (core) in the data center; 
(ii) an Internet service provider can control the edge routers and core routers within the same autonomous system \cite{sivaraman2020network}; 
(iii) a data center operator can concentrate its business in an area of the core of the metropolitan network, which it calls a regional network, and consists of several distributed data centers that form a mega data center \cite{filer2019low}. 
To aggregate computing resources from the devices at the edge of the network, there are fog computing nodes to perform critical calculations sensitive to the data, with the data from the analysis part sent directly to the cloud, for further processing, since traditional fog nodes have limited computing and storage capacity \cite{rejiba2019survey}.

Generally, the DCs at the core nodes of the metropolitan network is of less infrastructure and complexity, and are therefore called micro DCs (mDC). This configuration has recently been highlighted due to the greater ease of management and scalability, in addition to the reduced cost and lower latency in offering services to customers~\cite{thyagaturu2016software}. 
On the other hand, in the MA segment, there has been a constant integration of edge computing platforms, with simpler and less costly infrastructures, compared to mDC, with the task of providing a distributed architecture that brings computing and storage services closer to the end-user~\cite{rejiba2019survey}.

\subsubsection{Physical Perspective}\label{subsub:PhysicalPerspective}

\begin{figure}[H]
	\centering
	\includegraphics[width=0.99\linewidth]{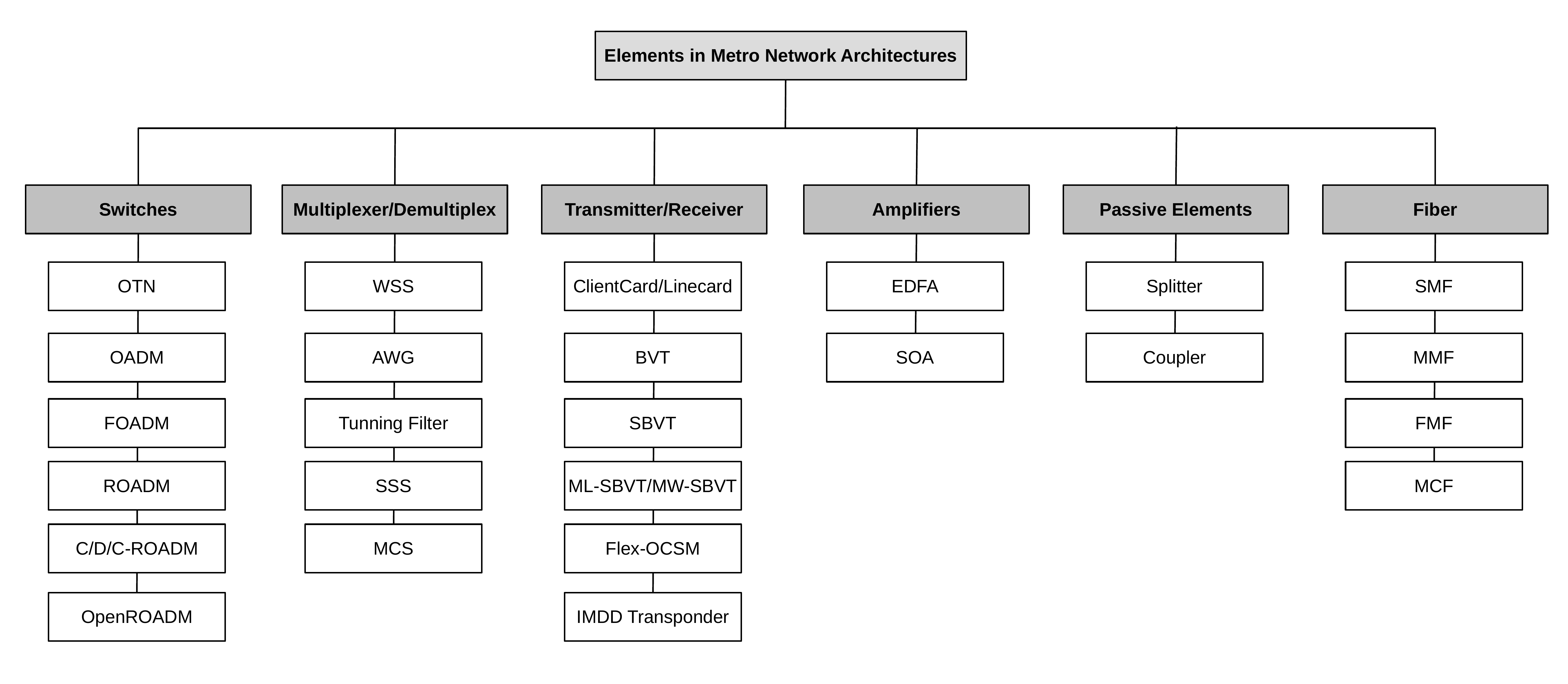}
	\caption{Main equipment for the construction of metropolitan optical networks.}
	\label{fig:arc-equipamentos}
\end{figure}

This subsection begins with the definition of some key equipment for the construction of optical network architectures, which, in general, are employed independently of the chosen transmission system. These elements are switches, multiplexers/demultiplexers, transmitters/receivers, amplifiers, passive elements, and the transmission media, that is, the optical fiber. Figure \ref{fig:arc-equipamentos} presents a list of the elements that will be highlighted below.

\begin{figure}
	\centering
	\includegraphics[width=0.7\linewidth]{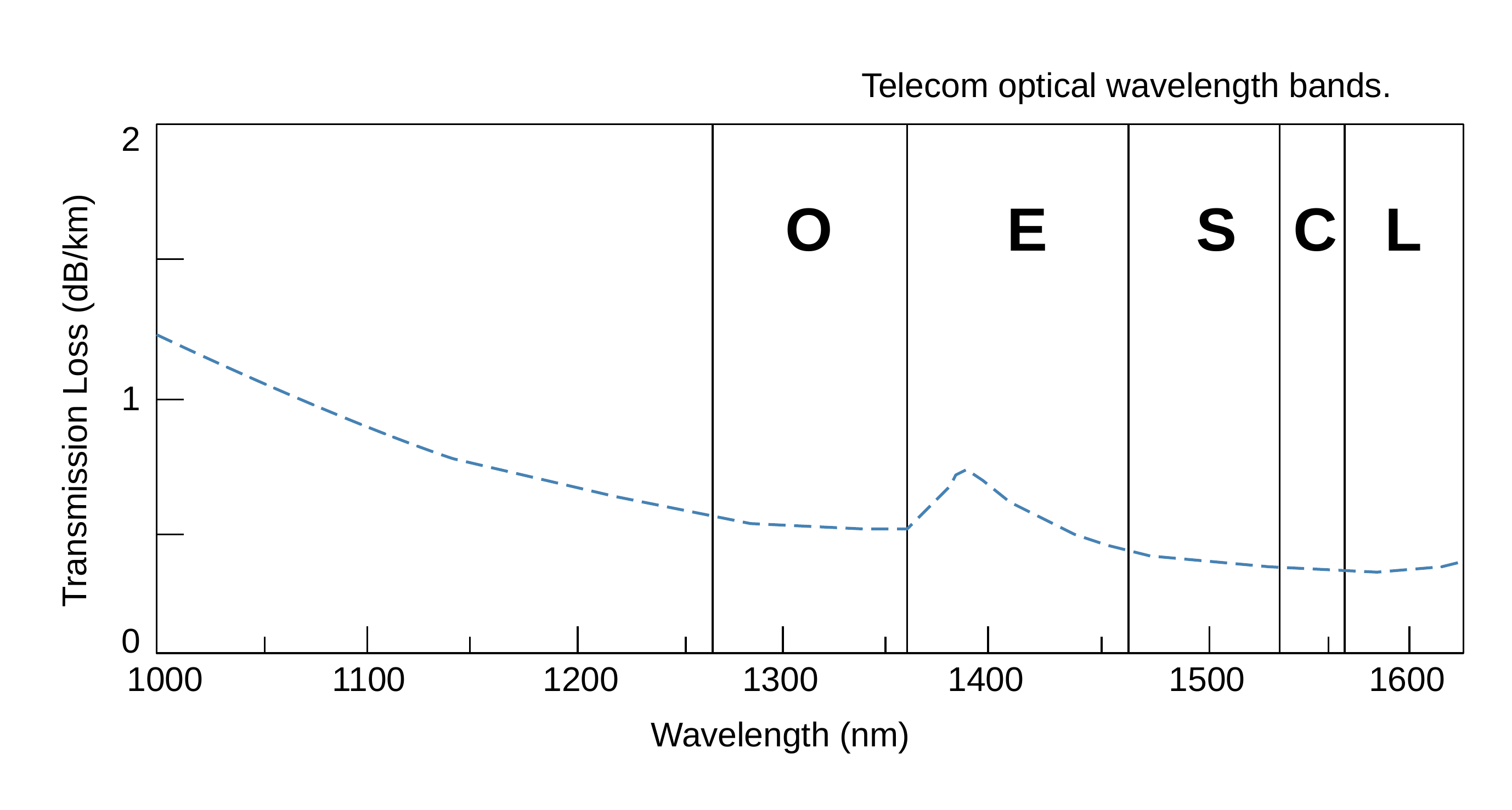}
	\caption{The fiber loss of different wavelength.}
	\label{fig:allBands}
\end{figure}

The optical spectrum in optical fibers is divided into several windows or bands for better use of the regions with lower attenuation. Figure \ref{fig:allBands} shows the optical transmission power loss, measured in decibels per kilometer ($dB/km$). The power loss varies according to the wavelength of the chosen light and the composition of the propagation material. The lowest loss occurs at the wavelength of $1550$ $nm$, inside the C band, which is commonly used for long distance transmissions, followed by the L-band. The O-band was one of the first to be used for telecommunications while the S-band is an alternative for passive network communications. Each transmission band has particular restrictions and requires specific equipment for its adoption.
%Some transmission systems that exploit these bands will be highlighted below.

\paragraph*{\textbf{Transponders}}

% Esse parágrafo abaixo é um exemplo de Transponder
Transponders are signal converters from the optical domain to the electronic domain, or vice versa. 
At each entrance door, there is a traffic receiving device (Rx) and at each exit door there is a traffic transmitting device (Tx) \cite{9041836}. In the literature, these devices are transponders~\cite{sambo2017sliceable} and transceivers~\cite{cugini2018flexibleSemiFL}.

\begin{figure}
	\centering
	\includegraphics[width=0.99\linewidth]{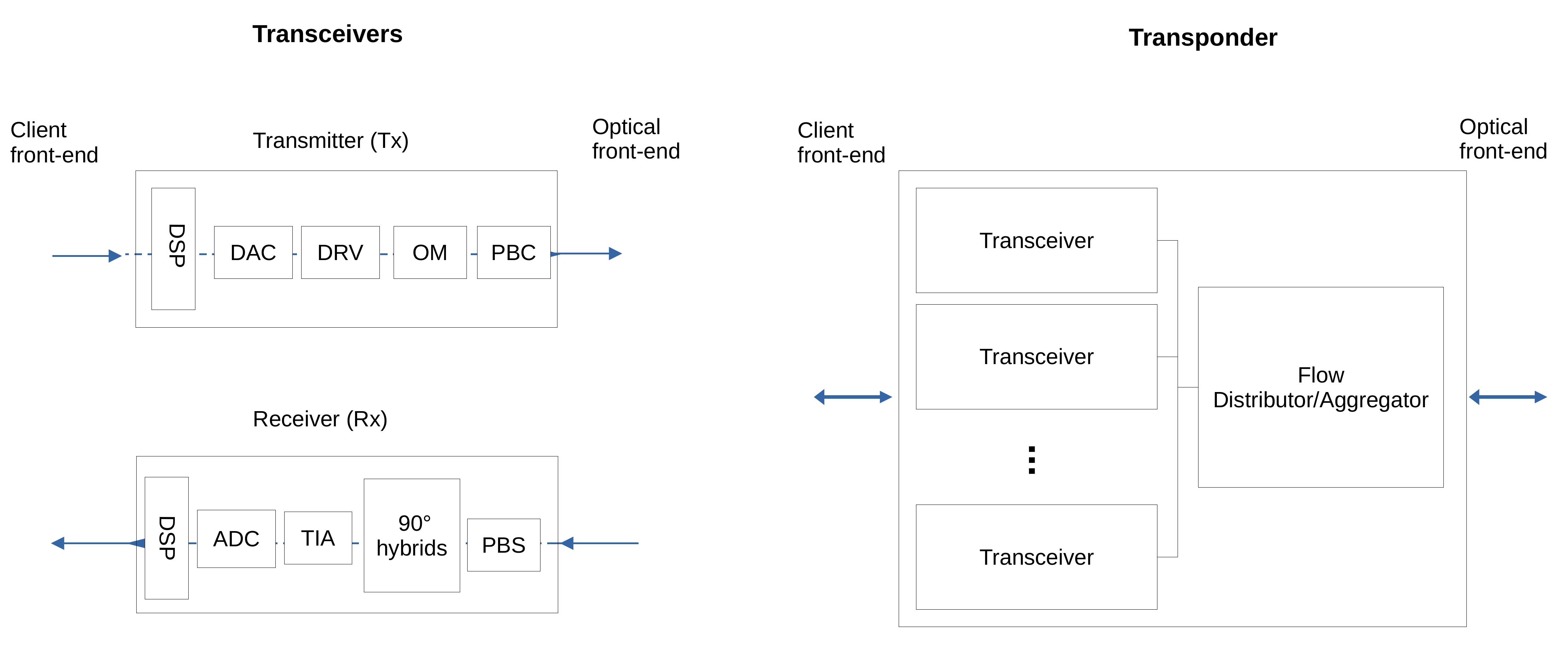}
	\caption{Schematic example of a Optical transponder.}
	\label{fig:transponder}
\end{figure}

The differences between a converter and optical filter, highlighted in the Subsection \ref{sec:SwitchesAndFilters} is that, while an OEO signal converter transforms an optical signal to an electronic one, as a way to regenerate it, and then converts it back into an optical signal so that it is possible to activate a laser transmitter capable of emitting at the desired wavelength \cite{shen2018ultra,uzunidis2018dufinet, Transceivers2017}, an optical filter is a device that allows the selection and passage of a specific portion of the spectrum while blocking the unselected part~\cite{Ayoub:22, ayoub2018filterless}. A scheme of transceiver is shown in the Figure \ref{fig:transponder}.

The building blocks of these elements were synthesized from \cite{shen2018ultra,Transceivers2017,Ou:15}
and are shown in Figure \ref{fig:transponder}.
At the transmitter, the signal is processed in the Digital signal processing (DSP) in a process capable of improve its efficiency. This signal is converted to analog signals using four digital-to-analog converters (DAC) and modulated on the modulator drivers (DRV)  to avoid a degradation in quality. Then, the signal is modulated by optical modulator (OM), in-phase and quadrature components for example, and their components are combined by polarization beam combiner (PBC) to be transmitted in the optical domain. At the receiver, the polarization beam splitter (PBS) splits the signal in components that are sent to $90^{\circ}$ hybrids to power dividing. Then, the signal is amplified with transimpedance amplifiers (TIA) to guarantee stability while makes the signal able to be converted and digitizing by the which then passes through an Analog-to-Digital Converter (ADC). In the sequence, the signal is sent to DSP to adapt the expected data rate. Transponders are formed by some transceivers modules and are able to combining and convert wavelengths \cite{Transceivers2017,Ou:15}.

In terms of Tx and Rx, both the simplest and the most complex DSP blocks can be used, that is, both coherent (CT) and intensity modulation direct-detection (IMDD) transponders, are well accepted in metropolitan network architectures~\cite{shen2018ultra,uzunidis2018dufinet}. As the fiber optic infrastructure with long routes is expensive to implement, and since operators want to take full advantage of this infrastructure by sending as much information as possible about each fiber, CT allows for various signal encoding formats with special modulations that increase spectral efficiency through signal constellations with several bits per symbol. These characteristics lead to higher data transmission rates and greater transmission range, although they are more expensive \cite{fabrega2020cost}. 
IMDD solutions are low-cost technologies, generally used in network segments with a lower transmission range, however, they need dispersion compensation methods (DCM) to tolerate signal losses during transmission, due to the chromatic dispersion (CD) problem. This problem occurs because different wavelengths of light travel at different speeds in the optical fiber, which generates a mixture of lights \cite{kosmatos2019building}. 
The DuFiNet architecture project (to be presented in the subsection \ref{subsec:fullyFL}) implements Tx and Rx both CT and IMDD, and all are devices with integrated tunable filter (TF) \cite{uzunidis2018dufinet}.

Another variety of Tx and Rx that will be adopted within the scope of metropolitan networks are the Bandwidth Variable Transceivers (BVTs) \cite{MILADICTESIC2019464}. This type of equipment is responsible for ensuring transmission flexibility property due to its ability to dynamically operate the transmission bandwidth, rate, and range. To exploit all the available bandwidth, the BVTs must be operated up to a maximum value of transmission bandwidth, referring to a channel. A traffic request lower than the maximum transmission bandwidth of the BVT results in an operation with a transmission rate below its maximum capacity, resulting in the waste of a part of the available bandwidth. To overcome the problem of wasted bandwidth and provide a greater degree of flexibility, Sliceable Bandwidth Variable Transceivers (SBVTs) or multi-stream transponders have been proposed. SBVTs allow multiple optical streams to be sent to different destinations, including simultaneously, with pre-selected signal through various transmission parameters, such as modulation format, encoding, and transmission range \cite{sambo2017sliceable}. 
The metropolitan network architecture presented in \cite{larrabeiti2020upcoming}, appropriate to support $5G$ mobile services, it provides for the implantation of SBVTs in its infrastructure to allow dynamic configuration of optical paths, to provide low latency services, and dynamic restoration.
 An SBVT device has been developed through the PASSION project~\cite{boffi2020multi} with a capacity of up to $8$ $Tb/s$ of transmission, based on VCSEL technology (vertical-cavity surface-emitting laser) to aggregate all the data volume expected for the future of MAN (edge, $5G$, HD-TV). SBVTs can receive two different classifications according to the type of optical carrier source that is used in the architecture. They are classified as SBVT Multi-Laser (ML-SBVT), a model that uses more than a conventional laser, and SBVT Multi-wavelength (MW-SBVT), which uses a single laser as a source of several wavelengths to generate the several carriers. MW-SBVT are more advantageous as they do not require the use of a guard band between channels in the creation of super-channels, but ML-SBVTs have greater freedom since each laser can be adjusted independently to tune a given frequency in a specific slot without the carrier spacing is necessarily equal~\cite{masood2020smart}.
 Also highlighted in the literature is an SBVT model called centralized flexible optical carrier source module (Flex-OCSM)~\cite{imran2016FlexOCSM} which is based on a centralized controller, responsible for supplying optical carriers to all transponders of a given node, thus allowing the sharing of carriers between implanted SBVTs, which is not possible in the case of ML-SBVTs and MW-BVTs.
 
 \paragraph*{\textbf{Splitters, Couplers and Blockers}} \label{subsubsubsec:splitters&Couplers&Blockers}
Splitter,  Couplers/Combiners  and  blockers are passive elements commonly used in MON mainly due to the low cost and simplicity \cite{7792281UDWSN, Zhang2018ExploitingEO}. The optical splitter divide the optical signal to more than one fiber. The output power of the signal in each of the branched fibers will depend on the split signal rate for each output. Optical couplers combine two or more wavelengths in the same fiber or distribute the signal strength of one fiber over several other fibers \cite{uzunidis2021Bidirecional}. Blockers are used in conjunction with couplers and splitters to limit the passage of certain wavelengths, ensuring that only the necessary signal is transmitted \cite{Dochhan2019FlexibleMetro}.

\begin{figure}[H]
\centering
\includegraphics[width=0.6\linewidth]{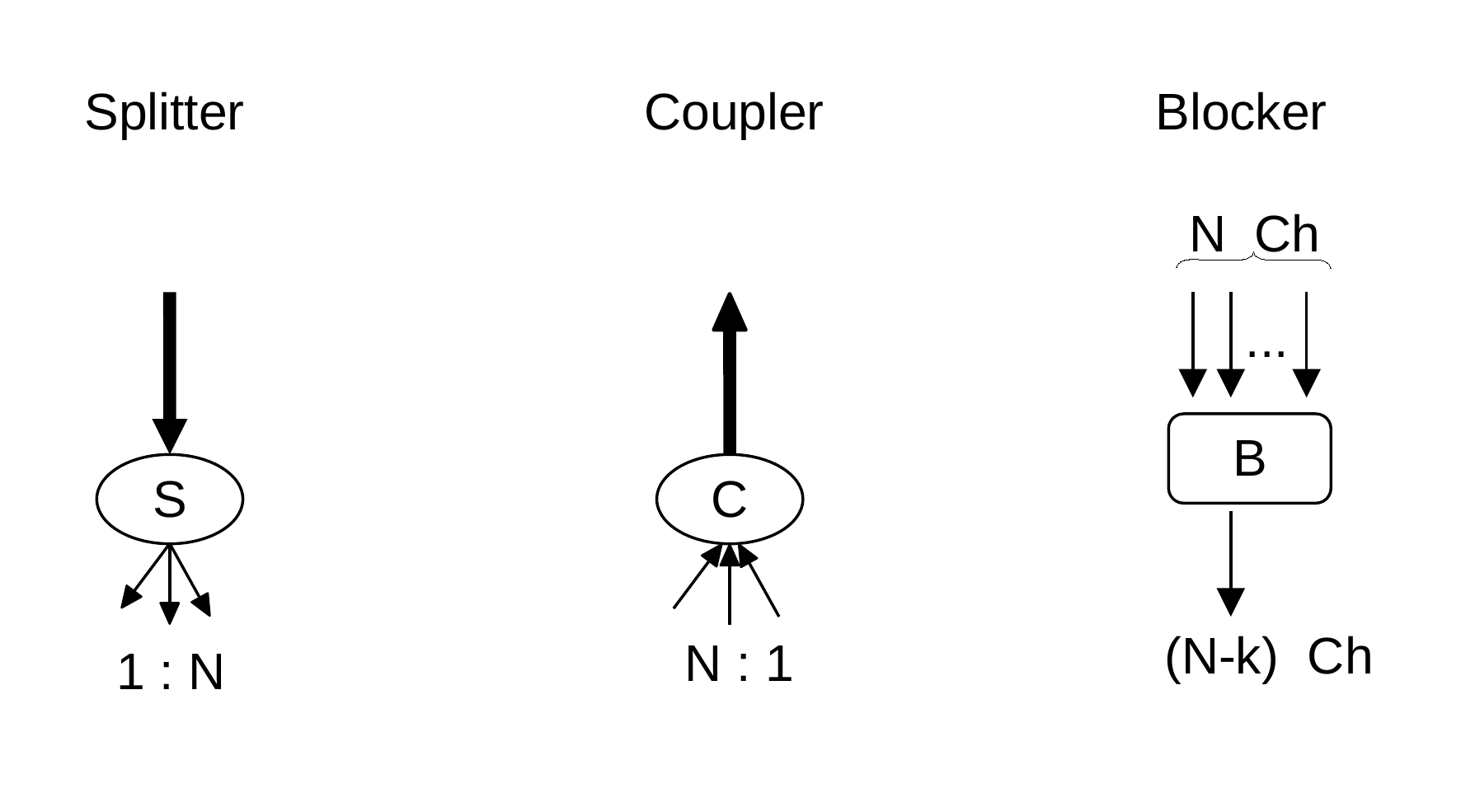}
\caption{Splitter, Coupler (or Combiner) and Blocker devices.}
	\label{fig:splitter_coupler}
\end{figure}

\begin{figure}
	\centering
	\includegraphics[page=1,width=0.99\linewidth]{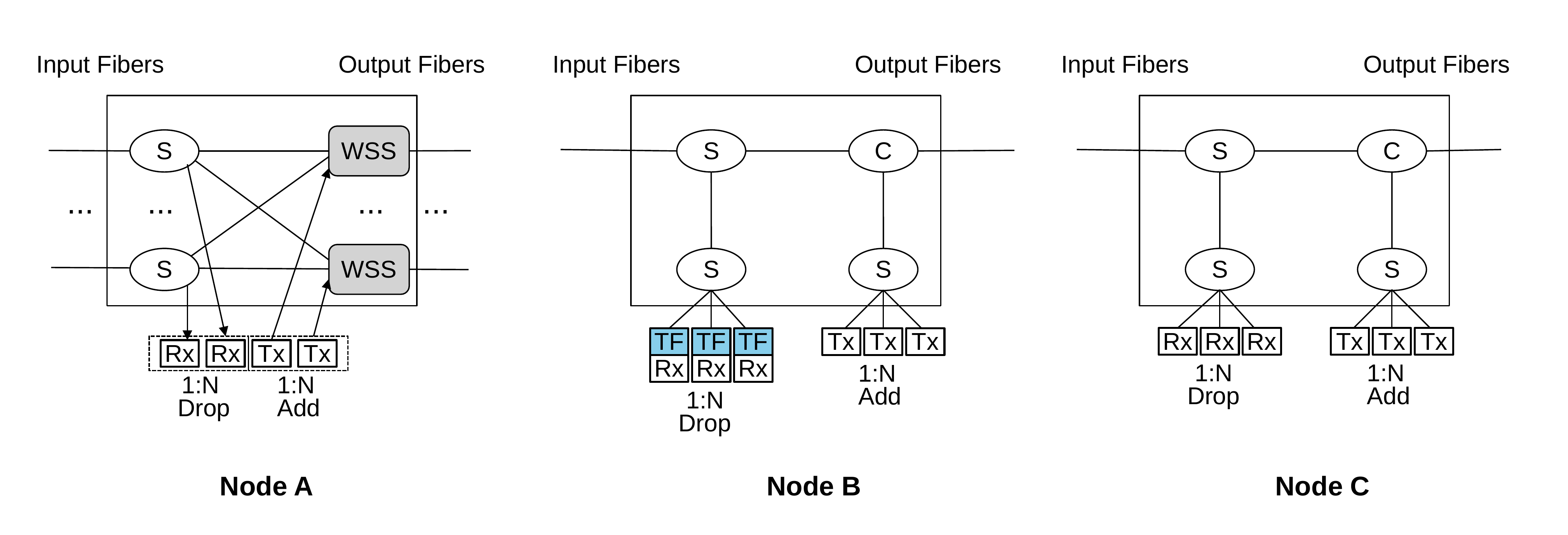}
	\caption{Optical node types: with filter (Node A), semi-filterless (Node B) and filterless (Node C)}
	\label{fig:nodes-model}
\end{figure}

When choosing these devices, it must take into account the type of operating bands and fiber mode that will be used, as single-mode fiber devices are different from multimode fiber devices. \cite{paolucci2020disaggregated}. An example of usage was shown in Figure \ref{fig:nodes-model}, as well as in more detail in Figure \ref{fig:splitter_coupler}. 

When couplers, splitters and blockers are used instead of filters, some side effects can be detected. Among them is the reduction in wavelength availability, the lower range of signals due to excessive noise and bit error rate (BER), and the accumulation of out-of-band noise generated by optical transmitters, and possible interference in adjacent channels \cite{paolucci2020disaggregated, 9042241, shen2018ultra, cugini2016receiver, 9041836}.

\paragraph*{\textbf{Switches and Filters}}\label{sec:SwitchesAndFilters}

In this topic, the main switching elements used in metropolitan optical networks will be highlighted, which are Optical Transport Network (OTN) and Optical add-drop multiplexer (OADM) swithes. The two main types of OADM that will be addressed are Fixed Optical Add/Drop Multiplexer (FOADM) and Reconfigurable Optical Add-Drop Multiplexer (ROADM). OADMS have important components that perform the selection of wavelengths, called filters. Thus, two types of filters used in switches will be highlighted, called Arrayed waveguide gratings (AWG) and Wavelength  Selective  Switch  (WSS).

\begin{figure}
	\centering
	\includegraphics[width=0.99\linewidth]{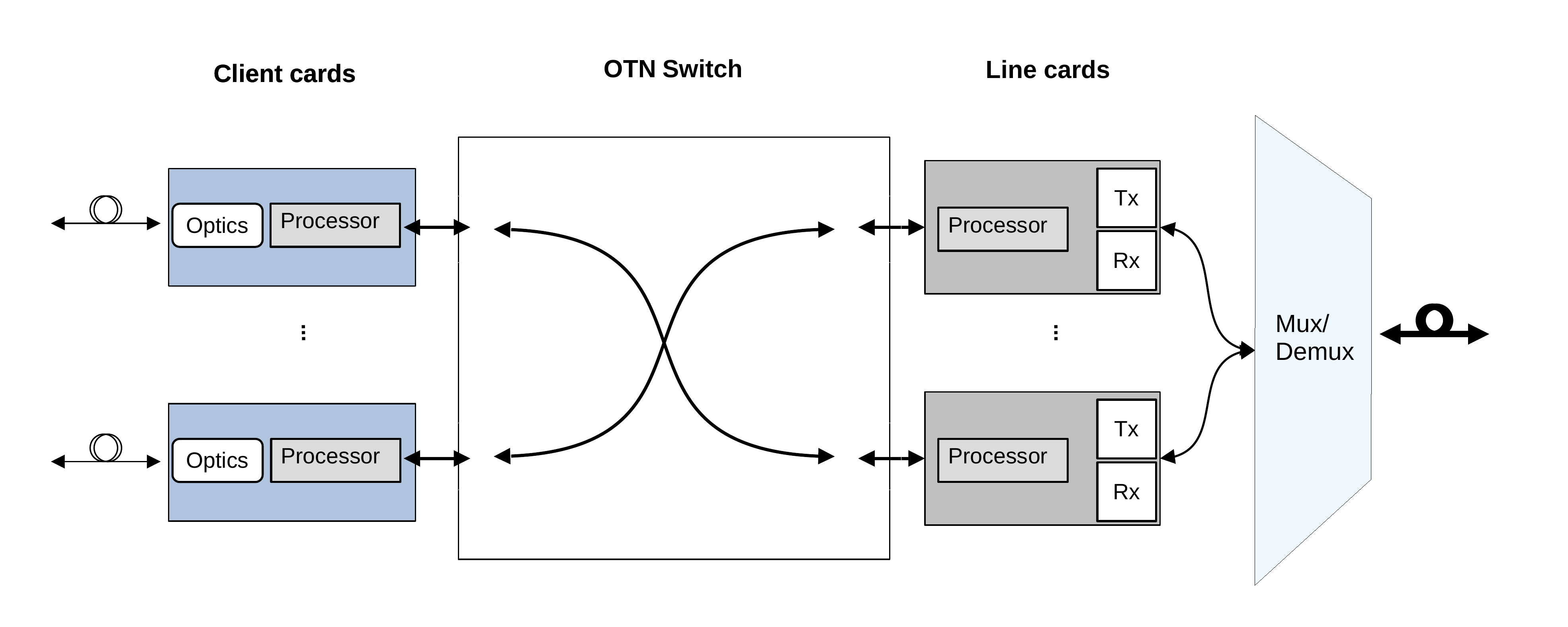}
	\caption{OTN nodes and their connections.}
	\label{fig:pureotn_node}
\end{figure}

From the physical point of view, the first MON architectures were built in a hybrid way, that is, with a mixture of the electronic devices and optical devices. In the nodes, electronic switches based on the OTN standard ITU-T $G.709$ with optical links based on WDM were used~\cite{katsalis2018towards}. 
This standardization defines the format of encapsulation, multiplexing, switching, management, supervision, and survivability of optical channels that carry the data flow.
Figures \ref{fig:pureotn_node} shows a schema of OTN nodes. Some client cards (on the client-side) and some line cards (on the optical line side) are shown as connection interfaces. Both on the client card and on the line card there is a processor that is responsible for managing data frame mapping functions. Traffic coming from the client-side is aggregated and mapped, on the respective client card, to an Optical Channel Payload Unit (OPU) frame. On the switch, the OPU frame is mapped to an Optical Channel Data Unit (ODU) container, which is then mapped to the Optical Channel Transport Unit (OTU) on the line card. In the line card, OTN standard performs digital encapsulation (Digital Wrapper) of various electronic data streams at wavelengths from each of its nodes and an OEO conversion procedure is required through transponders (Tx and Rx)~\cite{katsalis2018towards,MONIZ2019105608}. Transponders will be presented later in subsection \ref{subsec: transponders}. At the wavelength level, traffic is multiplexed to be transported by an optical path or demultiplexed to some lower-level ODU container. Multiplexing and demultiplexing is performed by optical filters, which will be highlighted later.  

An immediate technological advance for the implementation of OTN nodes in metropolitan networks was the introduction of more optical elements in the construction and establishment of the node. The node presented in the Figure \ref{fig:pureotn_node} is implemented based on stand-alone OTN node model since the OTN switch in the electronic layer and the optical layer elements are connected by short-range optical fiber, which generates many reverse fiber interconnections. This model evolved into the integrated node implementation model, in which the reverse fiber interconnections are removed and optical interfaces (Mux/Demux) are coupled in the same chassis as the OTN system that provides the switching functionality~\cite{DBLP:journals/corr/abs-1901-04301,da2019otn}. In WDM systems the device used as multiplexer/demultiplexer is the OADM.

Since the second generation of optical networks arrived, metropolitan networks have left the hybrid format and have become completely optical. The routing and optical switching nodes have been carried out by two types of OADM, named FOADM and ROADM. ROADM is also called by optical cross-connects (OXCs). Since OXC is a more generic way to refer to an optical switching device, both forms, ROADMs / OXC are often used interchangeably in the literature \cite{rottondi2013routing, 9041836}.

\begin{figure}
	\centering
	\includegraphics[width=0.8\linewidth,page=2]{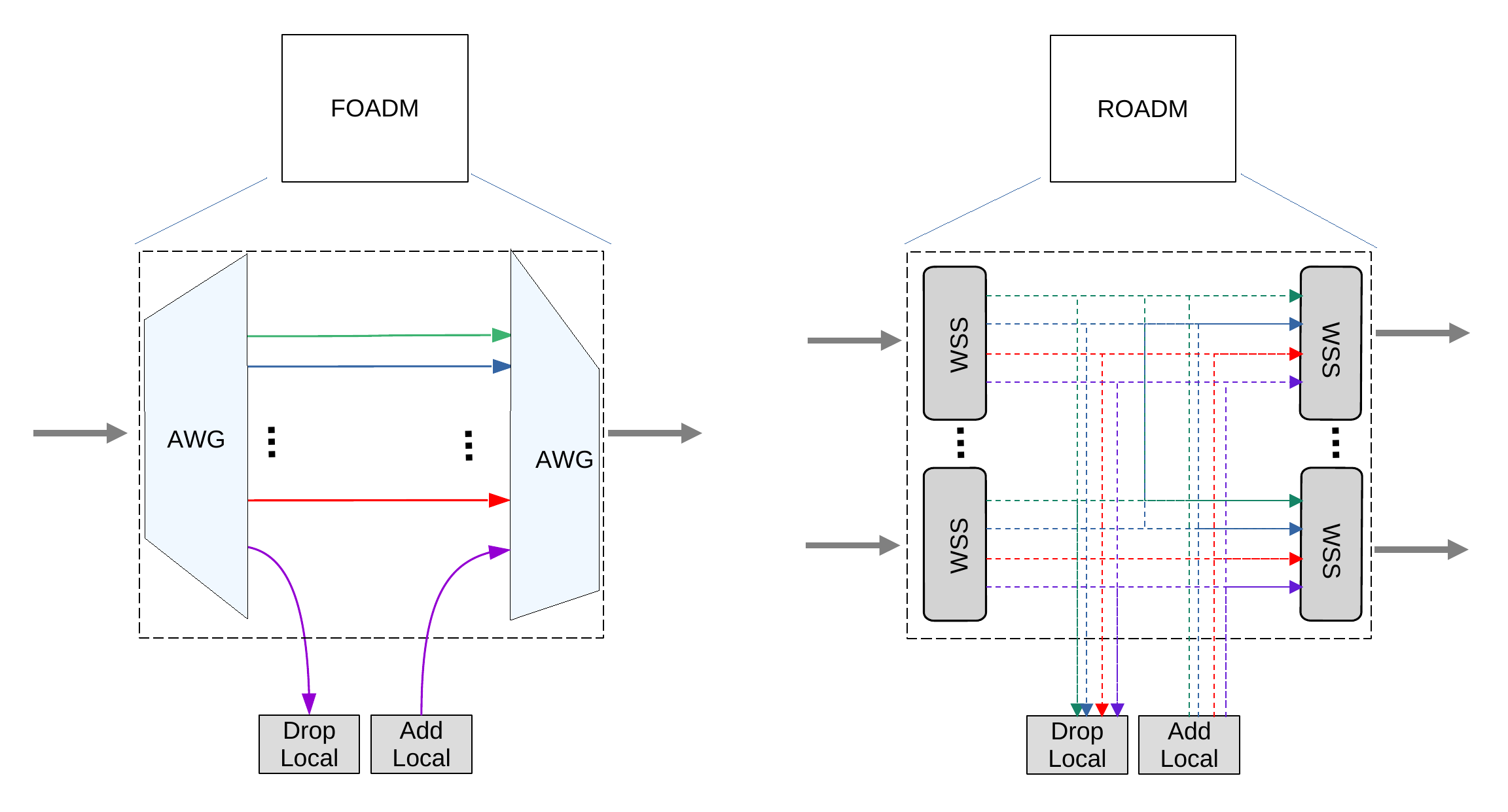}
	\caption{Simplified scheme of FOADM and ROADM.}
	\label{fig:foadm_roadm}
\end{figure}

While the main building block of ROADMs is the  WSS, the FOADM can be based on AWG \cite{nadal2019sdn}. Both are optical filters \cite{ji2019all}, but AWG~\cite{lin2019three,zervos2019new} can be used as an alternative to reduce network costs \cite{8025162OTN}. Figure \ref{fig:foadm_roadm} shows a simplified vision of both. FOADM allows one wavelength to be removed locally and reused, as well as allowing the same wavelength to be added to be transported in the opposite direction. ROADM allows any individual wavelengths, or multiples wavelength, to be redirected to other location, added and/or dropped at a location, and to adjust or change the add/drop setting if traffic changes occur.

FOADM allows static allocation of wavelengths, ROADM allows dynamic allocation. The configuration of FOADM is carried out manually in the place where the hardware is implanted, and the mappings of input and output ports are established for a considerable time. ROADM, being more flexible, allows hardware configurations to be made via software remotely. Also, with dynamic ROADMs, many other features can be implemented, including protection and restoration on the optical layer \cite{9042241}. In ecosystems with an EON transmission system, ROADM or OXC nodes are generally referred to as bandwidth-variable OXC (BV-OXC) \cite{rottondi2013routing} or BVXC \cite{8853968}

\begin{figure}
	\centering
	\includegraphics[width=0.8\linewidth]{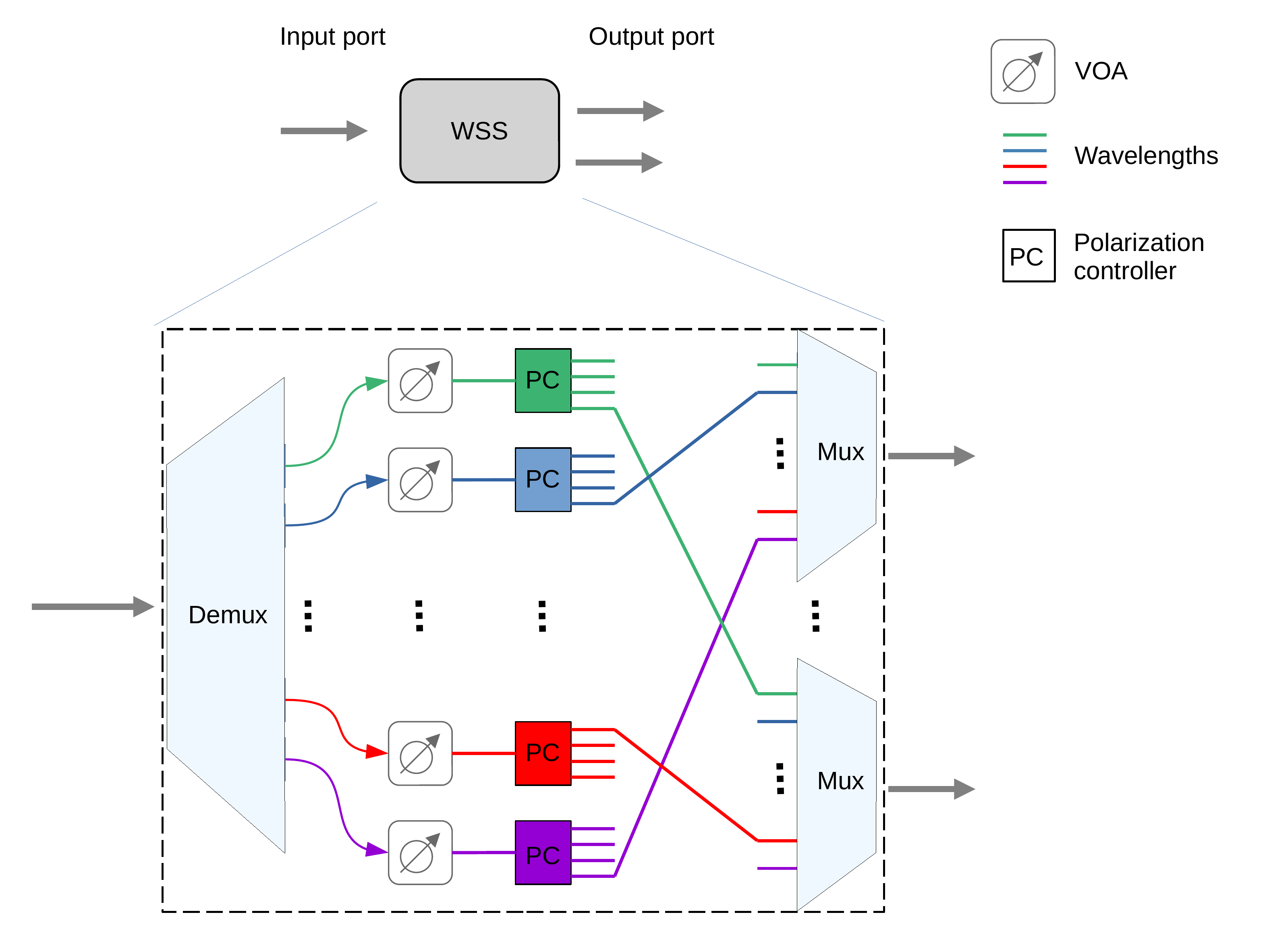}
	\caption{Simplified scheme of a wavelength selective switch (WSS). Featured is shown an enlarged view of the inner part.}
	\label{fig:WSS_vision}
\end{figure}

A schematic showing the composition of the WSS is highlighted in Figure \ref{fig:WSS_vision}. WSS receives all wavelength fiber through the common input port. The wavelengths are separated, routed to a variable optical attenuator (VOA) for equalizing the wavelengths power. After that, the wavelengths go through a polarization controller that redirects them to the respective output port, where they are multiplexed before being attached to the output fiber \cite{uzunidis2021Bidirecional}. Various others technologies are used in the construction of WSS. To meet EON, for example, WSS based on liquid crystal on silicon (LCOS) provides $12.5$ $GHz$ granularity~\cite{shiraki2020highly}

Based on the presence of filters in the composition of a switching node, this node can be said to be with filter (WF), semi filterless (sFL) or filterless (FL) architecture. These concepts are exemplified in the Figure \ref{fig:nodes-model} through $3$ nodes. They are identified as Node A for WF architecture, Node B for the sFL architecture, and Node C for the FL architecture. 
Input and output fibers provides connectivity from/toward other ROADMs while Add/Drop interfaces provides connectivity between transmitters with source/destination at the local node.
Node A is a ROADM with Broadcast-and-Select (BnS) architecture, composed of $N$ passive splitters (S) that connect the $N$ input fibers (left side of the figure), which direct the signal to each of the $N$ WSS filters on the right side of the figure. In this way, a signal that enters the node from an input fiber of the network can be sent to several output fibers. In this example, S is performing the transmission, while the WSS is selecting or filtering the signal to optimize performance. Although the passive splitter transmits each received signal, it does not normally divide power evenly between the WSS ports.

Node B has sFL architecture \cite{ayoub2018filterless}. In this case, the signals are broadcast in all passive element outputs, such as optical couplers that combine the signals from the other nodes and the signals added locally from the signal addition port (Add) where the Txs are located, while the Rx corresponds to a signal removal port (Drop) for pre-defined channels. Although there is no implanted WSS, a type of tunable filter (TF) of low complexity is integrated into the transponder, used to attenuate with limitations the effects of general cascading of the node. This filter model is simpler, generally referred to as "lite", and is responsible for making each Rx viable for medium distances, so that digital signal processing functions are adjusted for cost reduction and energy consumption~\cite{cugini2018flexibleSemiFL}.

Node C, of FL architecture \cite{ayoub2018filterless}, is built using a coupler, which only divides the signal into multiple outputs, consequently causing a reduction in power~\cite{shariati2020real}. Both Node B and Node C have Drop and Waste (DnW) architectures \cite{paolucci2020disaggregated}. 
In this architecture, the removed signal propagates in such a way as to occupy the entire spectrum in the direction posterior to the location of the removal stage (drop). Network architectures that employ these two types of nodes usually have ROADMs at the ends. Thus, for transmission, the signals propagate until they reach the ROADMs, but some optical signal may be lost or become useless.

As the architecture of nodes gets simpler due to the elimination of some deployed elements, its CAPEX and OPEX also tend to be reduced. This is due to the fact that the elements eliminated are active equipment, with higher acquisition cost and responsible for the high power consumption \cite{paolucci2020disaggregated, ayoub2018filterless}.

As it is more important to maintain the signal integrity of the passing traffic, to allow it to continue to the next node on its path, only a small part of the input signal strength is directed to each WSS port, with the rest directed to the network fibers.
In addition, there is usually amplification at a node, to help mitigate the loss of division. Several viable ROADM and WSS architectures for metropolitan optical networks are presented in \cite{liu2019space, 9041836} and \cite{zervos2019new}. 
Some of these architectures alter the layout of the devices on the left and right side of the figure, as is the case with the Route-and-Couple (RnC) architecture, which can be understood as the reverse of the BnS architecture, with WSSs in the left side position and S positioned on the right.
Many other configurations with WSS are possible since this is the equipment responsible for defining the degree of the ROADM node.

In the context of flexible networks, ROADMs are implemented using other specific optical components, such as Selective Spectrum Switches (SSSs), Multicast Switches (MCSs), and different transponder configurations. However, the selection and combination of these components are based on flexibility and performance requirements.
MCSs are deployed in Colorless, Directionless and Contentionless Reconfigurable Optical Add-Drop Multiplexers (CDC-ROADMs). In CDC-ROADM any wavelength of the input port can be switched to any output port, and this is because the MCS provides the copy of the signal. The MCS can be seen as a kind of NxM WSSs due to the number of connections, but it is composed of a series of splitters/couplers. Other variations with less degree of flexibility are Colorless ROADM (C-ROADM) and Colorless and Directionless ROADM (CD-ROADM) \cite{gangopadhyay20195g}. WSS can, without making copies of the signal, filter arbitrary bandwidth in the form of a frequency spectrum slot from an input port to a given output port, which allows flexibility for EON and SDM operations~\cite{sambo2017sliceable}.

The traffic volume characteristic of metropolitan networks requires several fibers between adjacent nodes and ROADMs / OXCs, which leads to a much larger number of ports than the equipment used at the backbone network level \cite{9042241}. For the future, ROADMs capable of operating in the optical spectrum without grids are expected \cite{Handbook_of_ON}. In addition, there are initiatives that seek to create a customized environment with optical elements from various manufacturers, including ROADMs. Cross-operation with maximum coupling between the parties is possible breaking down the barrier of proprietary software within this equipment, and implementing open platforms based on agreements between operators and industries. The Open ROADM Project \cite{openROADM} is a example of this iniciative.

%\paragraph*{\textbf{Transponders}}

% Léia, creio que na prática apenas as arquiteturas que fazem comutação eletrônica (OTN) usam diretamente Transceivers, no mais creio que usam transponders mesmo. VC só precisa deixar claro no início desta seção a diferença entre Transceiver e Transponder. Por exemplo, a Fig 7 é um exemplo de um Transceiver e não Transponder, seria melhor (creio eu) vc mostrar na figura um Transponder. Vejam como referência este link https://community.fs.com/blog/transceiver-vs-transponder-what-are-the-differences.html

%In both cases, FOADM and ROADM, traffic that enters through the arrival port can be routed without OEO conversion to fiber at the exit port. 

\paragraph*{\textbf{Amplifiers}} \label{subsec: transponders}

\begin{figure}
	\centering
	\includegraphics[width=0.99\linewidth]{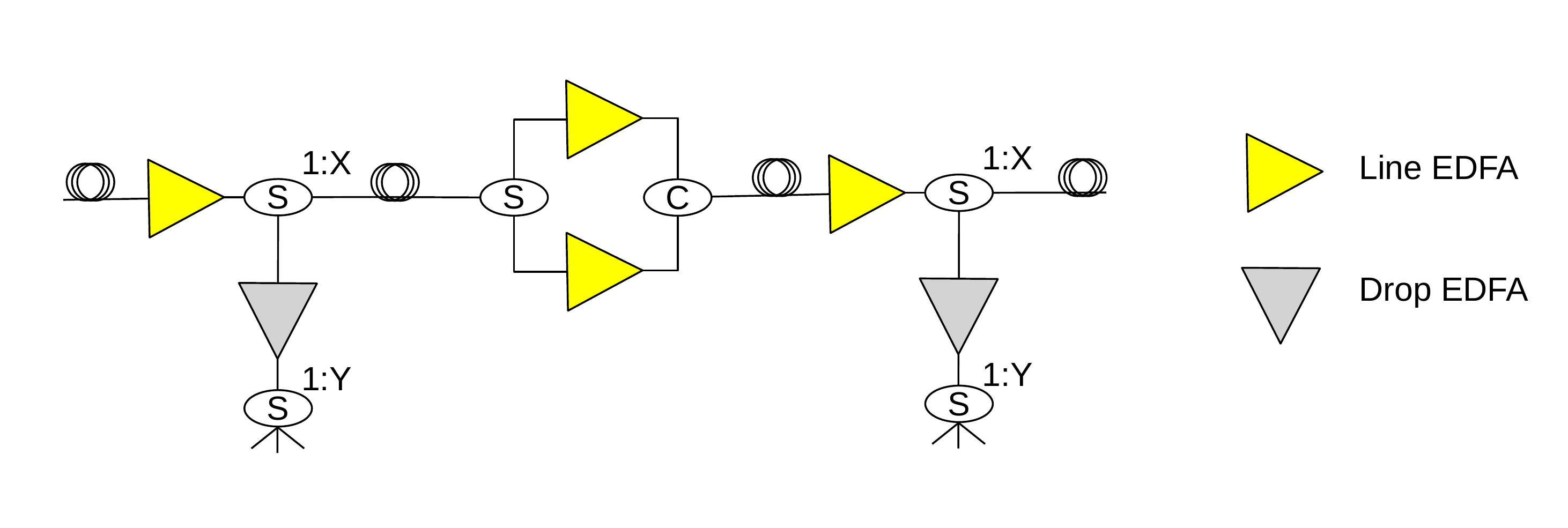}
	\caption{Amplification options in optical network}
	\label{fig:edfa}
\end{figure}

Amplifiers can be positioned on the input and output fibers to minimize the effects of end-to-end power loss due to cascading of multiple links when an optical path is established \cite{ayoub2019routing,ruffini2017access}.
Generally, metropolitan optical networks do not necessarily need to implement optical amplifiers or regenerators due to the average distance of the links being less than or equal to the signal span \cite{zheng2017metropolitan}. However, some projects foresee the use of erbium-doped fiber amplifier (EDFA), specifically when the C band is adopted by the architecture, still or linear semiconductor optical amplifier
(SOA), in multiband architecture \cite{ruffini2017access}. 
Two main types of EDFA amplifiers are adopted in networks without a filter in the C band, as shown in Figure \ref{fig:edfa}: 
Line EDFA and Drop EDFA. An Line EDFA has the function of compensating for the optical power losses that occur when the signal passes through the long-distance optical fiber. Drop EDFA is useful to compensate for the losses generated by the demultiplexer/disaggregator located near the optical receiver so that the level of the optical signal is improved before detection. The drop splitter present is capable of dividing the power between the local matrix of coherent receivers.

In FL network architectures and in some specific topologies, more than one amplifier may be required per node. In this case, they are used inline, over limited distances without the presence of repeaters, and a drop amplifier, which in the existence of a fiber branch, divides the signal strength between the various branches, each tuned at a given frequency \cite{paolucci2020disaggregated}. In addition to assessing the need for amplification or selecting the ideal technology for signal amplification in the MON architecture, another major challenge is to identify the network points for the implementation of these elements \cite{ayoub2019routing}.

\paragraph*{\textbf{Fibers}}

In terms of the spectral band, the MON is generally based on the C band and L band (multiband systems), operated in single-mode fiber (SMF) \cite{paolucci2020disaggregated,boffi2020multi,mitra2019effect}. A typical fiber has a capacity of at least $50$ $Tb/s$, which is much higher than the maximum electronic speed.
SMF has a lower occurrence of interference between transmission wavelengths, however, the increase in capacity in this fiber is limited due to the nonlinear limit of Shannon, with metropolitan networks already operating at about three times more than this limit~\cite{singleCarrier2016,Winzer2014SpatialMI}.
An alternative to increase the bandwidth and consequently the transmission capacity in metro networks is the implementation of Multi-core fiber (MCF), which has several independent cores in which the light propagates in the creation of optical paths, although these benefits come at the cost of the need to implement more complex and more expensive equipment.
Other types of fiber considered are Multi-Mode Fiber (MMF) and Few-Mode Fiber (FMF), both allowing different and independent modes of signal propagation in the core, being low cost and viable for short distances. In architectural network designs that include computing at the edge, these connection solutions can be employed to reduce the cost of the project in the access segment. Despite this, non-linear effects between the different modes can occur, and the increase in the number of modes results in increased latency in the DSP \cite{liu2019space}. Optical communication systems that exploit the capabilities of the various cores or the many modes are called multiple-input multiple-output (MIMO) \cite{Winzer2014SpatialMI}.

MIMO systems require filters, and for this reason, the ROADMs used in these ecosystems have Photonic Space Switch Matrix (PSM) modules, built based on WSS to connect all types of fibers. PSMs handle the traffic that arrives from other nodes or is sent to other nodes using MCF fibers. Local traffic, added or removed from the node itself, is operated using a Multi-Cast Switch (MCS) before being forwarded to the PSM \cite{calabretta2019photonic}.

\paragraph*{\textbf{OLTs and ONUs}}\label{subsubsec:PON}
Finally, in order to make this list of equipment complete, passive optical networks (PON) need to be considered, since there are MON architectures that includes the access layer. PON is an access network technology mainly based on passive components. It provide low cost of implementation and operation and provide more reliability given that it has no electronic devices in the field \cite{4150568}. PONs are built with Optical Line Terminals (OLT), splitters and Optical Network Units (ONU). 

The Central Office (CO) in a metropolitan network is capable of providing internet access to the access network segment. In COs, the OLT is a passive device that manages and distributes the optical signal at the provider. The optical signal that leaves the OLT is routed to a splitter, where it can be divided into different proportions, in accordance with the operator's services. Each signal resulting from this division can be routed to a ONU, destined for the subscriber \cite{muciaccia2019proposal, thyagaturu2016software}.

The ONU is a passive device that converts the optical signal received by the OLT terminal into an electrical signal to be distributed to the subscriber \cite{9145535}.
The signal to be converted is normally sent to the ONU from a AWG, hosted in some remote node (RN). The remote node can be a local internet provider. Different wavelength channels are defined for downstream and upstream traffic. On the subscriber side, each ONU can send data to an OLT where frame synchronization takes place \cite{lin2019three}.

\subsection{Metropolitan Optical Networks (MON) Topologies}\label{subsec:topologies}

\begin{figure}
	\centering
	\includegraphics[width=0.8\linewidth]{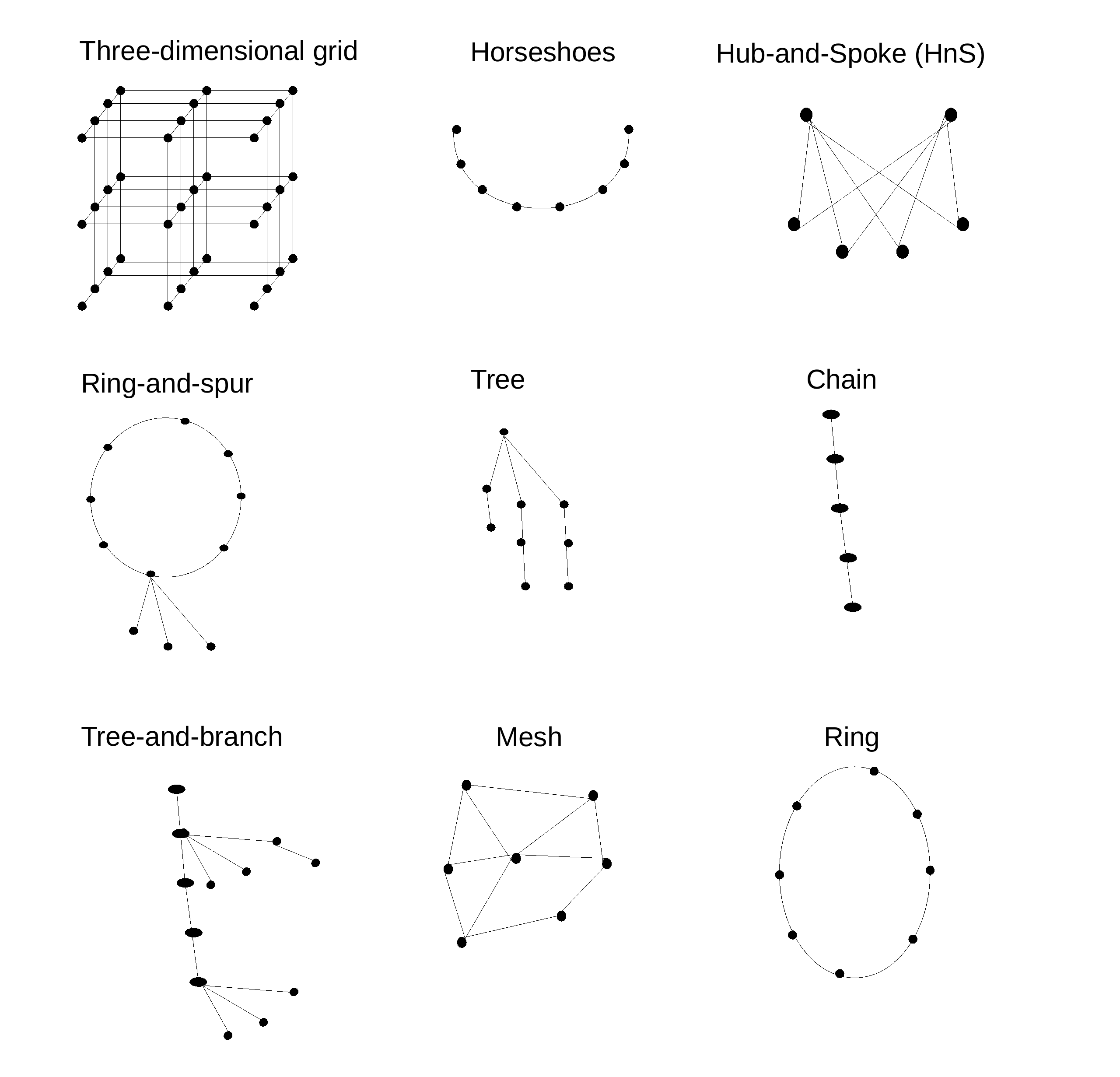}
	\caption{Example of types of network topologies.}
	\label{fig:topologies}
\end{figure}

In the Figure \ref{fig:hierarquia} it is also possible to observe some characteristics regarding the network topology. While in core and metropolitan networks, mesh and ring topologies are common, which are relatively more complex, in access networks it is common to organize nodes in chain, star, or tree topology. Some types of topologies that will be highlighted in this section are listed in Figure \ref{fig:topologies}.

These topologies considered more complex provide greater connectivity, a greater number of routes, and consequently more resources, which make it possible to reduce the latency for service provision. The reason they are considered complex is due to the physical performance limitations that impact the transmission of optical signals over greater distances, leading to the need to implement more equipment on the network to treat the signal and improve the transmission quality. Another direct reflection of the more complex topologies is the high cost of planning and operation, both due to the cost of the equipment and the energy consumption of this equipment when in operation. The ring model widely adopted in the past, today contrasts with the trend of implantation of the mesh topology \cite{ramachandran2019capacity}. 
The main reason for such changes is the need to adapt the infrastructure to deal with the increase in traffic and the provision of new types of services, such as artificial intelligence applications and 5G service, for example, and with the increasing number of subscribers as well as the data rate of these subscribers in the access networks \cite{yan2020area}. Recently, the three-dimensional grid topology for metro-access tracking was presented in \cite{lin2019three}. 
In this type of network topology, each node connects to two neighboring nodes along one or more dimensions. For reasons of reducing the cost of implantation and operation, the existence of chain and semi-ring or broken ring topologies, in particular, horseshoe, as is the case of the Croatian Telecom \cite{cugini2016receiver}. 

However, physical topology is not the only issue to be discussed among architectural trends for metropolitan networks. With an eye on the strategic functionality for the traffic of new services, the literature has shown that the metro and access segments have been planned together \cite{7792281UDWSN, metro-haulproject, kretsis2020armonia}. 
The Hub-and-Spoke (HnS) logical topology \cite{9145535} is a proposal in which some nodes are designated as a hub, and works as a communication vehicle and aggregator between the other nodes, which are then designated as spoke. In this case, the hub nodes are represented by metro-aggregation nodes that associate the spoke nodes represented as access nodes. The main advantage in adopting a logical topology is that, regardless of the physical topology, new policies and reconfigurations of operation can be implemented without the need to add new equipment to the underlying network infrastructure. From the point of view of traffic and services, the direct flow of communication between access nodes and nodes in the metropolitan network occurs with low latency, even though the physical topology of the metropolitan network is meshed.
However, this advantage is also a constraint regarding the scalability of the logical topology. Likewise, failures in the physical infrastructure can compromise the operation of the logical topology, leading to failures in communication.

These trends for new network topologies also drive the evolution of access networks. In the past, the most common topologies were tree and chain topologies (Figure \ref{fig:hierarquia}). Recently, the ring topology, Ring-and-spur, has been proposed for distances greater than $100$ $km$ in access networks \cite{abbas2013feeder}, whereas the tree-and-branch topology, previously considered a more inefficient topology because it requires many amplifiers and splitters, today, it can be thought of as a new way to take advantage of the greatest amount of available optical channels, since EON technology has led to reducing the spacing of the optical spectrum grid. These topologies are an important decision when choosing which network architecture to adopt. While tree topologies or chains, more traditional topologies, require amplifiers, splitters, and optical combiner to extend the tree, mesh topologies require photonic switching elements, such as ROADMs or OXCs to tune the transmission channel potential, and extensive bandwidth management. Although horseshoe and chain topologies look similar visually, structurally, when implemented, they can be different. This is because, in general, network architectures in horseshoe format have ROADMs at both ends, while chained topologies may have ROADMs at only one end. Chain topologies are common in the access networks segment \cite{ruffini2017access} while horseshoe topologies are mainly highlighted in the MC and MA segment \cite{kosmatos2019building}.

In \cite{Ayoub:22, nooruzzaman2015filterless}, the main restrictions for using filterless architectures are highlighted, for example: (\textit{i}) is the range of the transmission system used that will decide the maximum distance between the root and a leaf, in the case of tree topology, or the maximum distance between the two ends in the case of chain topologies; (\textit{ii}) as the transmission is of the broadcast-and-select type among all nodes, the capacity of the system limits the maximum number of wavelengths per fiber; (\textit{iii}) wavelength reuse limit in the “drop-and-continue” architecture of the line system, which leads to the accumulation of Amplified Spontaneous Emission (ASE) and creates unfiltered channels. As these problems are more present in long-distance systems, the literature has shown that in the metropolitan network environment, and with the implementation of coherent technology, it has been possible to achieve more efficient and low-cost solutions.

Many of the new architectures recently presented in the literature were thought of as a way to keep the cost reduced and improve the performance of the system as a whole. With the various types of services and applications that are being offered on metro networks, the trend is that the number of MA nodes will increase considerably to provide greater network capillarity to the various connections for end-users, as well as the massive migration of network structures. DCs for MC nodes in the metro segment, being offered in the form of fog or mDC \cite{thyagaturu2016software}.

Thus, the next generations of metro networks need new solutions to deal with the new requirements, and such solutions can be investigated both in the field of network engineering and in the field of traffic engineering. In the field of network engineering, new network architectures are proposed, which will be highlighted below, as well as new equipment capable of providing greater bandwidth capacity, with consistent technology and efficient use of the optical spectrum. The most recent metro network architecture suggestions will be highlighted in the Section \ref{sec:arquitetura}.

%%%%%%%%%%%%%%%%%%%%%%%%%%%%%%%%%%%%%%%%%%%%%%%%%%%%%%%%%%%%%%
%%%%%%%%%%%%%%%%%%%%%%%%%%%%%%%%%%%%%%%%%%%%%%%%%%%%%%%%%%%%%%
%%%%%%%%%%%%%%%%%%%%%%%%%%%%%%%%%%%%%%%%%%%%%%%%%%%%%%%%%%%%%%
\section{Metropolitan Optical Networks (MON) Architectures}\label{sec:arquitetura}
This section discusses some architectures for metropolitan optical networks identified in the optical transport networks literature. This work proposes a classification of the architectures presented for a further systematic discussion about transport technology trends in the next section.

In subsection \ref{subsub:PhysicalPerspective}, several components for the construction of networks were presented, and in subsection \ref{subsubsec:perspLogica} the roles of nodes in these networks were highlighted. This information will be used in this section to present the MON architectures.

\begin{figure}
	\centering
	\includegraphics[width=0.99\linewidth,page=2]{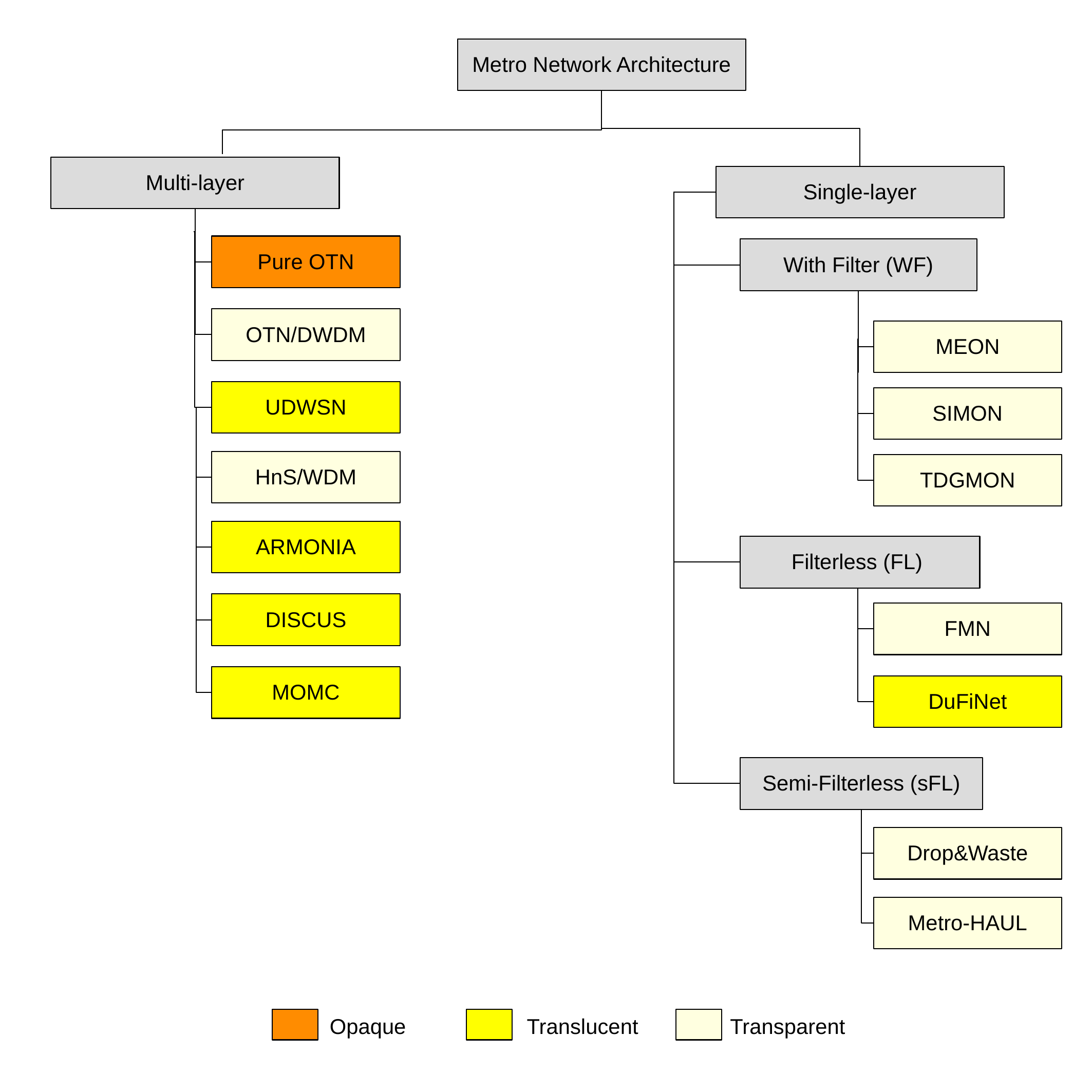}
	\caption{Classification of metro network architectures according to the respective required OEO conversion levels.}
	\label{fig:conversion}
\end{figure}

As shown in Figure \ref{fig:conversion}, the highlighted architectures were classified into two main classes, being multilayer and single layer, as presented in the reference scientific literature \cite{Handbook_of_ON}. To delimit the discussion, the architectures based on OTN and optical layer will be considered multilayer architectures, and the single layer architectures considered will be exclusively optical architectures.
Another existing classification is the electronic single-layer architecture or metro-Ethernet, that are based on exclusively electronic transmission and switching, in which the connections are Point to Point (P2P) and whose nodes are switches or routers. This type of networks are not the object of study in this work because they belong to the first generations of communication networks, being today only designated as legacy technology and not representing an alternative for the future \cite{Handbook_of_ON}. Nowadays, the use of these networks is generally intended for limited segments of the network \cite{katsalis2018towards}. 

Concerning single-layer architectures, a subclassification is made regarding the potential use of filters among their hardware equipment (also called degree of filtration). In this way, network architectures WF, FL, and sFL are identified. In Subsection\ref{subsec:equipaments} concepts were presented for WF, FL, and sFL nodes. WF architectures are composed of WF nodes. However, FL architectures are generally composed of a greater proportion of FL type nodes in the middle, and at most two WF nodes at their ends. As for the sFL architectures, as will be shown, they can be constituted in two ways: (\textit{i}) several nodes of type WF and type FL, or (\textit{ii}) a greater proportion of nodes of type sFL and, at most two WF nodes. A MON architecture requires at least some filtering nodes due to the considerable high number of nodes, distances and transmission quality that are common in the network segment that bridges the access networks and the core networks.

Due to the breadth of the subject of Section \ref{sec:arquitetura}, useful classification for conceptual comparison were formalized based on the (\textit{i}) number of layers considered; (\textit{ii}) existence or not of optical filters; and (\textit{iii}) degree of implementation of signal converters.

According to the attribute (\textit{i}), networks are classified as multi-layer architecture, with the optical and electrical domains, and as single-layer when only the optical layer domain is considered. Multi-layer networks will be highlighted in Subsection \ref{subsec:MultiLayerArchitecture} and single-layer will be discussed in the Subsection \ref{subsec:SingleLayerArchitectures}. 

By attribute (\textit{ii}), single-layer architectures were classified as type WF (Subsection \ref{subsubsec:MON_WF}), FL (Subsection \ref{subsec:fullyFL}) or sFL (Subsection \ref{subsubsec:metroSemiFL}). This classification was not performed in multi-layer architectures, since all multi-layer architectures implement filters. 

Regarding the attribute (\textit{iii}), an optical network can be classified as being transparent, opaque, or translucent. This classification is important because it demonstrates how optical communication is carried out in these networks. While in opaque networks fiber is used only as a robust transmission channel and all switching work is carried out in the electronic domain, in transparent networks all functions are performed exclusively in the optical medium. Translucent nets have opacity and transparency points. The following will be highlighted for each architecture.

\subsection{Multi-layer Architectures}\label{subsec:MultiLayerArchitecture}

The classification of multi-layer architecture fits the presented network architectures whose transport layer is multilayered, composed of electronic layer over optical layer. Architectures in this class are Pure OTN~\cite{katsalis2018towards} presented on the Subsection \ref{subsubsec:PureOTN}, OTN over DWDM (OTN/DWDM)~\cite{da2019otn} cited in Subsection \ref{subsec:OTN-WDM}, UDWSN~\cite{7792281UDWSN} in Subsection \ref{subsubsec:UDWSN}, in order of appearance in the literature, in addition to the HnS/WDM~\cite{she2017metro} in Subsection \ref{subsec:HnS_WDM}, ARMONIA~\cite{kretsis2020armonia} in Subsection \ref{subsubsec:ARMONIA}, DISCUS~\cite{ruffini2017access} in Subsection \ref{subsec:DiSCUSarchitecture} and MOMC~\cite{larrabeiti2019all} in Subsection \ref{subsubsec:MOMC}, that are not OTN based.

%This same order of citation for OTN-based architectures also indicates the proportion of existing OTN switches (Não entendi essa frase???): OK
While pure OTN and OTN/DWDM have electronic switches on all nodes, at UDWSN only a few optical nodes are defined for switch deployment overlapping OTN. Because they have this characteristic in common, these three architectures are compared in the literature, as will be highlighted later. Also, pure OTN and OTN/DWDM have been commercially available architectures for some time, while UDWSN was recently proposed and is not yet expected to be commercially available. In this way, it can be inferred that the trend for the future is to reduce the use of OTN in the transport layer and still push it to the edge of the network, at communication points that reduce the number of hops for communication, for example, the network edge at the front-haul layer of mobile networks. Another possible trend is the narrowing of the communication channels, with flexible grid and end-to-end communication. These alternatives can contribute to the reduction of latency in the establishment of optical paths, making the technology able to meet the requirements of the new services available at the edge of the network.

\subsubsection{Pure OTN}\label{subsubsec:PureOTN}

\paragraph*{\textbf{Composition and functioning}}

\begin{figure}
	\centering
	\includegraphics[width=0.6\linewidth,page=2]{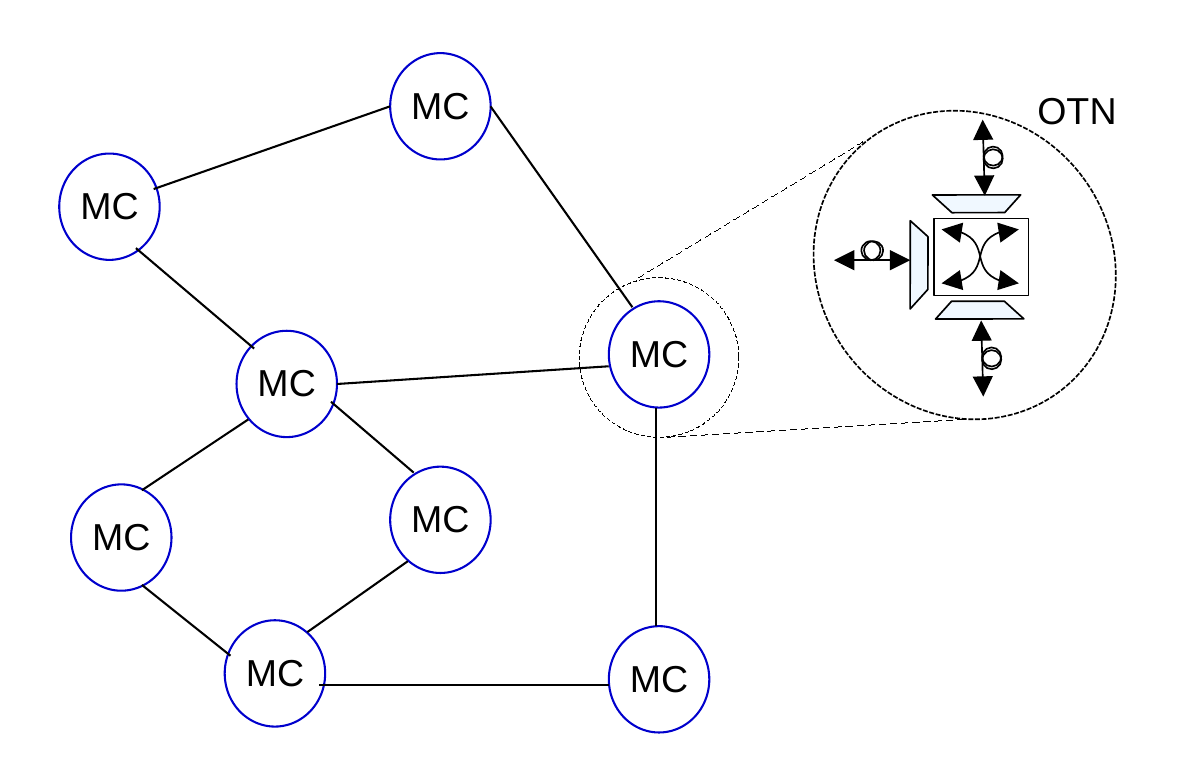}
	\caption{Network architecture of OTN}
	\label{fig:pureotn1}
\end{figure}

Metropolitan networks with pure Optical Transport Network (OTN) architecture, standardized by ITU-T $G.709$ \cite{katsalis2018towards}, have electronic OTN nodes, and their links and interfaces are optical, and the communication is point-to-point. Versatile, it allows the construction of several different topologies, from chain to mesh \cite{Handbook_of_ON}. The Figure \ref{fig:pureotn1} shows a representation of pure OTN architecture with its metro-core segment (Subsection \ref{subsubsec:perspLogica}) and mesh topology (Subsection \ref{subsec:topologies}). The detail on the right shows the physical composition of the node based on the elements presented in Subsection \ref{subsub:PhysicalPerspective}
\paragraph*{\textbf{Advantages}}
Pure OTN architecture is widely used in legacy networks due to the easy of updating and low cost of implementation while ensuring aggregation of traffic in the optical domain and reducing waste of wavelengths. For this reason, although it is one of the oldest architectures in use, it is still a viable solution alternative for the metropolitan network as it allows multi-service at low rates, such as sub-$10$ $Gb/s$~\cite{infinera2020cost}. Pure OTN is cited in several recent works \cite{shen2018ultra, 8025162OTN,Zhang2018ExploitingEO} for comparison with the most current technologies and which will be discussed later in the next sections. Another great advantage of this architecture is the facility for virtualizing operations using a software-defined network (SDN) and supporting new types of traffic of varying granularities due to its encapsulation property~\cite{8025162OTN}.

\paragraph*{\textbf{Disadvantages}}

However, pure OTN is still one of the architectures that most leads to increased costs related to the energy consumption of the equipment involved, since OEO conversions occur on all network nodes and the routing is point-to-point. The extensive use of OEO conversions results in a considerable and unnecessary increase in latency of connections~\cite{shen2018ultra}.

 \paragraph*{\textbf{Applications}}
 
For network providers, it is easier to leverage legacy infrastructure than to implement entirely new projects at the regional level due to size, cost, and complexity~\cite{singleCarrier2016, metro-haulproject}.
In this way, pure OTN remains a researched field today. The main developments identified in the literature for pure OTN are related to the development of mobile communication networks with $5G$ technology. The study group ITU-T Study Group $15$ (ITU-T $SG15$) and the International Mobile Telecommunications IMT-$2020$ ($5G$) have been working on the construction of technical specifications for the deployment of the $5G$ network with the support of optical networks~\cite{ITU-TSG15}. Also, new OTN equipment has been produced to include a common public radio interface (CPRI), making up a network segment called “full-stack” OTN \cite{8696394}, with communication interface with the mobile network. This trend has been highlighted as Mobile-Optimized Optical Transport Network (M-OTN), version of OTN to support $5G$ technology, carrying the client signal in the fronthaul communication layer~\cite{gangopadhyay20195g}, being implemented mainly in tree topologies, which in comparison with ring topologies, allows reducing access latency~\cite{liu2019evolution}.Some characteristics of Pure OTN are highlighted in Table \ref{Tab:PureOTNHighlights}.

\begin{table}[]
 \centering
	\caption{Pure OTN main features.}
	\label{Tab:PureOTNHighlights}
\begin{tabular}{|c|l|l|l|l|l|l|}
\hline
%\multicolumn{7}{|c|}{Pure OTN Architecture Highlights}        \\ \hline
\textbf{Advantages}                                                                               & \multicolumn{1}{c|}{\textbf{Disadvantages}}                                     & \multicolumn{1}{c|}{\textbf{Applications}}                                 & \multicolumn{1}{c|}{\textbf{Equipment}} & \multicolumn{1}{c|}{\textbf{Scope}} & \multicolumn{1}{c|}{\textbf{Topology}}&
\multicolumn{1}{c|}{\textbf{ToS}}\\ \hline

\multicolumn{1}{|l|}{\begin{tabular}[c]{@{}l@{}}Easy and low cost\\  implementation\end{tabular}} & \begin{tabular}[c]{@{}l@{}}OEO conversions on \\ all network nodes and \\high energy consumption\end{tabular} & \begin{tabular}[c]{@{}l@{}}Metro segments \\ interconnections\end{tabular} & Stand-alone OTN Switches    & MC & Many & P2P                              \\ \hline
\end{tabular}
\end{table}

\subsubsection{OTN / DWDM}\label{subsec:OTN-WDM}

\begin{figure}
	\centering
	\includegraphics[width=0.6\linewidth, page=2]{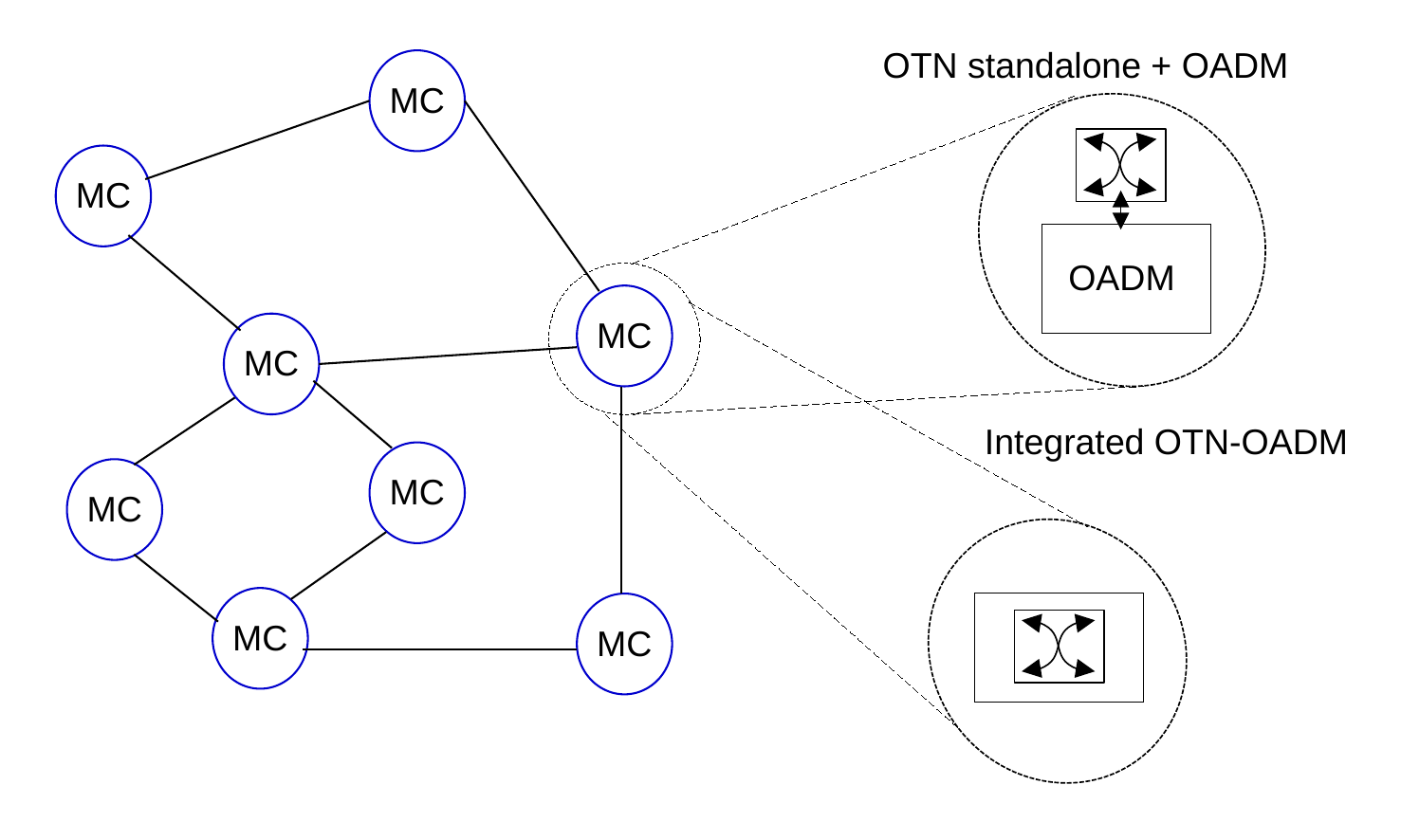}
	\caption{OTN over DWDM Network architecture in mesh topology}
	\label{fig:ArcOTN-DWDM}
\end{figure}

\paragraph*{\textbf{Composition and functioning}}
OTN combined with DWDM technology, defined by ITU-T G.872., was designed based on pure OTN (Subsection \ref{subsubsec:PureOTN}) to explore optical bypass, reducing the use of OEO conversions at each hop, and further reducing the use of WDM ports in the FOADM/ROADM. The structural organization of OTN / DWDM is similar to that of pure OTN architecture (Subsection \ref{subsubsec:PureOTN}), as shown in the Figure \ref{fig:ArcOTN-DWDM} but the physical organization has differences. OTN / DWDM represents an improvement over Pure OTN by reducing OEO conversion points for the same network segment. OTN / DWDM nodes can be either stand-alone OTN or OTN integrated with OADM, as shown in Section \ref{sec:background}. The architecture supports OADM of various types (FOADM or ROADM), due this, it was the first architecture where it was possible to perform end-to-end (E2E) optical switching.

\paragraph*{\textbf{Advantages}}
The E2E type of switching represents an economy in the use of expensive wavelength transponders, since OEO operations are only necessary at the origin and destination of the connections. Besides, other advantages are the potential reduction in energy consumption and network latency related to OEO conversion, which potentially occur to a lesser extent \cite{DBLP:journals/corr/abs-1901-04301,da2019otn}.
Another important point is that it is easy to handle the OTN layer or the optical layer separately, without major interference in the network. OTN over DWDM easily enables to mixed architectures composed of nodes with pure OTN and OTN/DWDM, especially to take advantage of legacy hardware equipment and implementing the concept of unbundled network architecture \cite{gangopadhyay20195g}.

\paragraph*{\textbf{Disadvantages}}
The cost of ROADMs is still an issue to be considered and has led to the study of new cost-effective design for nodes \cite{8204502WSS}, as well as filterless alternatives for metropolitan networks \cite{hernandez2020techno, cugini2018flexibleSemiFL}. Considering the high number of nodes in the metro network, the cost can be prohibitive. Due to the need for a greater number of channels to meet future demands, operators will need to increase investments in more robust equipment.

\paragraph*{\textbf{Applications}}
E2E connections on the OTN/DWDM network to the detriment of Pure OTN point-to-point connections, make this architecture an alternative solution for $5G$ technology due to the lower power consumption (least amount of OEO conversions). The most recent literature in terms of OTN/DWDM architecture has focused on optimizing the capacity of the transport network \cite{ramachandran2019capacity}, in adding coherent transponders \cite{filer2019low} and the use of this architecture for the feasibility of implementing $5G$ technology \cite{gangopadhyay20195g, katsalis2018towards}. As for capacity optimization, \cite{ramachandran2019capacity} proposes an optimization model that creates transmission tunnels whenever the traffic capacity between any pair of nodes is above a given limit defined by the operator. The advantage is that these tunnels are configured for each pair of nodes, using the path computing algorithm of multilayer transport networks of various technologies (SDH/SONET, IP/MPLS, OTN/DWDM), resulting in optimization between layers, that would not be possible if each layer or technology optimization was done independently.

The relationship between OTN/DWDM and $5G$ mobile networks derives from the industry consensus for this new sector since OTN/WDM technology should serve as the underlying physical layer infrastructure for 5G, allowing Ethernet services (in the form of FlexE) more dynamic \cite{katsalis2018towards}. To that end, 5G requires an infrastructure that is adaptable and resilient to multiple fiber failures, providing compromised service levels to end-users while significantly reducing the cost of the network compared to a traditional OTN switched network. Similarly, \cite{gangopadhyay20195g} describes an universal OTN switching model, which includes completely protocol-independent switching features, which seeks to aggregate traffic frame-by-frame across all ports on any layer of the system. Some characteristics of OTN/DWDM are highlighted in Table \ref{Tab:OTNDWDMHighlights}.

\begin{table}[]
 \centering
	\caption{OTN / DWDM main features.}
	\label{Tab:OTNDWDMHighlights}
\begin{tabular}{|c|l|l|l|l|l|l|}
\hline
%\multicolumn{7}{|c|}{Pure OTN Architecture Highlights}        \\ \hline
\textbf{Advantages}                                                                               
& \multicolumn{1}{c|}{\textbf{Disadvantages}}                                     
& \multicolumn{1}{c|}{\textbf{Applications}}                                 
& \multicolumn{1}{c|}{\textbf{Equipment}} 
& \multicolumn{1}{c|}{\textbf{Scope}} 
& \multicolumn{1}{c|}{\textbf{Topology}}&
\multicolumn{1}{c|}{\textbf{ToS}}\\ \hline
\multicolumn{1}{|l|}{\begin{tabular}[c]{@{}l@{}}Easy and low cost\\  implementation\end{tabular}} 
& \begin{tabular}[c]{@{}l@{}}Reduced OEO conversions and \\ number of WDM ports used\end{tabular} 
& \begin{tabular}[c]{@{}l@{}}Metro segments \\ interconnections\end{tabular} 
& \begin{tabular}[c]{@{}l@{}}stand-alone OTN \\ and ROADM \\ or OTN integrated \\ with ROADM \end{tabular}   
& MC/MA 
& Many  & E2E                             \\ \hline
\end{tabular}
\end{table}

\subsubsection{UDWSN} \label{subsubsec:UDWSN}

\paragraph*{\textbf{Composition and functioning}}
The Ultra-Dense Wavelength Switched Network (UDWSN) architecture, proposed in~\cite{zhang2016ultra}, is a special type of elastic optical network, a DEON, whose main difference is the implantation in metropolitan networks of specific WSSs with configurations of ultra-fine spectral granularities, below $12.5$ $GHz$, characteristic of EON \cite{shen2018ultra}. Thus, it is possible to experiment with four different types of spectrum granularity: $5$ $GHz$, $6.25$ $GHz$, $10$ $GHz$ and $12.5$ $GHz$~\cite{7792281UDWSN}. 
The architecture in question is based on low-cost optical equipment and passive elements, balanced and distributed with others of higher cost or that are still being designed by the manufacturers.
%OTN architectures would experience extensive optical-electronic-optical (OEO) conversions, which leads to high cost, high energy consumption, and increased connection latency due to electrical processes within a node.

\begin{figure}
	\centering
	\includegraphics[width=0.6\linewidth, page=2]{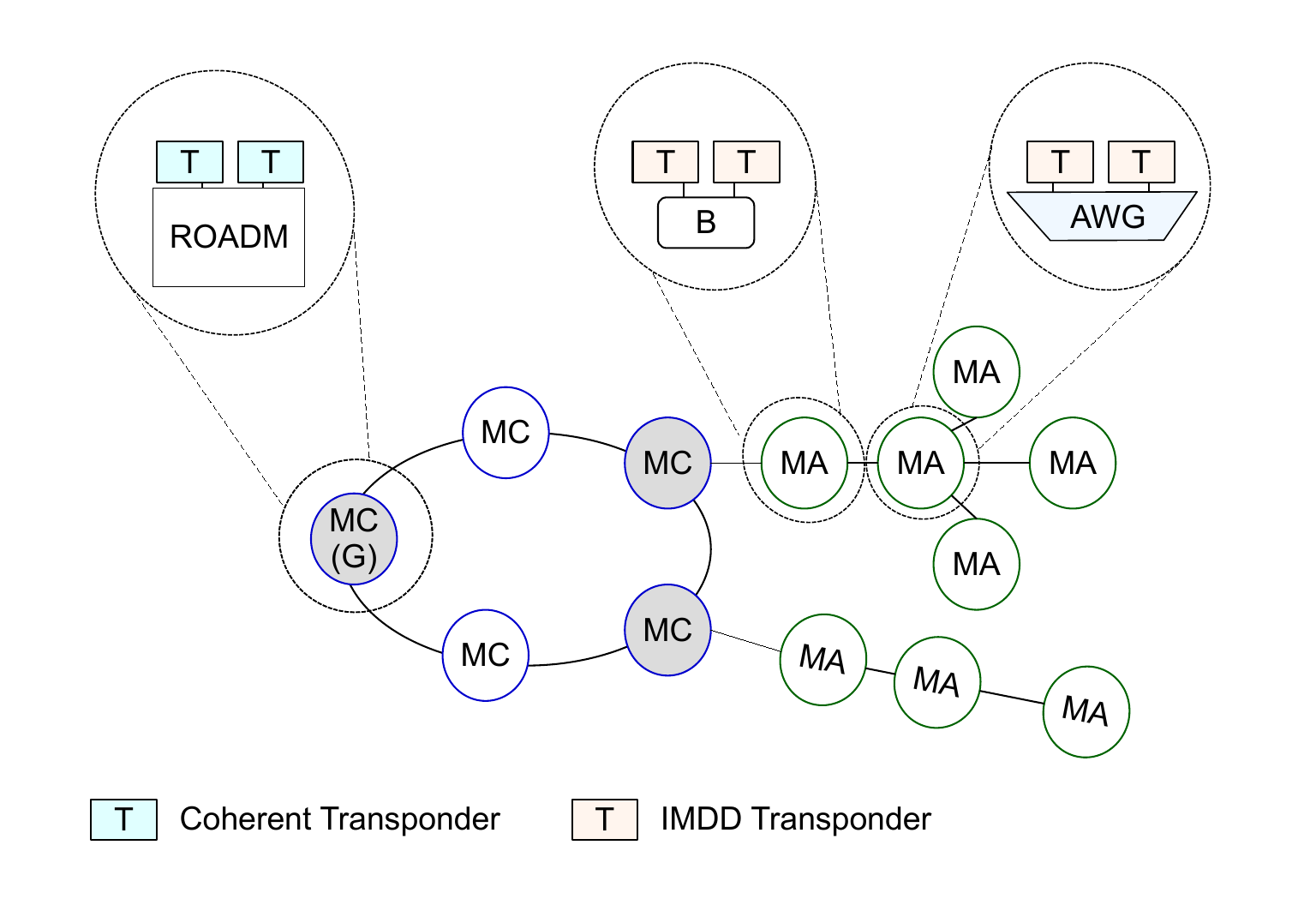}
	\caption{UDWSN architecture with metro-core (MC) and metro-access (MA) segments.}
	\label{fig:udwsn1}
\end{figure}

The Figure \ref{fig:udwsn1} describes the model of the UDWSN architecture. Each MC node in this architecture consists of WSS-based ROADM with $50$ $GHz$ channel spacing and coherent transponders (CT) with ultra-fine granularity. The MA nodes are formed by IMDD transponders and a passive element (PE), to be decided according to the sub topologies of the rest of the access network. PE were highlighted  in Subsection \ref{subsubsubsec:splitters&Couplers&Blockers}. As there may be sub topologies of nodes in the form of chains or trees, PE can be of two different types of multi/demultiplexer: UD-blockers in the chain branches and UD-AWGs in trees. In each aggregator MC  node there is an OTN switch installed that is responsible for aggregating the traffic demands of the access part of the network, which are numerous and of low granularity. 

The two network segments are also classified as symmetric (in the MC part) and asymmetric (in the MA part) in terms of the traffic characteristic of their respective regions. In the symmetrical traffic segment, origin and destination have the same characteristics, so the speed of traffic is the same in both directions. When the network segment has asymmetric traffic, the downlink data transmission speed is different from the uplink speed, and these are defined by different routes for each of these services, exactly because the origin and destination have different characteristics. Both, downstream and upstream flow can take different routes due the possibility of existing drop and add signals at two dedicated wavelengths. With the use of CT in the center of the network, where there is greater traffic, single equipment of this type can support many subcarriers transmitted to different MA nodes using OOFDM and, in the part of the access network, with IMDD transponders the direct modulation scheme is employed for MA nodes \cite{7792281UDWSN, Zhang2018ExploitingEO}. 

\paragraph*{\textbf{Advantages}}
According to \cite{shen2018ultra, 7792281UDWSN} and \cite{Zhang2018ExploitingEO}, the main advantages that support the UDWSN architecture are the following: (\textit{i}) the exploitable bandwidth granularities are so small that they reduce the underutilization of optical spectrum resources, since only enough spectrum is allocated to meet the demands; (\textit{ii}) the architecture allows the end-to-end optical switching of data traffic on the network, and this idea results in a reduction in energy consumption, due to the elimination of the need for OEO conversion; (\textit{iii}) Another advantage resulting from the elimination of OEO conversion in the intermediate nodes is the optimization of latency, since these operations consume extra fractions of time; (\textit{iv}) with narrow spectrum granularities, the number of channels (optical spectrum slots) that can be offered is quite high compared to the optical transport technologies in the metropolitan network today, which makes UDWSN architecture a good commercial solution; (\textit{v}) due to the short distances prevalent in the metro network segment, and by taking advantage of coherent transmission technology, more advanced modulation formats, and consequently better digital signal processing speed, can be used to improve the transmission efficiency.

\paragraph*{\textbf{Disadvantages}}

The authors \cite{Zhang2018ExploitingEO} affirm that it is also possible to implement other combinations of equipment using coherent technology in the network as a whole, to improve the performance of the network, however, this idea incurs an increase in the costs of implanting equipment, mainly concerning the access network part.
Some other noted disadvantages are: (\textit{i}) even if a granularity as fine as $5$ $GHz$ is possible to be adopted in a metropolitan network, serious wastes of optical spectrum resources will still occur, considering that the predominant majority of traffic demands are relatively small, ranging from sub-$1$ $Gb/s$ to $10$ $Gb/s$, and this quantity does not completely occupy the available bandwidth per channel. This problem is aggravated when using $10$ $GHz$ or $12$ $GHz$ granularities. Thus, data aggregation is required; (\textit{ii}) there are technological limitations to be considered in the process of developing optical equipment with spectral resolution as thin as $5$ $GHz$, (\textit{iii}) WSS equipment with bandwidth as small as for the highlighted granularities ($5$ $GHz$ or $6.25$ $GHz$, for example) need to be manufactured and its cost, which is dependent on this measure of bandwidth, is relatively high \cite{8204502WSS}; 
(\textit{iv}) as well as in core networks with EON technology of optical transport, the application performance can be impacted by the fragmentation of the optical spectrum, with the UDWSN architecture, this problem can occur more frequently, since most demands in this network scope have origin and end in the metropolitan network itself.

\paragraph*{\textbf{Applications}}
UDWSN is viable for deployment of integrated metro and access networks. It becomes an alternative to meet the demands of the metro segment as it represents a flexible approach while offering greater capacity for different types of services \cite{Zhang2018ExploitingEO}. This architecture is proposed as a low cost solution to replace pure OTN and OTN / DWDM. Some characteristics of UDWSN are highlighted in Table \ref{Tab:UDWSNHighlights}.

\begin{table}[]
\centering
	\caption{UDWSN main features.}
	\label{Tab:UDWSNHighlights}
\begin{tabular}{|c|l|l|l|l|l|l|}
\hline
%\multicolumn{7}{|c|}{Pure UDWSN Architecture Highlights}  \\ \hline
\textbf{Advantages}                                                                                                         & \multicolumn{1}{c|}{\textbf{Disadvantages}}               
& \multicolumn{1}{c|}{\textbf{Applications}}                
& \multicolumn{1}{c|}{\textbf{Equipment}}             
& \multicolumn{1}{c|}{\textbf{Scope}} 
& \multicolumn{1}{c|}{\textbf{Topology}} 
& \multicolumn{1}{c|}{\textbf{ToS}}
\\ \hline
\multicolumn{1}{|l|}{\begin{tabular}[c]{@{}l@{}}Ultra-fine spectral\\ granularities, \\ enables mesh \\architecture on Metro\end{tabular}} 
& \begin{tabular}[c]{@{}l@{}}Some elements\\ with high cost\\ and high impact \\on the spectrum \\ fragmentation\end{tabular} & \begin{tabular}[c]{@{}l@{}}Metro segments \\ interconnections\end{tabular}
& \begin{tabular}[c]{@{}l@{}}Stand-alone OTN Switchs\\ AWG\\ IMDD transponder\\ Coherent Trasnponder\end{tabular} 
& MC/MA 
& \begin{tabular}[c]{@{}l@{}}Mesh \\ Tree \\Chain \end{tabular}
&E2E\\\hline
\end{tabular}
\end{table}

\subsubsection{HnS / WDM}\label{subsec:HnS_WDM}

\paragraph*{\textbf{Composition and functioning}}
The Dual Hub-and-Spoke over WDM (HnS/WDM) architecture is a multi-layer approach composed by Ethernet switch over mesh WDM network. In the optical layer all nodes have ROADMs. In the eletronic layer,
if the node plays the role of metro-core, an Ethernet Core Switch is implemented. A Ethernet Edge Switch is implemented on nodes that are metro-access \cite{she2017metro}. 
While DWDM offers protocol independence, so operators can carry any data traffic, provide storage and TDM services, Ethernet guarantees low transfer rates that match many of the services that are offered. The HnS logical topology can be implemented on such physical architectures to contribute to the improvement of the communication time made possible by reducing hierarchical levels.

\begin{figure}
	\centering
	\includegraphics[width=0.8\linewidth, page=2]{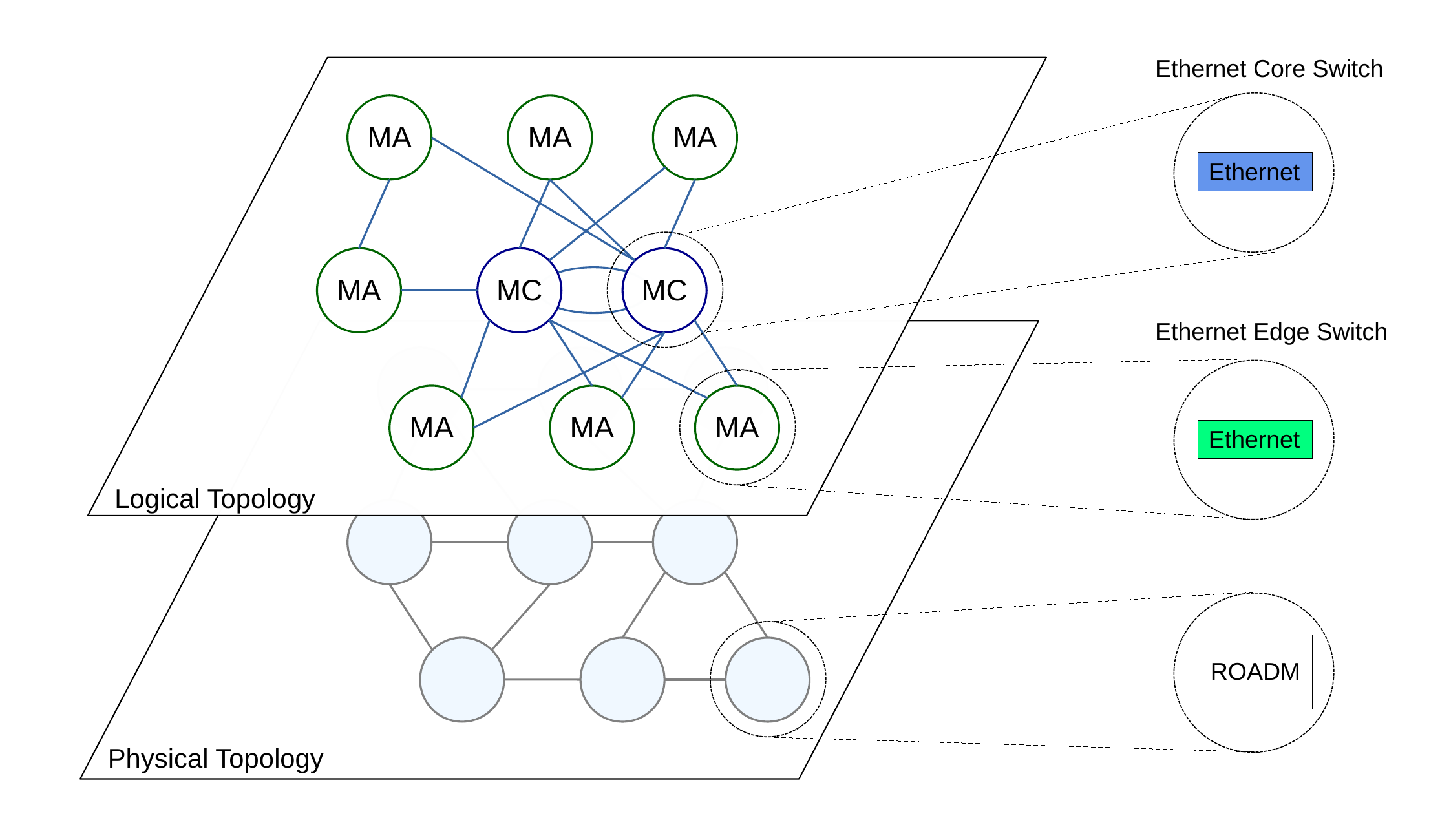}
	\caption{Logical Dual Hub-and-Spoke over physical WDM Architecture.}
	\label{fig:hns}
\end{figure}

Figure \ref{fig:hns} is a schematic representation of the HsN/WDM architecture. The logical topology shows the two MC nodes that are the destination hubs of the metropolitan network in which they are located PoPs, and also aggregate the MA segment data, which are mainly CO to which the network operators provide connectivity. In the optical layer, end-to-end lightpaths are established between the MC and MA nodes, since every MA is directly connected to at least one MC.

\paragraph*{\textbf{Advantages}}
This heterogeneously organized environment is favorable to energy savings compared to a metropolitan network environment in which all nodes implement large switches. The main differential of this architecture for the others presented is the focus on logical topology following the HnS model. Physical meshed or hierarchical ring architectures incur multiple levels of communication between the access segment and the hub that is the gateway to the backbone network level. These multiple levels end up leading to longer routes and, consequently, multiple hops. The logical topology determines the origins and destinations of the requests, and only at these points will OEO conversion occur in the case of lightpaths that are established. This strategy helps to reduce latency in the establishment of services. Ethernet is especially important for meeting numerous fine-grained data requests because it reduces wasted resources.

\paragraph*{\textbf{Disadvantages}}

Ethernet frames associated with a specific service are usually transported on the same optical path, as they are fixed, there are not many possibilities for alternative paths. Also, it should be noted that the use of optical-electrical components required in a hybrid architecture such as HnS/WDM incurs an increase in latency due to the extra processing that is required in the task of translating information between the two layers. In the same way, an increase in energy consumption is expected with this infrastructure, since electronic switches are used in all nodes, in addition to amplifiers that may be needed in WDM links.

\paragraph*{\textbf{Applications}}
The authors point out that this architecture is especially interesting for 4K/VR and cloud video services, both because it is capable of providing bandwidth on demand, and because of the shortening of response times due to traffic paths with fewer hops \cite{she2017metro}. Some characteristics of HnS/WDM are highlighted in Table \ref{Tab:HnSWDMresume}.

\begin{table}[]
\centering
	\caption{HnS / WDM main features.}
	\label{Tab:HnSWDMresume}
\begin{tabular}{|c|l|l|l|l|l|l|}
\hline
%\multicolumn{7}{|c|}{HnS/WDM Architecture Highlights}   \\ \hline
\textbf{Advantages}                                                                                                         & \multicolumn{1}{c|}{\textbf{Disadvantages}}               
& \multicolumn{1}{c|}{\textbf{Applications}}                
& \multicolumn{1}{c|}{\textbf{Equipment}}             
& \multicolumn{1}{c|}{\textbf{Scope}} 
& \multicolumn{1}{c|}{\textbf{Topology}} 
& \multicolumn{1}{c|}{\textbf{ToS}}
\\ \hline
\multicolumn{1}{|l|}{\begin{tabular}[c]{@{}l@{}}Low latency\\ to meet requests.\\Reduces the \\amount of\\ hierarchical levels.\end{tabular}} 
& \begin{tabular}[c]{@{}l@{}}Increase in\\ energy consumption.\end{tabular} &
\begin{tabular}[c]{@{}l@{}}Metro segments \\ interconnections\end{tabular}
& \begin{tabular}[c]{@{}l@{}}ROADM\\ Ethernet Core Switch\\ Ethernet Edge Switch\end{tabular} 
& MC/MA 
& \begin{tabular}[c]{@{}l@{}} HnS \end{tabular}
&E2E\\\hline
\end{tabular}
\end{table}

\subsubsection{ARMONIA}\label{subsubsec:ARMONIA}

\paragraph*{\textbf{Composition and functioning}}

The ARMONIA architecture is an architecture that comprises the unification of the core segment (with MC nodes) and aggregation segment (with MA nodes) of the metropolitan network, as well as the access segment, which is self-configurable due to the implementation of monitoring with telemetry service~\cite{kretsis2020armonia}. 

The main strategy of this architecture is to consider the unification of these various segments to facilitate the dynamic operation of the network while taking advantage of the use of coherent transponders for high capacity transmission and to obtain data about the transmission that are used by the telemetry agents. ARMONIA can be deployed in a distributed or centralized manner and is usually organized in multiple layers. The number of layers is defined according to the network size or operator objectives.

\begin{figure}
	\centering
	\includegraphics[width=0.7\linewidth,page=2]{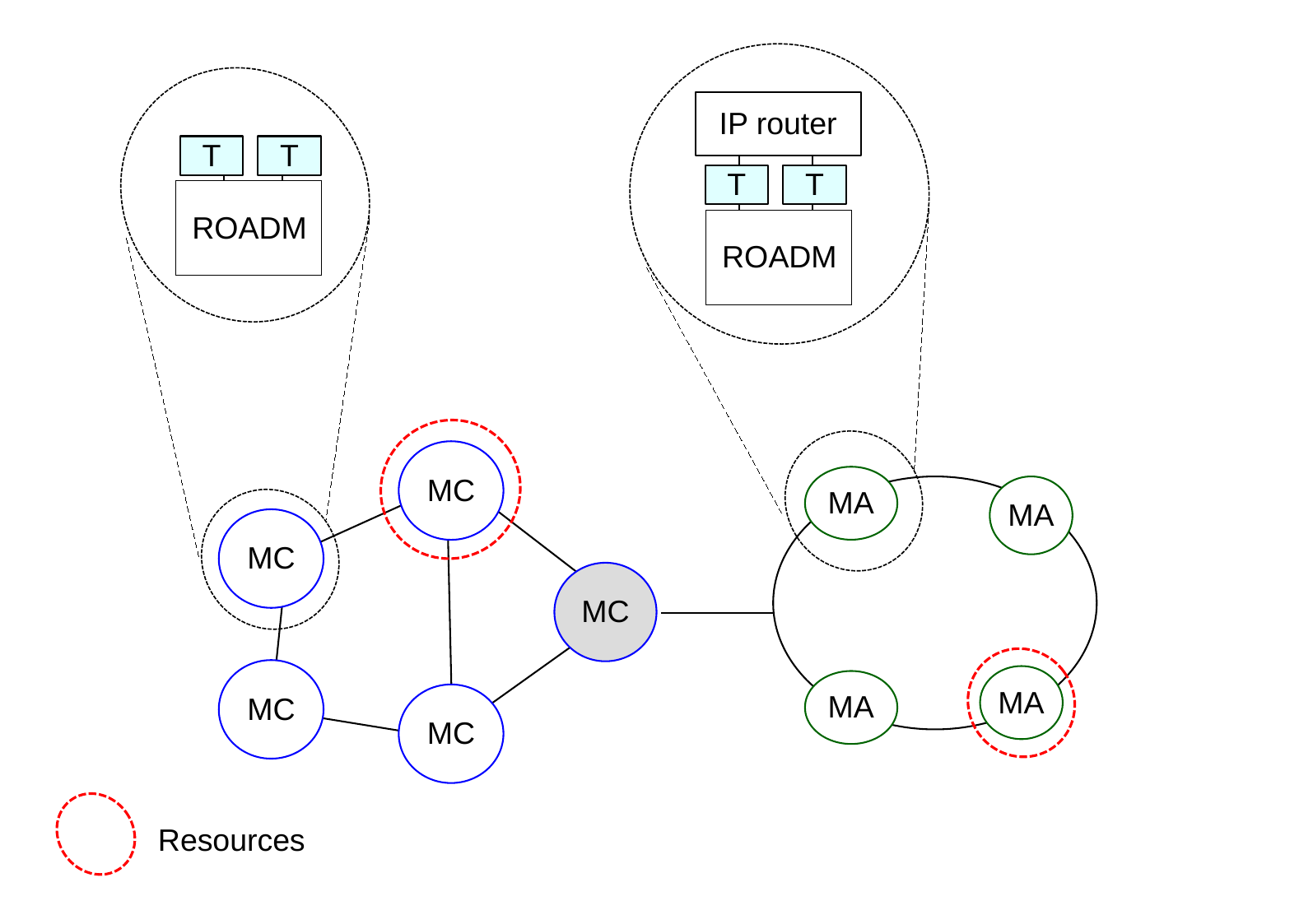}
	\caption{ARMONIA multilayer metropolitan network architecture.}
	\label{fig:arc-armonia}
\end{figure}

Figure \ref{fig:arc-armonia} shows a schematic representation of ARMONIA architecture. The MC node segment is considered the upper layer of ARMONIA. The aggregation node MC, colored in gray, is considered as an intermediate layer and can be deployed at various levels of aggregation, according to the network size or operator objectives. The MA node segment is considered the lower layer. The layers of MC and MA nodes are built based on ROADMs and coherent transponder, and implement NWDM, a flexible transmission system. Amplifiers are used when needed. The services are provided on an end-to-end basis. The MA node segment, considered as the edge of the network, contain IP routers installed to collect and aggregate access network traffic (fixed and mobile) and better take advantage of transmission resources. Each MC node connects PONs and mobile network. The PON are based on OLT-based components to ensure cost savings at strategic points to support edge computing services. In Figure \ref{fig:arc-armonia}, some nodes marked by a red circle represent locations where additional resources are installed, such as computation and storage resources. The resources available on some MC node are of greater capacity than the resources available on some MA node.

In addition to an integrated infrastructure, ARMONIA also has integrated control and management. The ARMONIA architecture also has a QoT estimator for optical paths and traffic predictors. The QoT estimator is based on linear regression and Machine Learning (ML) and uses information from the location of the implanted amplifiers and links to analyze the validity of an optical path before its establishment. Traffic predictors are based on Long Short-Term Memory (LSTM) networks, a recurrent neural networks (RNNs) algorithm, and Prophet, a Facebook time series forecasting algorithm. Traffic forecasts provide solutions for more efficient resource allocation and it is a differential of ARMONIA in relation to all other architectures covered in this work. The solutions of this architecture are designed mainly to serve Network Slice services and virtual network services. 

\paragraph*{\textbf{Advantages}}

The flexible approach of the ARMONIA architecture and the distribution of resources across the network ensure greater agility for high-priority and low-latency services, especially for intra-metro connections. Unified operations across network segments ensure the establishment of end-to-end paths in the optical layer, and also contribute to reduced latency.
Intelligent algorithms that help control and manage the network are extra tools, useful to improve operator decision making. The availability of extra resources and some network points reduces the capital and operational cost for the operator, freeing him from equipping all nodes equally. Furthermore, the amount of resource may increase over time according to the demand for the services.

\paragraph*{\textbf{Disadvantages}}
The ARMONIA architecture foresees a greater amount of hierarchical levels along the unified segments to increase the aggregation points. The increase in the number of levels represents a greater amount of hops for the establishment of optical paths. Longer paths can potentially increase the blocking rate of requests, in addition to increasing latency. This is an issue that must be taken into account by the operator, as metro networks tend to have many nodes. Another big challenge is related to unified control and management. Globally, maintenance and updates may be required frequently, and because of this, the operator needs to ensure that the behavior of the network is not affected.

 \paragraph*{\textbf{Applications}}

The ARMONIA architecture serves the entire metropolitan network segment and provides agility and dynamism, especially for mobile network services, access networks, cloud computing and edge computing. Table \ref{Tab:ARMONIAHighlights} summarizes the main features of the ARMONIA architecture.
 
 \begin{table}[]
 \centering
	\caption{ARMONIA main features.}
	\label{Tab:ARMONIAHighlights}
\begin{tabular}{|c|l|l|l|l|l|l|}
\hline
%\multicolumn{7}{|c|}{ARMONIA Architecture Highlights}        \\ \hline
\textbf{Advantages}                                                                               
& \multicolumn{1}{c|}{\textbf{Disadvantages}}                                     
& \multicolumn{1}{c|}{\textbf{Applications}}                                 
& \multicolumn{1}{c|}{\textbf{Equipment}} 
& \multicolumn{1}{c|}{\textbf{Scope}} 
& \multicolumn{1}{c|}{\textbf{Topology}}
& \multicolumn{1}{c|}{\textbf{ToS}}\\ \hline

\multicolumn{1}{|l|}{\begin{tabular}[c]{@{}l@{}}Monitoring with telemetry service;\\  Customizable number of layers\end{tabular}} 
& \begin{tabular}[c]{@{}l@{}}Higher latency due\\ to potentially higher \\number of layers\end{tabular} 
& \begin{tabular}[c]{@{}l@{}}Metro segments \\ interconnections\end{tabular} 
& \begin{tabular}[c]{@{}l@{}}ROADM \\ CO-BVT \\ OLT\\IP Router \end{tabular}   
& MC/MA 
& \begin{tabular}[c]{@{}l@{}}Mesh \\ Ring \end{tabular}      &  E2E                      \\ \hline
\end{tabular}
\end{table}
 
 %possui  nós OLT organizados como topologia em anel ou malha. Roteadores IPs  nesse segmento agregam os dados da rede de acesso com suas granularidades específicas.

%Voltada para o atendimento de serviços de network slicing, edge computing e redes móveis, a ARMONIA possui plano de controle baseado em SDN. 

%A arquitetura é composta por camada IP sobreposta à camada óptica com tecnologia de transporte EON.

\subsubsection{DISCUS}\label{subsec:DiSCUSarchitecture}

\begin{figure}
	\centering
	\includegraphics[width=0.7\linewidth]{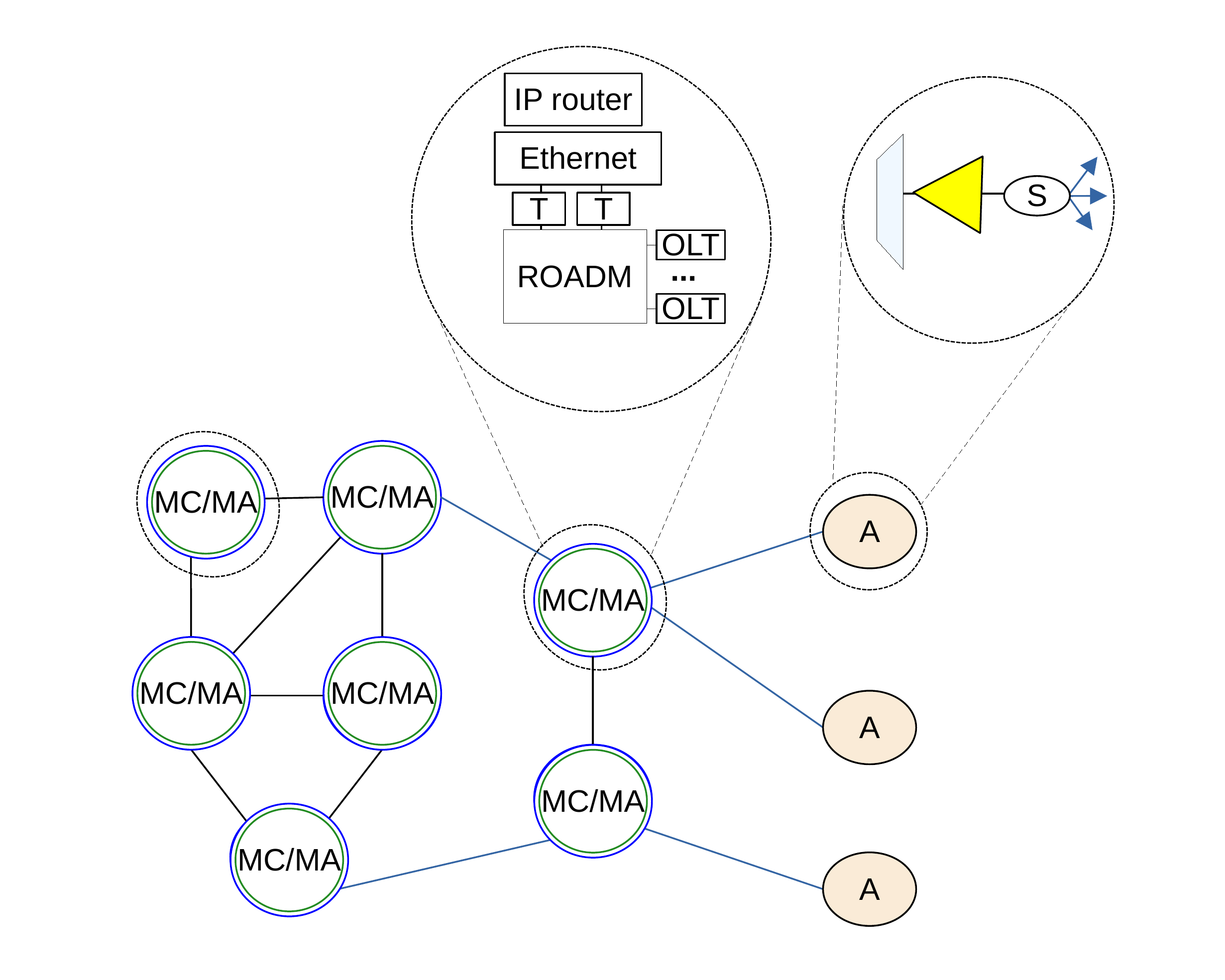}
	\caption{DISCUS architecture comprising a meshed flat optical access and metro networks.}
	\label{fig:arc-discus}
\end{figure}

\paragraph*{\textbf{Composition and functioning}}
The architecture highlighted by the DISCUS FP7 consortium \cite{ruffini2017access} is also a solution combining the metro and access segments, implementing multilayer with IP router, Ethernet switch and ROADM for DWDM in the metro nodes. In access nodes Long-Reach Passive Optical Networks (LR-PONs) is implemented. A simplified representation of the DISCUS architecture is shown in Figure \ref{fig:arc-discus}. MC/MA nodes are Clos optical switches\cite{glkabowski2020simulation} multistage allocated in CO, with large number of ports, and implement DWDM to tolerate multiple wavelengths in the access segment. The MC/MA node segment is a translucent network where IP routers are implemented, with OEO conversion on some of the nodes, generating islands of transparency. The presence of IP routers is mainly used to communicate to and from other metropolitan networks. 
This segment is called by the authors as flat network. In turn, the LR-PON in the access segment connects the access nodes and nodes with amplifiers to the MC nodes. There may be several stages of signal amplification and splitting due to LR-PON until the service reaches subscribers. Every LR-PON starts and ends at an MC/MA (dual-homed) node for protection reasons. The sharing of resources in the access network is done through the implanted splitters, which requires nodes with amplifiers to support the bandwidth division rates at the division points.
The metro segment can be designed with mesh topology, while the access segment follows the chain topology model or "open ring" (horseshoe). However, the logical interconnection at the access nodes is done through a tree topology.

Although the authors emphasize the presence of MC nodes, a segmentation of MA nodes is not mentioned, as occurs in other architectures (such as the architecture in Subsection \ref{subsubsec:ARMONIA}, for example). However, due to the nature of the LR-PON, which combines the metro network and the access network in a single network and having connections established directly, MCs nodes also accumulate MA nodes functions, such as switching, routing, and data aggregation, since the main objective of the architecture is to reduce the number of nodes in the topology.

\paragraph*{\textbf{Advantages}}

The DISCUS architecture aims to reduce the number of nodes and OEO conversion points in the metro segment, being a bet for regions of convergence of fixed and mobile networks in the metro segment. The main strategy of this architecture to reduce deployment costs in the long term is the sharing of nodes and equipment by as many customers as possible. Additionally, the reduced number of hierarchical levels contributes to lower latency.

\paragraph*{\textbf{Disadvantages}}

Some disadvantages of the architecture can be highlighted. With OEO implemented in all nodes, it is not possible for multiple line rates to coexist. Also, the reduction in the amount of nodes implies greater traffic congestion at peak times. Due to the functions of the electrical layer, energy consumption can be considerably high by metro network standards when considering future traffic trends. In addition to the increase in power consumption, the electronic layer adds extra latency to services.

\paragraph*{\textbf{Applications}}

According to authors \cite{ruffini2017access}, the DISCUS architecture is especially advantageous for residential and 
mobile backhauling clients, due to the variety of services offered, such as point-to-point,  point-to-multipoint (P2MP) and multipoint-to-multipoint (MP2MP) services connections. 
Table \ref{Tab:DISCUSHighlights} summarizes the main highlights of the DISCUS architecture.

\begin{table}[]
 \centering
	\caption{DISCUS main features.}
	\label{Tab:DISCUSHighlights}
\begin{tabular}{|c|l|l|l|l|l|l|}
\hline
%\multicolumn{7}{|c|}{DISCUS Architecture Highlights}        \\ \hline
\textbf{Advantages}                                                                               
& \multicolumn{1}{c|}{\textbf{Disadvantages}}                                     
& \multicolumn{1}{c|}{\textbf{Applications}}                                 
& \multicolumn{1}{c|}{\textbf{Equipment}} 
& \multicolumn{1}{c|}{\textbf{Scope}} 
& \multicolumn{1}{c|}{\textbf{Topology}}
& \multicolumn{1}{c|}{\textbf{ToS}}\\ \hline

\multicolumn{1}{|l|}{\begin{tabular}[c]{@{}l@{}}Node multifunctional\\as MC/MA(Lower capex);\\Low cost due accumulation\\of functions by nodes\end{tabular}} 
& \begin{tabular}[c]{@{}l@{}}Possible bottlenecks due\\to the accumulation of\\functions in some nodes in\\the transparency islands\end{tabular} 
& \begin{tabular}[c]{@{}l@{}}Metro segments \\ interconnections\end{tabular} 
& \begin{tabular}[c]{@{}l@{}} ROADM\\CO-BVT\\PON equip.\\IP Router\\Eth. Switches \end{tabular}   
& \begin{tabular}[c]{@{}l@{}} MC/MA\\Access \end{tabular}
& \begin{tabular}[c]{@{}l@{}} Mesh\\Ring \end{tabular}
& \begin{tabular}[c]{@{}l@{}} P2P\\P2MP\\MP2MP \end{tabular} \\ \hline
\end{tabular}
\end{table}

 %limitações com relação a distância dos enlaces e latência de transmissão podem ser um impeditivo para a adoção da proposta em cenários nos quais uma infraestrutura maior é requerida. 
 
 \subsubsection{MOMC}\label{subsubsec:MOMC}
 
 \paragraph*{\textbf{Composition and functioning}}
 
The Modular Optical Metro-Core (MOMC) architecture is highlighted in \cite{larrabeiti2020upcoming, calabretta2019photonic} and \cite{larrabeiti2019all}.
The motivation behind MOMC is to propose an architectural solution to meet the traffic services and new applications that will be required by 2030, according to current forecasts for the coming decades \cite{larrabeiti2019all}.

MOMC modular approach makes it possible to scale the architecture, following the "pay-as-you-grow" concept and, although it consists of many hierarchical levels, the main idea is to increasingly implement transparent circuits along with these levels, to reduce the need for OEO conversion and reduce latency in service delivery. The SDN-based control plane provides flexibility in adjusting components, channels and power consumption.
 
\begin{figure}
 	\centering
 	\includegraphics[width=0.99\linewidth]{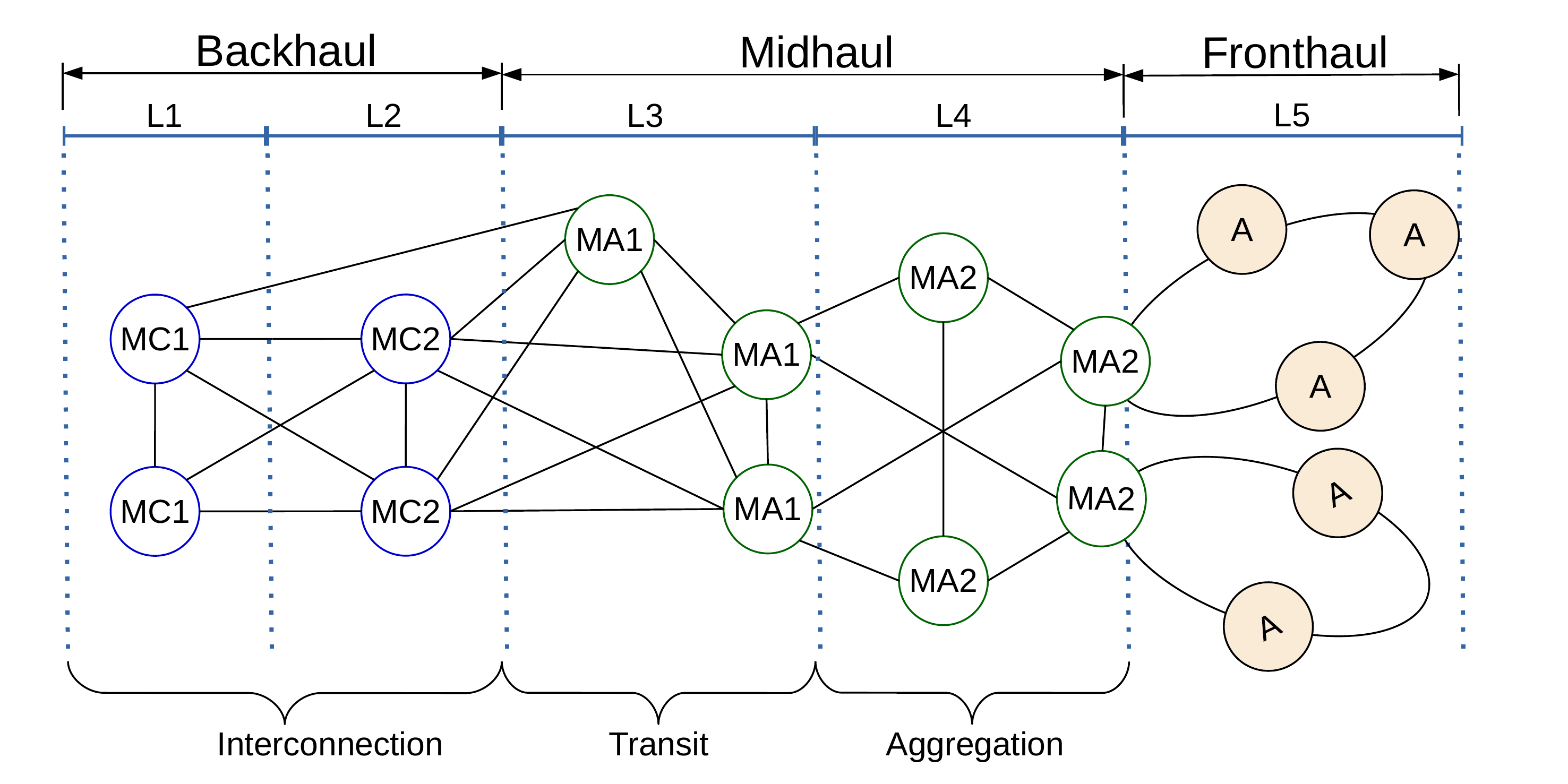}
 	\caption{SDN enabled metro core and metro access network architecture with modular node architecture based on WDM.}
 	\label{fig:momc}
 \end{figure}

A MOMC architecture scheme is shown in Figure \ref{fig:momc} according \cite{larrabeiti2019all}. This architecture is based on WDM optical transport technology and at all levels, nodes are based on ROADMs. However, there is a difference between the switches implemented at each level, mainly regarding the degree in ROADM. The 5 hierarchical levels are L1, L2, L3, L4, and L5. 
The L1 level is closest to the core networks, with MC1 nodes, represents the core-segment edge MC and comprises nodes that are gateways for WAN. L2, on the other hand, with MC2 nodes, represents the border between the MC-MA segments and comprises nodes that are gateways to services such as IPTV and Content Delivery Network (CDN), for example. The MA segment is divided into two aggregation stages, as MA1 and MA2,highlighted by levels L3 and L4, respectively. MA2 in L4 is responsible for aggregating access network traffic, while MA1 in L3 handles aggregated traffic. The nodes presented at levels L1, L2, and L3 can be of the same architecture and capacity and can implements CDC-ROADMs, $25$ $GHz$ bandwidth granularity, specific coherents SBVTs with large capacity for aggregate traffic of the order of $Tb/s$~\cite{boffi2020multi} and  multinucleated fibers (MCF) in the SDM perspective and with modern equipment capable of climb easily.  The L4 level can be implemented with low-cost 2-degree photonic integrated mini-ROADMs.
 
However, due to the cost and the better use of legacy architectures, architectural restrictions are imposed on nodes in L3, which represents critical points due to their higher aggregation rate compared to the other levels, routes the traffic directly to L1 or L2 (Figure \ref{fig:momc}). Finally, L5 represents the access network, where it is located the edge network with the layer of client devices, commercial VPN, telecommunications network antennas (Remote Radio Head - RRH), OLTs, among other elements. A large number of levels was designed to distribute functions across network scopes. The electronic layer at all levels is composed of Ethernet switch, but the goal is to make the connections between layers L2, L3 and L4 end-to-end, reducing the need for OEO conversion.

To make MOMC cost-effective and share resources with 5G mobile networks, in \cite{larrabeiti2020upcoming} is proposed a mapping of the locations of CU, DU and RRU in the hierarchy of MAN nodes. 
The mapping serves to indicate the locations where the most needed resources, such as processing and storage, can be added so that cost and latency are reduced. In this mapping, the core segment of metro (L1 and L2 levels) operates as Backhaul for 5G network interconnecting Next Generation Core to Centralized Unit (CU). Midhaul segment is the interface between CU and Distributed Unit (DU) operating in the MAs, that is, L3 and L4 levels, with different functionality. The Fronthaul segment is mapped in L5 and interconnects DU to Remote Radio Unit (RU).
However, there are other function distribution models for the $5G$ mobile network in the metro. This models takes into account the bandwidth and latency of transport, as well as the inherent processing complexity \cite{yu2020isolation}. Such models will not be discussed here as they are outside the scope of the work.
 
 \paragraph*{\textbf{Advantages}}
The "pay as you grow" approach in architecture allows components to scale without major disruptions in services. The high-capacity elements implemented in the network topology allow efficient establishment of services, even in a scenario of competing network services. Furthermore, most levels work transparently, establishing end-to-end paths with less latency and reduced power consumption.

\paragraph*{\textbf{Disadvantages}}
Since most operations are implemented as packet-based switching, further investigation is needed to determine the power consumption for various types of services, where OEO conversions are required.
The various layers of the network can represent more hops and higher latency for some types of high-priority applications, which can result in blocked service. These issues deserve more attention, especially when the applications in question are related to IoT and mobile networks, due to their strict requirements.

 \paragraph*{\textbf{Applications}}
 
 The MOMC architecture was specially designed to serve and share resources for mobile network, C-RAN, edge-cloud computing, access segment and IoT applications. The main features of the MOMC architecture are highlighted in Table \ref{Tab:MOMCHighlights}.

\begin{table}[]
 \centering
	\caption{MOMC main features.}
	\label{Tab:MOMCHighlights}
\begin{tabular}{|c|l|l|l|l|l|l|}
\hline
%\multicolumn{7}{|c|}{MOMC Architecture Highlights}        \\ \hline
\textbf{Advantages}                                                                               
& \multicolumn{1}{c|}{\textbf{Disadvantages}}                                     
& \multicolumn{1}{c|}{\textbf{Applications}}                                 
& \multicolumn{1}{c|}{\textbf{Equipment}} 
& \multicolumn{1}{c|}{\textbf{Scope}} 
& \multicolumn{1}{c|}{\textbf{Topology}}
& \multicolumn{1}{c|}{\textbf{ToS}}\\ \hline

\multicolumn{1}{|l|}{\begin{tabular}[c]{@{}l@{}}Low cost\\ 'pay-as-you-grow;\\  Provides services for\\edge computing, \\mobile and access\\ networks\end{tabular}} 
& \begin{tabular}[c]{@{}l@{}}Many  hierarchical  levels; \\ Requires OEO; Both\\ increase the latency; \\\end{tabular} 
& \begin{tabular}[c]{@{}l@{}}Metro segments \\ interconnections\end{tabular} 
& \begin{tabular}[c]{@{}l@{}} CDC-ROADM\\WSS\\CO-SBVT\\IP Router \end{tabular}   
& MC/MA 
& Ring  
& \begin{tabular}[c]{@{}l@{}}E2E \\ P2P\end{tabular}\\ \hline
\end{tabular}
\end{table}

\subsection{Single-layer Architectures}\label{subsec:SingleLayerArchitectures}

In this section, the new single-layer metropolitan network architecture will be presented. Moreover, it will present the WF, sFL, and FL classifications. The main inferences that can be made in the light of this classification are as follows: (\textit{i}) WF networks are generally more expensive and more robust than the FL and sFL variations; (\textit{ii}) the reduced cost of implementing and operating the FL and sFL variations is a great motivator for network operators and can be a good alternative to support 5G technology and its progressive increase in capillarity, through the implementation of mid and fronthaul infrastructure.

However, while the FL and sFL networks suffer from the limitation in the optical range of the signal due to the removal of the optical filter. In this scenario, each transmitted signal occupies the entire network, and when passing through the splitters, there is a greater insertion loss (IL) than would occur if filters were being used. Also, an increased consumption of a larger portion of the frequency spectrum occurs due to the presence of unfiltered signals. In the WF networks, when optical signals pass through several of these filters, the equivalent bandwidth of the resulting channel can be significantly reduced, resulting in spectral distortions and distortions resulting from optical signal to noise (OSNR) \cite{jin2016improved}.

\centerline{\rule{0.4\textwidth}{0.4pt}}
\bigskip

\textbf{MON With Filter (WF)}\label{subsubsec:MON_WF} - Optical filters are devices that select a segment of the optical spectrum while blocking unwanted parts of that optical spectrum. These devices do not require OEO conversion, and therefore do not incur an increase in electricity consumption when optical bypass is applied for not requiring additional transceivers. Some types of optical filters have been presented in Subsection \ref{sec:background}.
%(Atenção, não é bem assim, um componente óptico pode ser passivo ou ativo. Então tem que ficar claro que pode haver gasto energético sim. Certo???): OK. 

\subsubsection{MEON} \label{subsubsubsec:MEON}

\paragraph*{\textbf{Composition and functioning}}

Pure OTN architectures (Subsection \ref{subsubsec:PureOTN}) and OTN/WDM (Subsection \ref{subsec:OTN-WDM}) are based on WDM transmission technology. In contrast, Metro Elastic Optical Network (MEON) are optical networks with EON transmission technology. MEON was first mentioned in \cite{rottondi2013routing}, in $2013$, and have recently been highlighted in \cite{yan2018tidal, yan2020area, 8734478} and \cite{8853968}. Whereas pure OTN and OTN/WDM make use of several intermediate amplifiers in their respective optical links, for MEON the use is dispensed under the argument that the distances are sufficiently short. The few amplifiers employed are located at the input and output flow of the nodes, and only to ensure efficient power for signal processing steps at the nodes \cite{8734478}. 

\begin{figure}
	\centering
	\includegraphics[width=0.5\linewidth, page=2]{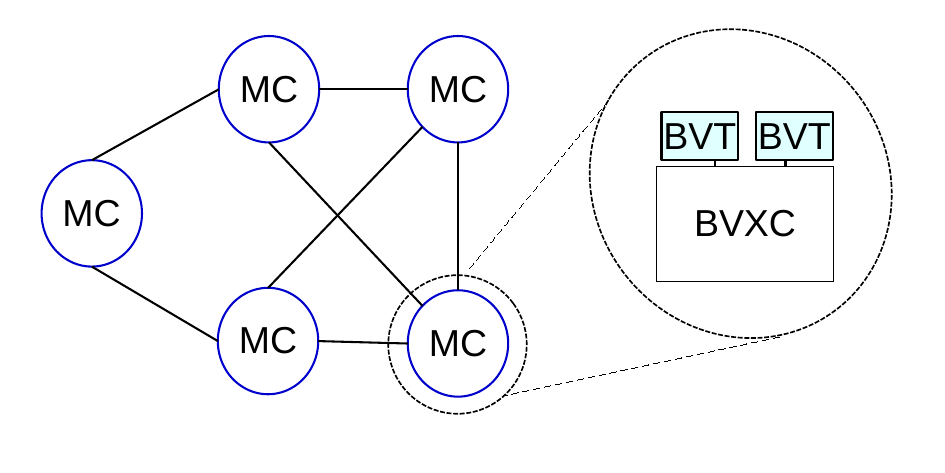}
	\caption{Metro Elastic Optical Network (MEON) architecture in mesh.}
	\label{fig:meon1}
\end{figure}

In the MEON architecture, as shown in Figure \ref{fig:meon1}, the nodes in the metro-core segment are composed of BVXC and several BVTs operating in an arbitrary and continuous set of spectrum slots with a granularity of $12.5$ $GHz$ in the frequency spectrum of the optical fiber, as highlighted in Subsection \ref{subsec:techMetro}. The flexibility required in this architecture can be achieved using different technologies, such as OOFDM, Nyquist WDM, and time-frequency packing (TFP)
 \cite{8596108, liu2018joint, 8999006}. While the architecture presented in \cite{rottondi2013routing} has a ring topology, the topology implemented in \cite{yan2018tidal, yan2020area, 8734478} and \cite{8853968}, for example, are mesh.

The MEON architecture is identified with both DEON and EON transmission systems. In the MEON architecture presented in \cite{8734478}, the nodes do not have a buffer (buffer-less cross-connect) and the fiber bandwidth of $4$ $THz$ is divided into $640$ frequency slots with a granularity of $6.25$ $GHz$, thus it is a DEON.
Additionally, some works approach MEON from the traditional EON with FSU of $12.5$ $GHz$, as in \cite{yan2018tidal,8596108,liu2018joint,wu2019analysis, iyer2020investigation}.

With this configuration, the node will have a lower cost in comparison with the cost of equipment of the same technology implemented in core networks. However, due to the lack of buffers, a request can be blocked immediately because there are no slots or BVT available, which results in the generation of a minimum delay in the network configuration. On the other hand, in \cite{8853968}  there are buffers at the nodes at the edge of the network. The transponder is fed by a packet queue in the limited size buffer in the electrical domain, where the traffic shaping process takes place. The network control plan, based on SDN, performs a three-stage process to configure the modulation format, spectrum width, and central frequency of these transponders so that the highest bit rates are attributed to the shortest transmission distances. Note that the electrical domain is considered only for the nodes that originate the transmission.
 Already in \cite{yan2020area} and \cite{yan2018tidal} the EON architecture is explored in the metropolitan network and the use of amplifiers is not mentioned, and $100$ slots are still considered for links with a bandwidth of $12.5$ $GHz$. These differences in attributes signal that, while a specific architecture is not defined, many technologies can be tested for performance analysis. 
 
\paragraph*{\textbf{Advantages}}
The large number of channels that can be configured is an advantage that can represent new business opportunities for operators, since the available capacity of spectral resources can surpass those of other architectures. The fine spectral granularity is ideal for serving a multitude of low data rate services that are common in metro networks. At the same time, with MEON, superchannels can be formed to meet large aggregated demands of underlying network segments. Reduced latency is also expected due to the establishment of transparent and  end-to-end lightpaths.

\paragraph*{\textbf{Disadvantages}}
The biggest challenge for the implementation of MEON architecture may be the high cost related to the acquisition and installation of new equipment, since the technology represents a major paradigm shift compared to the infrastructure in operation today.
Furthermore, the problem of spectral fragmentation resulting from the inefficient exploration of spectral capacity in the constant arrivals and departures of new connections, which generates holes in the spectrum, can be more complex to be solved in metropolitan networks, due to the greater number of service connections.

 \paragraph*{\textbf{Applications}}
 
MEON is designed for metro-core segment due the high capacity and spectral granularity capacity \cite{yan2018tidal, yan2020area}. However, the technology is likely to be applied in long-distance networks, especially for implementing various formats of modulation of the signal adaptable to distance \cite{din2019rbcmlsa}. A summary of MEON architecture highlights is shown in Table \ref{Tab:MEONCHighlights}.

\begin{table}[]
 \centering
	\caption{MEON main features.}
	\label{Tab:MEONCHighlights}
\begin{tabular}{|c|l|l|l|l|l|l|}
\hline
%\multicolumn{7}{|c|}{MEON Architecture Highlights}        \\ \hline
\textbf{Advantages}                                                                               
& \multicolumn{1}{c|}{\textbf{Disadvantages}}                                     
& \multicolumn{1}{c|}{\textbf{Applications}}                                 
& \multicolumn{1}{c|}{\textbf{Equipment}} 
& \multicolumn{1}{c|}{\textbf{Scope}} 
& \multicolumn{1}{c|}{\textbf{Topology}}
& \multicolumn{1}{c|}{\textbf{ToS}}\\ \hline

\multicolumn{1}{|l|}{\begin{tabular}[c]{@{}l@{}}Largest number of \\channels available\\ Transparency\end{tabular}} 
& \begin{tabular}[c]{@{}l@{}}Higher CAPEX; \\ Higher spectral \\fragmentation; \\\end{tabular} 
& \begin{tabular}[c]{@{}l@{}}Metro segments \\ interconnections\end{tabular} 
& \begin{tabular}[c]{@{}l@{}} BV-ROADM\\CO-BVT \end{tabular}   
& MC/MA 
& From Ring to Mesh       
& E2E       \\ \hline
\end{tabular}
\end{table}

\subsubsection{SIMON}

\paragraph*{\textbf{Composition and functioning}}

\begin{figure}
	\centering
	\includegraphics[width=0.8\linewidth]{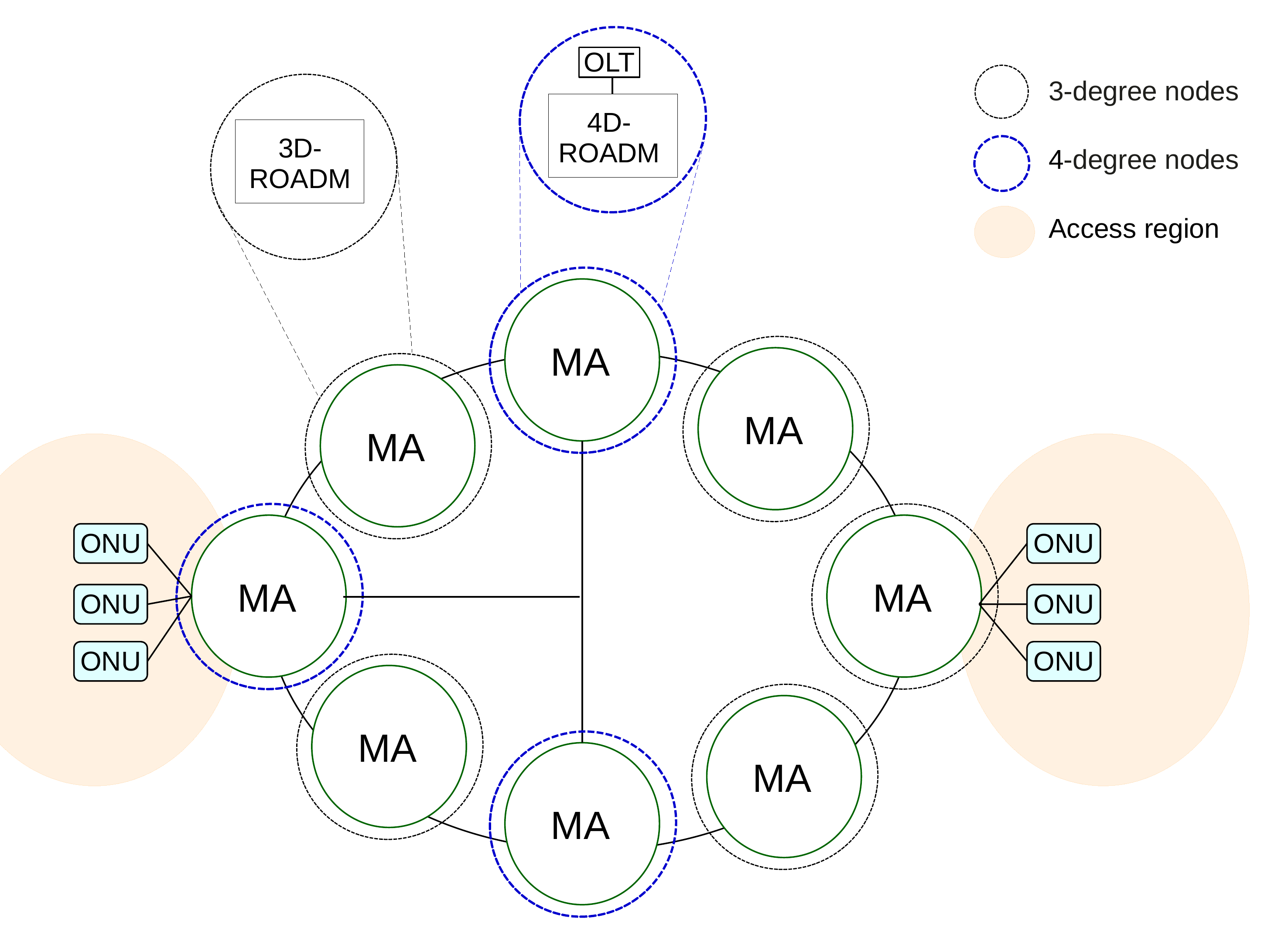}
	\caption{Star-in-ring topology of the proposed SDN-like Innovative Metro-Access Optical Network (SIMON) architecture.}
	\label{fig:simon}
\end{figure}

Described in \cite{muciaccia2019proposal}, the SDN-like Innovative Metro-Access Optical Network (SIMON) architecture, with a star-in-ring topology, was also designed to support the deployment of edge computing. As shown in Figure \ref{fig:simon}, a scheme is proposed to explain the composition of the SIMON architecture. The DWDM transmission system is implemented from the metro edge, that is, MA nodes, to the access segment region. All MA nodes are based on ROADMS. However, to balance cost and capacity, some of these nodes are based on $3$-degree , and others are $4$-degree ROADMS. Only nodes with higher degree connect directly to OLTs. The central topology is ring and has about $100$ DWDM channels. Because of this, each tree composed by ONU in the access region created from an MA node, can have up to $100$ units of ONU equipment, in transparent point-to-multipoint connections, based on DWDM-PON, in which each ONU corresponds to a wavelength.
A SDN controller provides centralized management. The role of the controller is to assign channels dynamically as requests from the ONU arrive to establish a possible optical path.

In Downstream (DS) direction, signals are allocated in the S and C band, while the upstream (US) signals are allocated in the O band. The bit rate that can be achieved on each channel can be configured at up to $10$ $Gb/s$ with one channel spacing of $50$ $GHz$ bandwidth and $100$ $GHz$.

\paragraph*{\textbf{Advantages}}
SIMON architecture is capable of facilitating resource management for dynamic traffic environments, and the centralized controller enables programmable operations. According to the authors \cite{muciaccia2019proposal}, the topology of this architecture can be modified to mesh at a low cost. The establishment of circuits takes place without the need for OEO conversion. Thus, the architecture guarantees reduced latency, especially for applications that require resources in the access segment, where physical distances are smaller.

\paragraph*{\textbf{Disadvantages}}

The number of ONUs defines the number of channels arranged in the ring network segment. This architecture is scalable to up to $100$ DWDM channels, that is, $100$ ONUs connected to a ROADM. If the number of ONUs is greater than this limit, then denser packing of wavelengths would be necessary. Due to this limitation and considering the fixed spectral grid transmission system, the scalability of the system in the long term and with reduced cost can be a difficult objective to be achieved. In this way, link overload can compromise the performance in the nodes of the smallest degree. The installation of new ONUs may require the implantation of new fibers, superimposed on the ring topology.

 \paragraph*{\textbf{Applications}}
The SIMON architecture was designed to bring mesh topology to access networks at a low cost. Even with the high CAPEX arising from the massive deployment of ROADMs, the architecture may be able to ensure the trade-off between cost and efficiency. Thus, this infrastructure can especially support mobile computing and edge computing services and applications. A summary of SIMON architecture highlights is shown in Table \ref{Tab:SIMONHighlights}.

\begin{table}[]
 \centering
	\caption{SIMON main features.}
	\label{Tab:SIMONHighlights}
\begin{tabular}{|c|l|l|l|l|l|l|}
\hline
%\multicolumn{7}{|c|}{SIMON Architecture Highlights}        \\ \hline
\textbf{Advantages}                                                                               
& \multicolumn{1}{c|}{\textbf{Disadvantages}}                                     
& \multicolumn{1}{c|}{\textbf{Applications}}                                 
& \multicolumn{1}{c|}{\textbf{Equipment}} 
& \multicolumn{1}{c|}{\textbf{Scope}} 
& \multicolumn{1}{c|}{\textbf{Topology}}
& \multicolumn{1}{c|}{\textbf{ToS}}\\ \hline

\multicolumn{1}{|l|}{\begin{tabular}[c]{@{}l@{}}Support to edge computing;\\ Costs reduced;\end{tabular}} 
& \begin{tabular}[c]{@{}l@{}}Limited scalability \\to 100 OLT/DWDM\\ channels; \\\end{tabular} 
& \begin{tabular}[c]{@{}l@{}}Metro segments \\ interconnections\end{tabular} 
& \begin{tabular}[c]{@{}l@{}} MD-ROADM\\OLT\\ONU \end{tabular}   
& MA 
& Star-in-Ring  
& P2MP\\ \hline
\end{tabular}
\end{table}

\subsubsection{TDGMON}

 \paragraph*{\textbf{Composition and functioning}}

\begin{figure}
	\centering
	\includegraphics[width=0.8\linewidth, page=2]{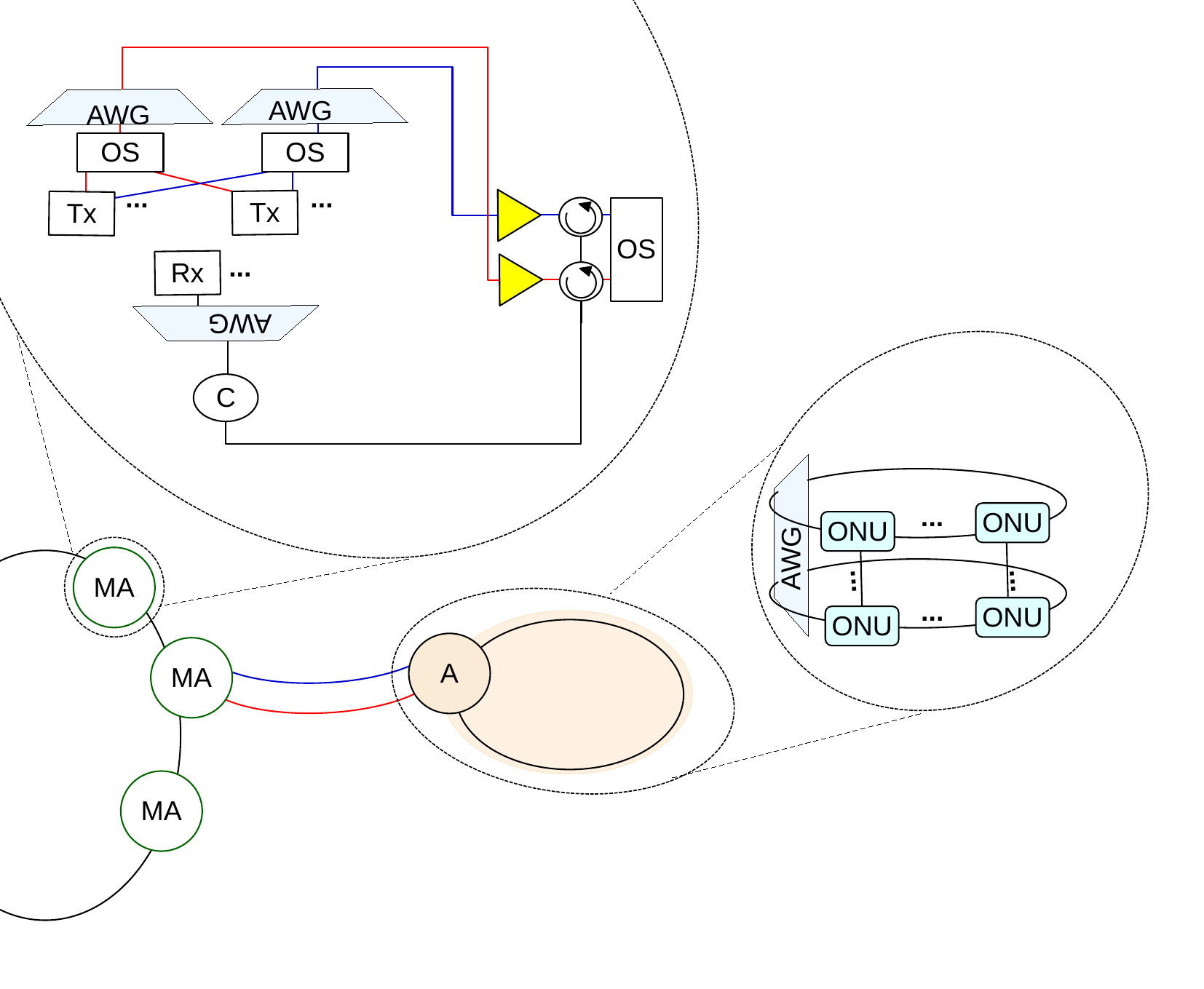}
	\caption{Representative scheme of Three-dimensional Architecture of the Grid Network, adapted from \cite{lin2019three}.}
	\label{fig:grid}
\end{figure}

In \cite{lin2019three} the Three-dimensional Grid Metro-Access Optical Network (TDGMON) architecture that includes the metro-access segment and the access segment is presented. The idea is to expand the number of lines available in the access segment with the implementation of nodes in a three-dimensional grid. In the MA node, the CO is located, while in node A, seen as a remote node (RN), several transversal and longitudinal rings are implanted along with the ONUs, as shown in Figure \ref{fig:grid}. In the connections shown between both nodes, two optical fibers are used (red and blue colors), one of which is the protection link. In the CO there are tunable Tx and Rx, which use differential phase-shift keying (DPSK) modulation format, ROADM in $1:2$ configuration, in addition to optical circulators, EDFA, MUX, and DEMUX amplifiers. This node configuration is very similar to that of other architectures shown earlier.
Node A is based on NxN AWG in CWDM connecting several rings with ONUs. In optical fiber interconnections between ONUs, the optical spectrum is divided into spectrum intervals called free spectral ranges (FSRs), with the wavelength allocation occurring periodically in fixed paths with peer-to-peer communication. Simultaneously, signal transmission can occur in US (from right to left), DS (from left to right) and in direct communication, in the transversal and longitudinal directions.

\paragraph*{\textbf{Advantages}}

The architecture also implements four levels of protection, described as follows: (\textit{i}) protection against fiber failure of the wavelength feeder, in cases where the failure occurs in the working fiber between CO and RN and a port exchange is performed on the optical switch so that the transmission to the protection fiber is switched and the tunable transmitter emits the wavelengths on that designated fiber; (\textit{ii}) Distribution fiber failure protection, similar to the previous mechanisms, but guaranteeing resilience to multiple distribution fiber failures while maintaining power; (\textit{iii}) Protection against fiber interconnection failure, in the case where the cross interconnection fiber between two adjacent ONUs breaks, and this point of failure in the ring then marks the existence of an upper half and a lower half .
The ONUs of the upper half of the ring continue to function in normal work mode, with signal transmission clockwise, and the ONUs of the lower half of the ring functions in protection mode, with signal transmission in a counterclockwise direction. The reverse of these transmission directions is also possible, and besides, these rules still apply to rings formed in the longitudinal direction, this being the fourth level of protection, since the paths are fixed.

\paragraph*{\textbf{Disadvantages}}

The authors guarantee that the proposed architecture can support a theoretical limit of up to $512$ ONUs and that it has a high degree of reliability due to the four existing protection levels. However, some additional studies are needed, including on the energy consumption of this infrastructure.

 \paragraph*{\textbf{Applications}}

The SIMON architecture was designed to bring mesh topology to access networks at a low cost. Even with the high CAPEX arising from the massive deployment of ROADMs, the architecture may be able to ensure the trade-off between cost and efficiency. Thus, this infrastructure can especially support mobile computing and edge computing services and applications. A summary of TDGMON architecture highlights is shown in Table \ref{Tab:TDGMONHighlights}.

\begin{table}[]
 \centering
	\caption{TDGMON main features.}
	\label{Tab:TDGMONHighlights}
\begin{tabular}{|c|l|l|l|l|l|l|}
\hline
%\multicolumn{7}{|c|}{TDGMON Architecture Highlights}        \\ \hline
\textbf{Advantages}                                                                               
& \multicolumn{1}{c|}{\textbf{Disadvantages}}                                     
& \multicolumn{1}{c|}{\textbf{Applications}}                                 
& \multicolumn{1}{c|}{\textbf{Equipment}} 
& \multicolumn{1}{c|}{\textbf{Scope}} 
& \multicolumn{1}{c|}{\textbf{Topology}}
& \multicolumn{1}{c|}{\textbf{ToS}}\\ \hline

\multicolumn{1}{|l|}{\begin{tabular}[c]{@{}l@{}} Four levels of protection\\ against failure\\Support theoretical limit\\ of up to 512 ONUs;\end{tabular}} 
& \begin{tabular}[c]{@{}l@{}}More EDFAs must \\be employed in the\\CO (increasing costs); \\\end{tabular} 
& \begin{tabular}[c]{@{}l@{}}Metro segments \\ interconnections\end{tabular} 
& \begin{tabular}[c]{@{}l@{}} ROADM\\AWG\\EDFA\\CO-Transponders\\ONU \end{tabular}   
& MA 
& Star-in-Ring
& P2MP\\ \hline
\end{tabular}
\end{table}

\centerline{\rule{0.4\textwidth}{0.4pt}}
\bigskip

\textbf{Filterless (FL) MON}\label{subsec:fullyFL} - FL network solutions are built based on passive elements such as amplifiers, splitters, and couplers (covered in Subsection \ref{subsubsec:PON}) that are in charge of creating branches with optical nodes and interconnecting network links while giving up ROADMs and WSSs on intermediate nodes~\cite{paolucci2020disaggregated} or considerably minimize the number of these devices in use in the architecture \cite{Mantelet:13}. 
These elements are already used frequently in the context of access networks, as well as in the transport segment and underwater networks~\cite{paolucci2020disaggregated}. 
The main advantages of filterless networks are the low cost and simplicity of the infrastructure and topologies. However, in networks without filters, a greater number of wavelengths must be used for communication in comparison with networks with filters, since it is natural for the power fluctuations of the conjugated signal of these wavelengths to occur while crossing a cascade of amplifiers, route in which there are no devices to provide equalization and control in power levels.

Recently, some MON architectures have been highlighted in \cite{paolucci2020disaggregated, kosmatos2019building} and, in particular, a specific architecture called Dual
Fibre Network (DUFiNet) is presented in \cite{uzunidis2018dufinet}, while the Broadcast-and-Select (BnS) architecture is presented in \cite{ayoub2018filterless}. 

\subsubsection{FMN} \label{subsubsec:FMN}

\paragraph*{\textbf{Composition and functioning}}

\begin{figure}
	\centering
	\includegraphics[width=0.8\linewidth, page=3]{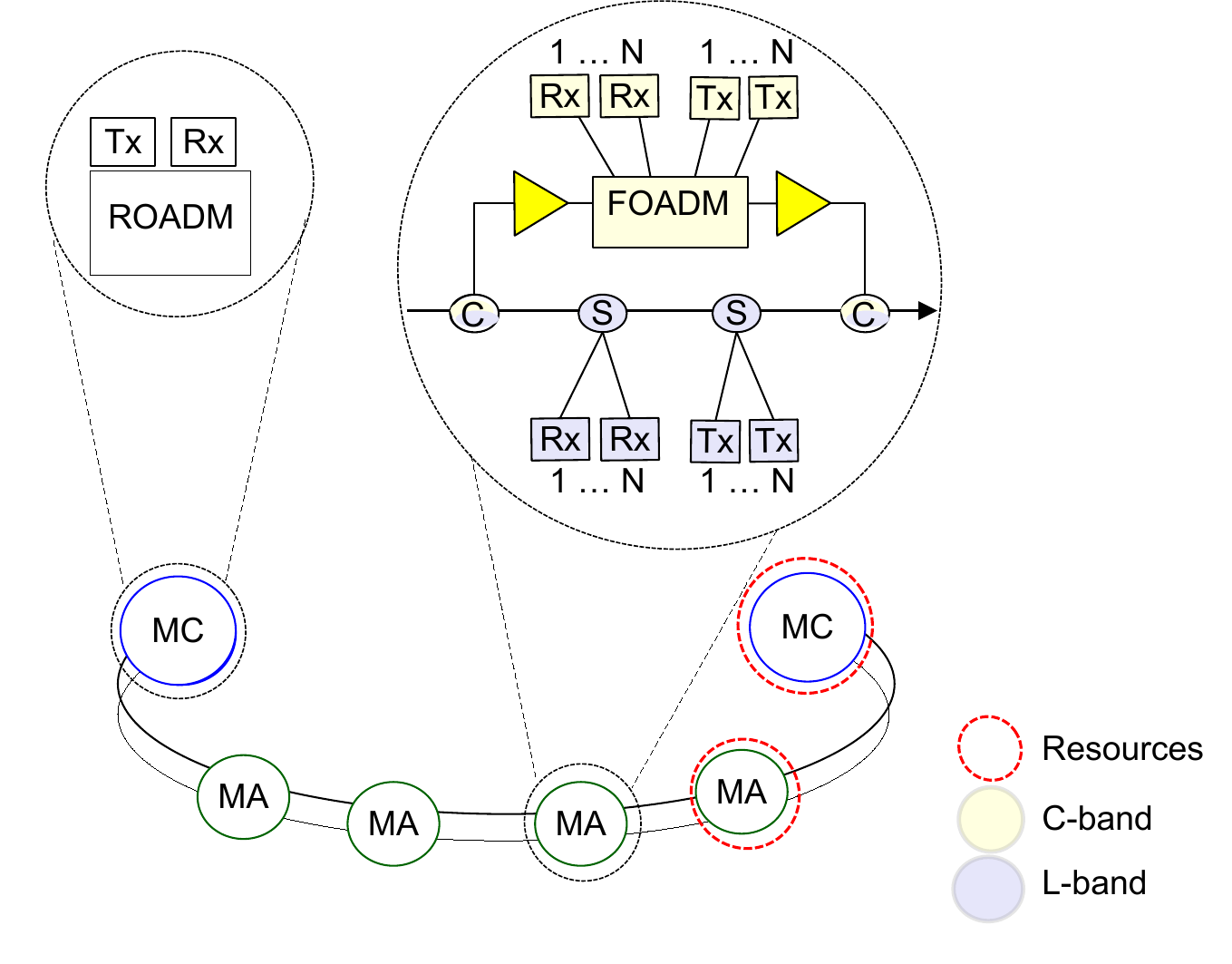}
	\caption{Filterless Metropolitan Network (FMN) architecture.}
	\label{fig:fmn}
\end{figure}

The Filterless Metropolitan Network (FMN) architecture is presented in \cite{paolucci2020disaggregated}, and illustrated in Figure \ref{fig:fmn}. The architecture works in two distinct frequency bands: C band and L band. With respect to L-band, FMN is considered filterless, but with respect to C-band, FMN is considered with filter.
Nodes from the MC and MA segments in horseshoe topology are shown. The MC nodes on the edge of the horseshoe are interconnected by means of two fibers representing paths for US and DS communication. In any of these directions, there are already pre-configured fixed channels at a certain frequency for each of these nodes. Coherents transponders (Tx and Rx) are implemented in all nodes providing channels with different bitrates, modulation formats, and FECs.
MC nodes have filtering properties because they are WSS-based ROADMs, while MA nodes, considered transit nodes, have FOADM for C-band (in yellow in the picture) but it does not have any type of filter for the L-band (in purple in the figure). Couplers are the only devices that can be shared by both bands. MA nodes additionally have splitters and EDFA amplifiers to handle the signal at this point. MC nodes are interfaces to the core network (backbone), and also perform broadcast communication with MA nodes. The red dotted line highlights that some of the nodes in the architecture can be equipped with additional processing and storage resources, and thus, the FMN is able to handle services that require low latency.

FMN has strong potential to extend the transmission capacity beyond the frequency C band, increasing the number of channels through the inclusion of the L band, and can be modeled as a set of point-to-point optical line systems (OLS).
In \cite{paolucci2020disaggregated} two possible types of multiband architecture are proposed for FMN: C + L-FMN Single Region and C + L-FMN Dual Region. The basic difference is that, as the development process of these two new architectures is mainly based on updating EDFA amplifiers to meet the C and L bands, the deployment of equipment of this nature can occur in only part of the network or the network as a whole. 

% Figura removida
%\begin{figure}
%	\centering
%	\includegraphics[width=0.99\linewidth]{Figure/EDFA}
%	\caption{Amplification options in optical network}
%	\label{fig:edfa}
%\end{figure}

In the C + L-FMN Single Region network architectures, all the amplifiers present (line and drop EDFA) are hybrid equipment, that is, they operate in both the C and L bands, and for this reason, they allow signal transmission on the network entire. In the C + L-FMN Dual Region architecture, the network is divided into two band reception domains, one with amplifiers for the C band and the other for the L band, making it possible to reduce the total cost of the upgrade. As the load on the network is continuously increasing, hybrid amplifiers are a promising technology and widely used to amplify large numbers of channels with narrow spacing and better performance.  

In \cite{uzunidis2021Bidirecional}, FMN is covered with C-band filterless configuration, where MA nodes do not have filters. Instead, employing bidirectional transmission in FMN using a single fiber for US and DS traffic.

\paragraph*{\textbf{Advantages}}

In general, FMN can provide $100$ $Gb/s$ and $400$ $Gb/s$ connections at a reduced cost. Multiband architectures are capable of providing a greater number of channels and, therefore, can represent cost-effective solutions over time.

\paragraph*{\textbf{Disadvantages}}

Choosing one of of possible FMN configurations should consider the cost impact to fully upgrade the network by deploying new hybrid amplifiers, splitters, and couplers, or performing this task in only one domain while taking advantage of the remaining legacy equipment. 

 \paragraph*{\textbf{Applications}}

The FMN architecture serves the MC and MA segment, providing point-to-point communication between the MA nodes and end-to-end, between the MA and MC nodes (or vice versa). A summary of FMN architecture highlights is shown in Table \ref{Tab:FMNHighlights}.

\begin{table}[]
 \centering
	\caption{FMN mais features.}
	\label{Tab:FMNHighlights}
\begin{tabular}{|c|l|l|l|l|l|l|}
\hline
%\multicolumn{7}{|c|}{FMN Architecture Highlights}        \\ \hline
\textbf{Advantages}                                                                               
& \multicolumn{1}{c|}{\textbf{Disadvantages}}                                     
& \multicolumn{1}{c|}{\textbf{Applications}}                                 
& \multicolumn{1}{c|}{\textbf{Equipment}} 
& \multicolumn{1}{c|}{\textbf{Scope}} 
& \multicolumn{1}{c|}{\textbf{Topology}}
& \multicolumn{1}{c|}{\textbf{ToS}}\\ \hline

\multicolumn{1}{|l|}{\begin{tabular}[c]{@{}l@{}} Multiband\\Low-latency for\\some services; \end{tabular}} 
& \begin{tabular}[c]{@{}l@{}}More equipment \\ to be employed \\ (increasing costs); \\\end{tabular} 
& \begin{tabular}[c]{@{}l@{}}Metro segments \\ interconnections\end{tabular} 
& \begin{tabular}[c]{@{}l@{}} ROADM\\FOADM\\EDFA\\CO-Transponders\\Couplers\\Splitters \end{tabular}   
& MC and MA 
& Horseshoe
& P2P\\ \hline
\end{tabular}
\end{table}

\subsubsection{DuFiNet}

\paragraph*{\textbf{Composition and functioning}}

\begin{figure}
	\centering
	\includegraphics[width=0.6\linewidth, page=2]{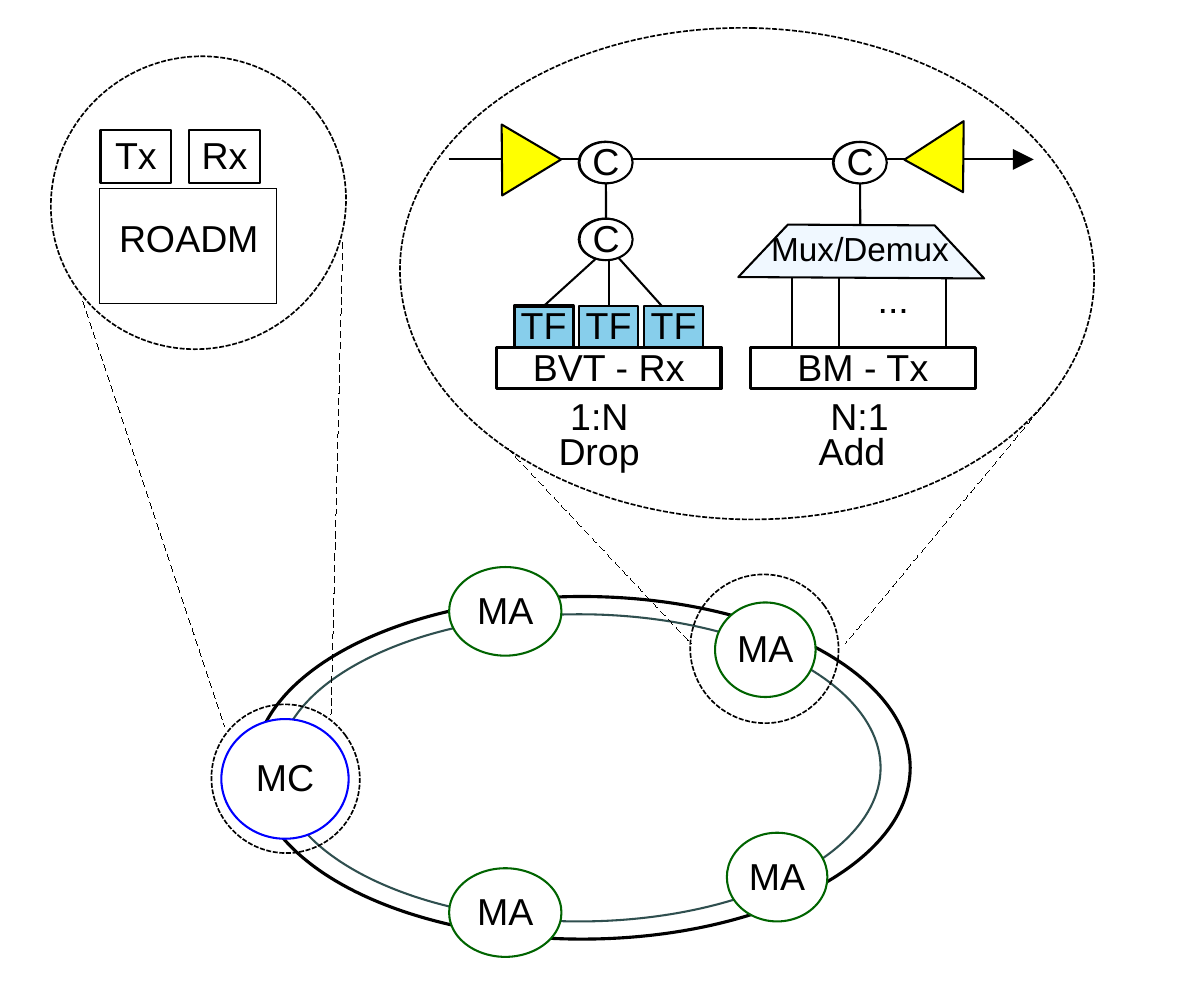}
	\caption{Representation of the filterless Dual
		Fibre Network (DuFiNet) architecture.}
	\label{fig:figura1}
\end{figure}

In \cite{uzunidis2018dufinet} the Dual Fibre Network (DuFiNet) architecture with ring topology and point-to-multipoint communication (P2MP) is proposed. It is a filterless solution becouse the wavelength carriers cannot be
reused. PON elements are reused from legacy architecture and SDN-based control plan. It consists of nodes from the segments MC and MA, connected by two fibers (US and DS), which are necessary at the edge of their respective segments, acting as gateway and aggregators, as shown in Figure \ref{fig:figura1}. 
The MC is a WSS-based ROADM while the MA couplers, small tunable filters (TF), mux/demux for a type of communication (DS), amplifiers and coherent transponders, called BVT for receiving, and Burst Mode (BM) for transmission. BM transponder is used for dynamic TDMA-sharing. BVT transponder is a "lite" model for coherent $100$ $G$ channels.
In this architecture, there is no direct connectivity between the MA nodes. For the communication between two MAs to occur, traffic at the destination is added in a specific channel in the US direction, which is concentrated at the MC node, and then forwarded on the DS to the destination MA.

In the MC node, there are two modes of operation, called opaque and transparent. In opaque mode, the channels are terminated at the MC, are regenerated and aggregated before being forwarded to the core tracking or DS fiber. DuFiNet is easily scalable because it tolerates the addition of new MA rings by leveraging and sharing the MC node infrastructure. In this case, when a metropolitan network is formed by more than one DuFiNet, this metropolitan network becomes a semi filterless network, due to the sharing of the node by all defined DuFiNet.

The main advantages of implementing DuFiNet are the low cost of deploying and operating the nodes without a filter, as well as the facility to update the infrastructure with the addition of new equipment. However, effects of signal degradation can be seen, which is typical of unfiltered network ecosystems. The main similarity between DuFiNet and the FMN architecture (Subsection \ref{subsubsec:FMN}) is that both concentrate the MC and MA segments and are multiband. However, while FMN has a horseshoe topology and operates in the C and L band, DuFinet has a ring topology and can transmit in the C, O, E, S, and L. bands.

\paragraph*{\textbf{Advantages}}

The DuFiNet architecture is a low-cost solution for medium-haul. It is based on transparent and opaque transport mode and is a viable multiband solution for communicating over a large portion of the optical spectrum (far beyond the C-band). Although it does not have intermediate node filters, the signal performance is kept within limits with the use of lite tunable filters.

\paragraph*{\textbf{Disadvantages}}

Although it is a PON-based unfiltered architecture, dufinet implements opaque communication modes and provides channels of maximum $100$ $G$. A careful cost-performance analysis regarding the potential energy consumption due to increased traffic is necessary. BM technology is a specific solution for DuFiNet that cannot be supported from other filterless solutions of filtered solutions, and can imply the acquisition of new equipment for the implementation of the architecture.

 \paragraph*{\textbf{Applications}}
 
 DuFiNet is feasible for metro-access interconnections, being appropriate for medium-haul, while support future 5G services. It can be easily scaled to increase the number of connected MAs. A summary of DuFiNet architecture highlights is shown in Table \ref{Tab:DuFiNetHighlights}.

\begin{table}[]
 \centering
	\caption{DuFiNet main features.}
	\label{Tab:DuFiNetHighlights}
\begin{tabular}{|c|l|l|l|l|l|l|}
\hline
%\multicolumn{7}{|c|}{DuFiNet Architecture Highlights}        \\ \hline
\textbf{Advantages}                                                                               
& \multicolumn{1}{c|}{\textbf{Disadvantages}}                                     
& \multicolumn{1}{c|}{\textbf{Applications}}                                 
& \multicolumn{1}{c|}{\textbf{Equipment}} 
& \multicolumn{1}{c|}{\textbf{Scope}} 
& \multicolumn{1}{c|}{\textbf{Topology}}
& \multicolumn{1}{c|}{\textbf{ToS}}\\ \hline

\multicolumn{1}{|l|}{\begin{tabular}[c]{@{}l@{}} Multiband\\Low-cost\\Upgradeability \end{tabular}} 
& \begin{tabular}[c]{@{}l@{}}Opaque mode \\ Only 100G per Ch; \\\end{tabular} 
& \begin{tabular}[c]{@{}l@{}}Metro segments \\ interconnections\end{tabular} 
& \begin{tabular}[c]{@{}l@{}} ROADM\\EDFA\\CO-Transponders\\(BVT and BM)\\Couplers\\Mux-Demux \end{tabular}   
& MC and MA 
& Ring
& P2MP\\ \hline
\end{tabular}
\end{table}

\centerline{\rule{0.4\textwidth}{0.4pt}}
\bigskip

\textbf{Semi-Filterless (sFL) MON}\label{subsubsec:metroSemiFL} - While filterless architectures are particularly suitable for metropolitan network scenarios of limited topological complexity, without recirculation of optical energy \cite{cugini2018flexibleSemiFL}, and given that fully filtered architectures are economically costly, semi-filterless architectures can represent the trade-off among reach, infrastructure complexity and costs.

Two different definitions for sFL optical networks are found in the literature:
\begin{itemize}
	\item (i) sFL networks are networks composed of the combination of WF networks and FL networks, that is, some nodes have no filter while others are equipped with a filter. For this reason, optical network architecture with these characteristics is also called heterogeneous architecture~\cite{ayoub2018filterless};
	\item (ii) sFL networks are networks in which all traffic nodes have an adjustable optical filter (OF) integrated into the receiver interface, with low cost, capacity, and complexity. Due to its simplicity, a node with OF cannot be compared to a robust node based on AWG or WSS, for example~\cite{8346157SemiFilterless}. According to this concept, an sFL network has improved performance in terms of BER, compared to FL networks, while still ensuring an adequate cost-benefit ratio.
\end{itemize}

The architectures identified according to such definitions are highlighted below, referred to as Metro-Haul Architecture\ref{Metro-Haul_SemiFL} and Drop-and-Waste (DnW) Architecture \ref{DropNWaste_SemiFL}.

\subsubsection{Metro-Haul}\label{Metro-Haul_SemiFL}

\paragraph*{\textbf{Composition and functioning}}

\begin{figure}
	\centering
	\includegraphics[width=0.8\linewidth, page=2]{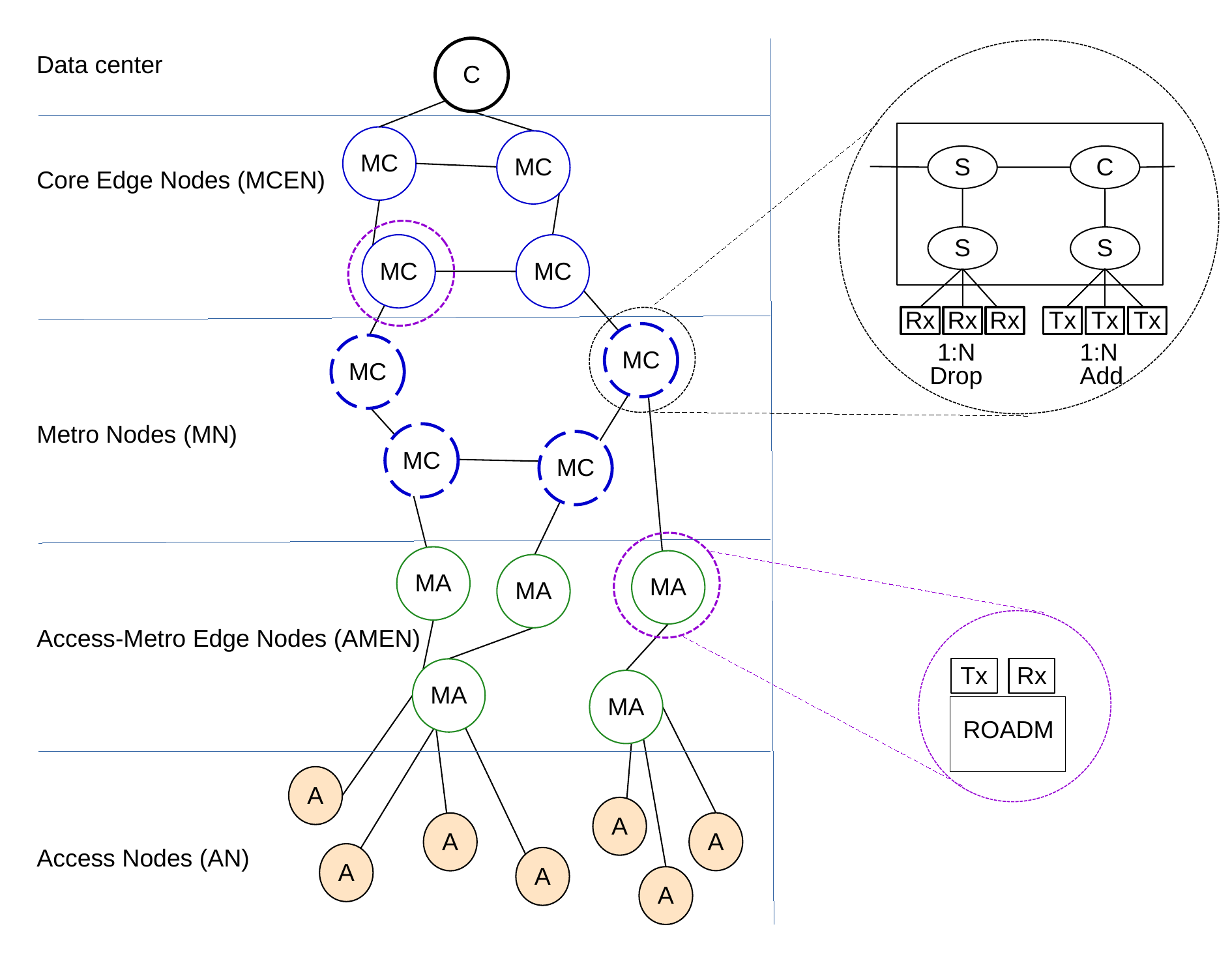}
	\caption{Representation of the semi-filterless Metro-Haul Architecture.}
	\label{fig:MetroHaulArchitecture}
\end{figure}

An sFL architecture in line with the definition (\textit{i}) is highlighted in \cite{ayoub2018filterless}, and it is proposed in the context of MetroHAUL project \cite{metro-haulproject}, as cited by the authors.
 The architecture presented is based on an EON transmission system, being shown as a flexible solution to meet the requirements of metropolitan networks for the future while ensuring cost-effectiveness. The Metro-Haul architecture is shown in Figure \ref{fig:MetroHaulArchitecture}. Four categories of nodes are observed in this architecture: MCEN, MN, AMEN and AN. These categories are used to specifically define the role of a node. AMENs and MCENs work as mini and regional data centers, respectively. While ANs are interconnected to AMENs in tree topologies, AMENs are interconnected to MN in ring topology, and these interconnect MCEN into mesh topologies. The entire architecture implements coherent transponders. The figure shows a node configuration in which the MN segment is equipped with passive and unfiltered elements, while the other nodes MCEN and AMENs are based on ROADM. However, other combinations were tested such as filterless AMENs.
 
 \paragraph*{\textbf{Advantages}}
 
 In the study that addresses Metro-Haul architecture \cite{ayoub2018filterless}, three different network construction scenarios are compared, two of which are partially unfiltered, and a third scenario is FL. The conclusion obtained is that the greater the number of nodes that implement architecture without filter, the lower the cost of that architecture. Also, a partially unfiltered architecture performs better in the use of spectrum resources than FL architectures. 

\paragraph*{\textbf{Disadvantages}}

According to the authors \cite{ayoub2018filterless}, in a hierarchical network with mesh topology, the main problem related to the construction of the infrastructure is to identify the ideal locations to accommodate nodes with filters and nodes without filter. Depending on the traffic and organization of the network architecture, the simple reduction in the number of WSS deployed can lead to a decrease in performance. 

 \paragraph*{\textbf{Applications}}
  
The Metro-Haul architecture meets all hierarchical levels, making it viable for edge computing and cloud computing solutions, as it provides direct end-to-end communication between the various types of nodes and available resources.A summary of Metro-Haul architecture highlights is shown in Table \ref{Tab:MetroHaulHighlights}.

\begin{table}[]
 \centering
	\caption{Metro-Haul main features.}
	\label{Tab:MetroHaulHighlights}
\begin{tabular}{|c|l|l|l|l|l|l|}
\hline
%\multicolumn{7}{|c|}{Metro-Haul Architecture Highlights}        \\ \hline
\textbf{Advantages}                                                                               
& \multicolumn{1}{c|}{\textbf{Disadvantages}}                                     
& \multicolumn{1}{c|}{\textbf{Applications}}                                 
& \multicolumn{1}{c|}{\textbf{Equipment}} 
& \multicolumn{1}{c|}{\textbf{Scope}} 
& \multicolumn{1}{c|}{\textbf{Topology}}
& \multicolumn{1}{c|}{\textbf{ToS}}\\ \hline

\multicolumn{1}{|l|}{\begin{tabular}[c]{@{}l@{}} Low-cost;\\Efficient use \\of spectrum; \end{tabular}} 
& \begin{tabular}[c]{@{}l@{}}Requires ideal locations\\to accommodate nodes;\\Spectral Fragmentation \\\end{tabular} 
& \begin{tabular}[c]{@{}l@{}}Metro segments \\ Interconnections\end{tabular} 
& \begin{tabular}[c]{@{}l@{}} ROADM\\CO-Transponders \\Couplers\\splitters \end{tabular}   
& MC and MA 
& Many
& E2E\\ \hline
\end{tabular}
\end{table}

\subsubsection{DnW}\label{DropNWaste_SemiFL}

\paragraph*{\textbf{Composition and functioning}}

\begin{figure}
	\centering
	\includegraphics[width=0.6\linewidth]{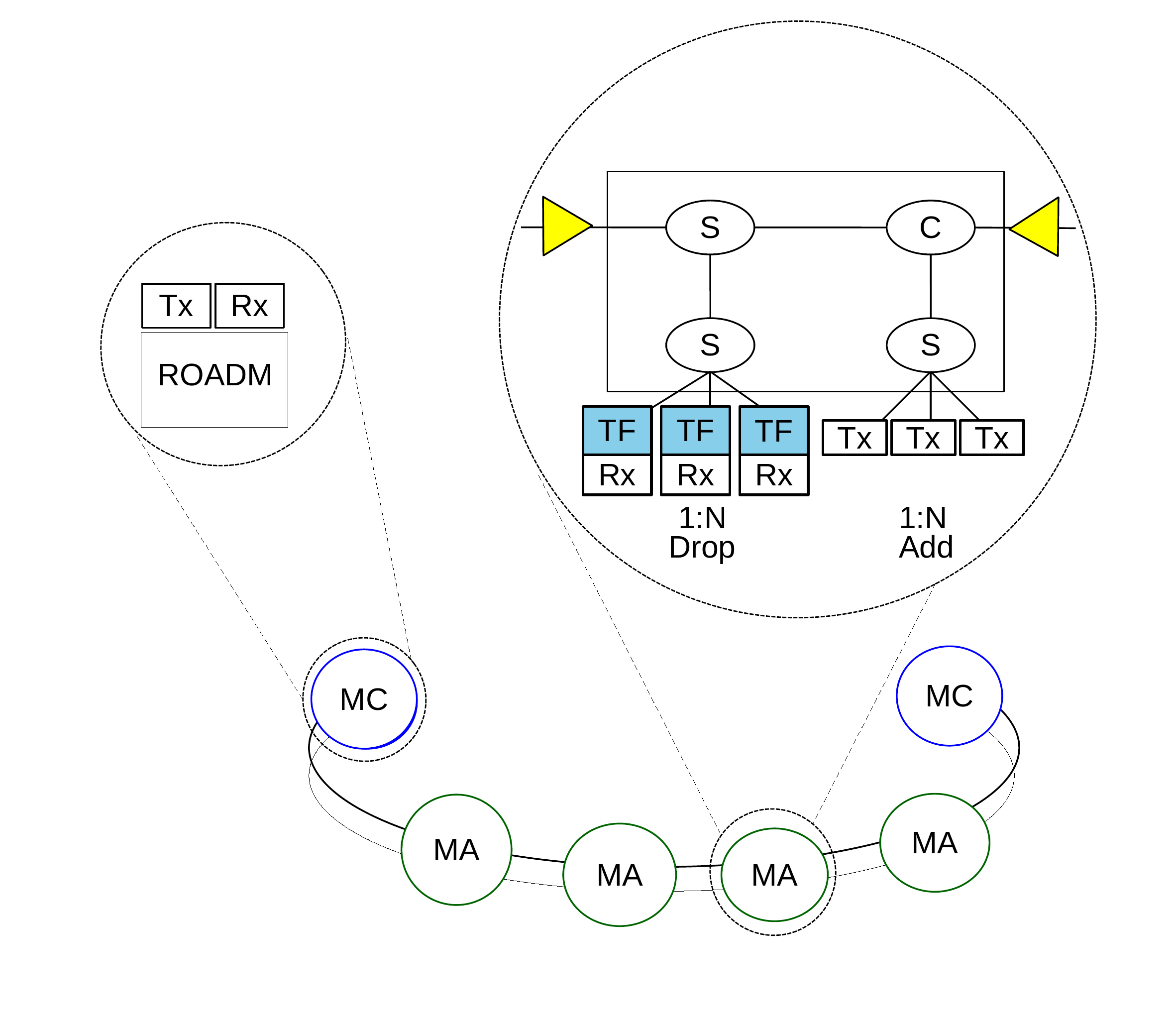}
	\caption{Representation of the Drop-and-Waste (DnW) Architecture.}
	\label{fig:DnWArchitecture}
\end{figure}

In \cite{8346157SemiFilterless} an sFL network architecture is presented according to the definition (\textit{ii}).The Drop-and-Waste (DnW) architecture featured is based on DWDM transport technology, has $80$ DWDM channels in the C band with $50$ $GHz$ and horseshoe topology. A schematic of the DnW architecture is shown in Figure \ref{fig:DnWArchitecture}. The MC nodes are equipped with ROADMs and coherent transponders. The MA nodes are equipped with typical passive elements and coherent transceivers are equipped with lite-type tunable optical filters, which constitutes the differential of this architecture in relation to the other highlighted architectures.

The tunable optical filter are low-cost device integrated to the receiver, being based on silicon-on-insulator (SOI) photonic technology able to reduce the signal strength at reception, keeping it within acceptable limits of restriction for OSNR and BER.

According to the article, the transmission rate in this architecture is $25$ $Gb/s$ (with OOK modulation) and $56$ $Gb/s$ (with QPSK modulation). On the receiving side, the signal passes through the optical filter before being detected by the coherent receiver. The integrated filter limits the optical power entering the receiver to an average of $9$ channels, instead of $96$ as in the case of nodes that have a filter. This semi-filterless system is successful in a scenario with up to a maximum of $12$ adjacent nodes with the adjustable optical filter implanted and a span of $80$ $km$, having a maximum optical signal range of around $960$ $km$ between the two ends of the topology~\cite{cugini2016receiver}.  Services are connected in a point-to-point manner as each MA connects to the MC nodes via fixed wavelengths..

\paragraph*{\textbf{Advantages}}

DnW \cite{8346157SemiFilterless} gets its name due to the architecture of its transit nodes. These nodes have a low cost, low power adjustable optical filter inserted into the Rx interface and this module has little impact on the total cost of the network.

\paragraph*{\textbf{Disadvantages}}
The range restrictions and available rates may not represent a good compromise between cost and performance, especially in the metropolitan network segment, which is composed of a large number of nodes. The main impacts arising from these issues are the impossibility of scaling the architecture, as well as the potential risk of spectral bandwidth waste due to wide and fixed channels, which do not match the main demands of the metropolitan network.

 \paragraph*{\textbf{Applications}}

The DnW architecture can satisfactorily serve the interconnections of the MA segments, being a viable and low-cost way to offer service for the applications that exist today. A summary of DnW architecture highlights is shown in Table \ref{Tab:DnWHighlights}.

\begin{table}[]
 \centering
	\caption{DnW main features.}
	\label{Tab:DnWHighlights}
\begin{tabular}{|c|l|l|l|l|l|l|}
\hline
%\multicolumn{7}{|c|}{DnW Architecture Highlights}        \\ \hline
\textbf{Advantages}                                                                               
& \multicolumn{1}{c|}{\textbf{Disadvantages}}                                     
& \multicolumn{1}{c|}{\textbf{Applications}}                                 
& \multicolumn{1}{c|}{\textbf{Equipment}} 
& \multicolumn{1}{c|}{\textbf{Scope}} 
& \multicolumn{1}{c|}{\textbf{Topology}}
& \multicolumn{1}{c|}{\textbf{ToS}}\\ \hline

\multicolumn{1}{|l|}{\begin{tabular}[c]{@{}l@{}} Low-cost;\\Tunable lite filter; \end{tabular}} 
& \begin{tabular}[c]{@{}l@{}}Few channels;\\
Fixed grid channels;\\
Range limitation; \end{tabular} 
& \begin{tabular}[c]{@{}l@{}}MA segments; \\ Interconnections\end{tabular} 
& \begin{tabular}[c]{@{}l@{}} ROADM\\CO-Transponders \\Couplers\\splitters \end{tabular}   
& MC and MA 
& Horseshoe
& P2P\\ \hline
\end{tabular}
\end{table}

%%%%%%%%%%%%%%%%%%%%%%%%%%%%%%%%%%%%%%%%%%%%%%%%%%%%%%%%%%%%%%
%%%%%%%%%%%%%%%%%%%%%%%%%%%%%%%%%%%%%%%%%%%%%%%%%%%%%%%%%%%%%%
%%%%%%%%%%%%%%%%%%%%%%%%%%%%%%%%%%%%%%%%%%%%%%%%%%%%%%%%%%%%%%
\section{Comparative Analysis of Featured Architectures}\label{sec:comparingArch}

\begin{figure}
	\centering
	\includegraphics[width=0.8\linewidth]{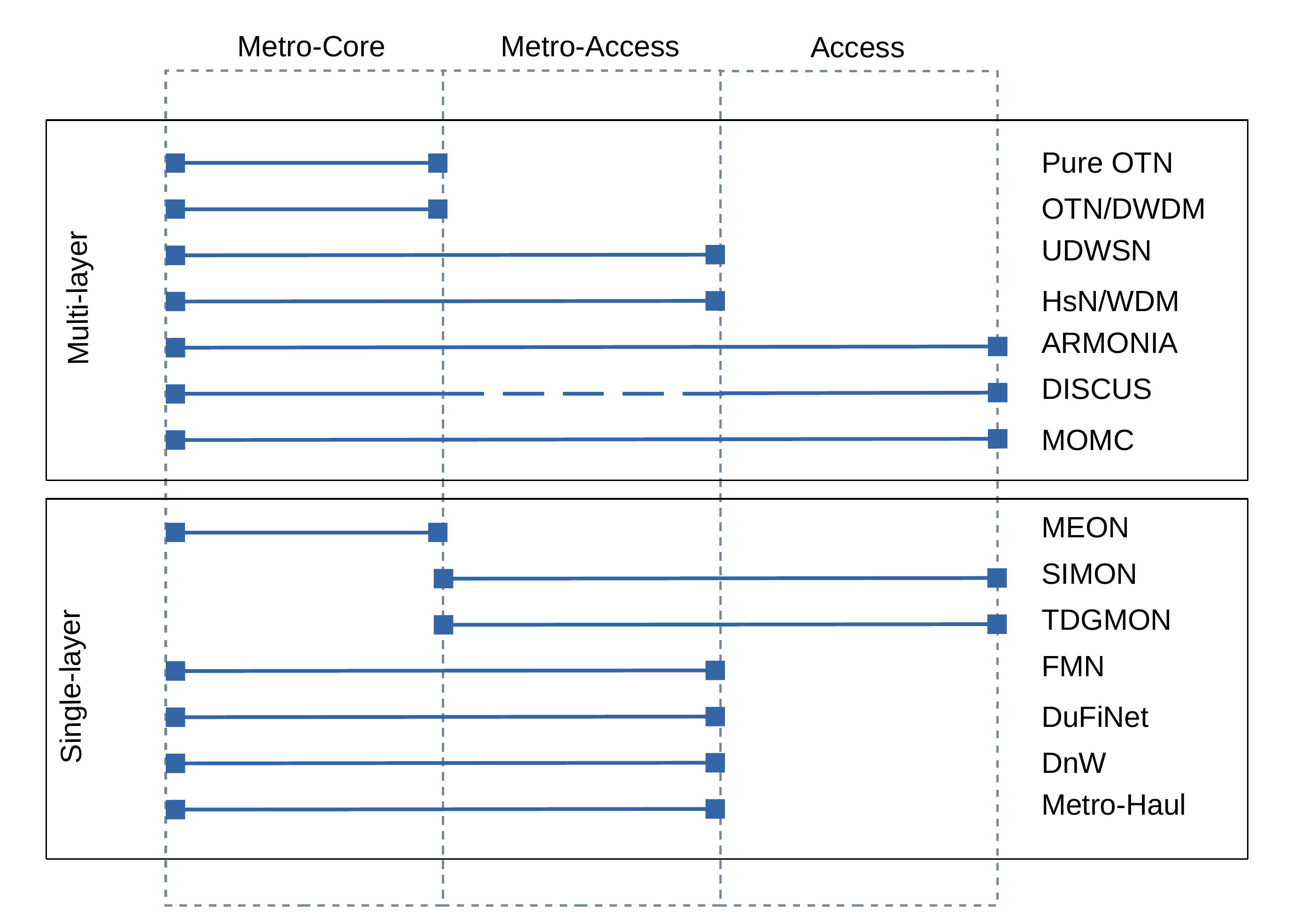}
	\caption{Network segment scope of the highlighted architectures.}
	\label{fig:mapeamento}
\end{figure}

This section aims to list the network architectures presented and to highlight common points for a comparative analysis. The selection of architectures was made based on the state of the art literature on metropolitan optical networks, identifying and highlighting the solutions of the last $8$ years, as shown in the Table \ref{tab:papersByArchitecture}. 
Legacy architectures, such as Pure OTN, for example, are not recent but are often cited as an alternative for comparison with the latest architectural solutions. Then, after the listing of the main architectures, the first most complete works were selected regarding each architecture, named seminal papers, although, for some of the solutions, there is in fact only one publication so far. Thus, each architecture and its respective seminal paper is shown in Table \ref{Tab:ComparingArchitectures}, along with a description of their characteristics. But first, it will be discussed the network scope of each architecture, as is shown in Figure \ref{fig:mapeamento}.

With the mapping by scope covered by the architectures it is possible to see that most of these architectures are concentrated in the metro segment. Likewise, it is possible to contribute to an alignment of the generic nomenclature used in that literature, as well as to draw attention to unexplored segments.

%Conforme foi percebido, as redes metropolitanas são segmentadas em vários níveis e seus nós possuem diferentes funções e perspectivas lógicas (Seção \ref{subsubsec:perspLogica}), de acordo com o que é apresentado nos papers seminais referenciados na Tabela \ref{Tab:ComparingArchitectures}. As delimitações de escopo para cada arquitetura (Figura \ref{fig:mapeamento}) contribuem para um alinhamento da nomenclatura genérica utilizada nessa referida literatura,  bem como chamam a atenção para segmentos não explorados. 
%Com isso, é possível um alinhamento da nomenclatura genérica utilizada para referir-se a cada segmento identificado, o que resolve uma questão  

%(??? Sugiro remover a coluna "Core" desta figura... Além de identificar claramente o grupo Híbrido e Puro.): OK

In Figure~\ref{fig:mapeamento}, it is possible to observe that only three of the architectures do not include the metro-access segment, a portion of the metro network that has become important, especially due to the need to deal with edge computing services in the future. The DISCUS architecture is represented as a partially dashed line in the metro-access segment, and this is because the authors do not specifically mention this part of the network, which can be explained by the technology adopted, LR-PON, interconnecting the segment of access to MC nodes, for a strategy to reduce infrastructure costs. ARMONIA and MOMC architectures are the most sectioned in several layers. While the first allows several intermediate layers of aggregation between the segment of MC nodes and MA nodes, the second has aggregation at two different levels, between MC and Access.
The SIMON and TDGMON architectures discuss solutions for access networks that are located from the metro-access segment. Although access networks are not the main focus of this work, we chose to include them as a way of showing recent technological developments at the edge of the network, an issue that is well explored in the ARMONIA, DISCUS, and MOMC architectures, which have as main philosophy the unified vision of metro and access networks. The study of these architectures combined suggests that the metro-access segment has become more robust, with storage and processing capacity, and in this way the access segment has been prepared to explore these potentialities, making it feasible to offer new types of computing, processing, and storage services.
 
Only HnS/WDM is a logical architecture, while the rest are physical architectures.
As for the MEON architecture, shown in the mapping by scope as an architecture concentrated in the metro-core segment, it is possible to expect more comprehensive solutions, with an end-to-end view, given a large number of more recent publications in the literature compared to the others architectures, as shown in Table
\ref{tab:papersByArchitecture}.

\begin{table}
	\caption{List of publications citing the metropolitan network architectures listed in this work.}
	\label{tab:papersByArchitecture}
	%\begin{tabular}{|*{11}{p{1.11cm}|}}
	%\begin{tabular}{|C|C|C|C|C|C|C|C|C|C|C|}}
	\adjustbox{max width=\textwidth}{%
		\centering
		\begin{tabular}{|l|l|l|l|l|l|l|l|l|l|l|l|}
			\hline
			&Architectures & 2013 & 2014 & 2015 & 2016 & 2017 & 2018 & 2019 & 2020\\ 
			\hline
			\multirow{7}{*}{\rotatebox{90}{Multi-layer}}
			&Pure OTN &  & & &  & \cite{8025162OTN} & \cite{katsalis2018towards}\cite{shen2018ultra}\cite{Zhang2018ExploitingEO} & & \cite{infinera2020cost}  \\ 
			&OTN/DWDM &  & & &   & \cite{8025162OTN} & \cite{katsalis2018towards} &
			 \cite{MONIZ2019105608} \cite{da2019otn}\cite{DBLP:journals/corr/abs-1901-04301} \cite{gangopadhyay20195g}\cite{ramachandran2019capacity} &  \\ 
			
			&UDWSN    &  & & & \cite{zhang2016ultra} & \cite{7792281UDWSN} \cite{8025162OTN} & \cite{shen2018ultra} \cite{Zhang2018ExploitingEO} &  & \\ 
			&HnS/WDM  & &  & & & \cite{she2017metro}  & & & \\ 
			&ARMONIA  & &  & & &  &  &    & \cite{kretsis2020armonia} \\ 
			&DISCUS   & &  & & &\cite{ruffini2017access}  & & &\\ 
			&MOMC     & & & & &  & & \cite{moreolo2019spectrum} \cite{larrabeiti2019all} \cite{calabretta2019photonic}&  \cite{larrabeiti2020upcoming} \cite{boffi2020multi} \\ \hline
			
			\multirow{7}{*}{\rotatebox{90}{Single-layer}}
			&MEON &\cite{rottondi2013routing} & & & &\cite{7847391} &\cite{moreolo2018modular} \cite{yan2018tidal} &\cite{ayoub2019routing} \cite{8734478} \cite{wu2019analysis} \cite{8853968}  & \cite{8853968} \cite{yan2020area} \\ 
			&SIMON    & & & & & & &  \cite{muciaccia2019proposal}& \\ 
			&TDGMON   & & & & & & & \cite{lin2019three}  & \\ 
			&FMN   	 & & & & & &&  \cite{paolucci2020disaggregated}&\\
			&DuFiNet  & & & & & & \cite{uzunidis2018dufinet}& \cite{kosmatos2019building}& \\
			&DnW      & & & & &  \cite{8346157SemiFilterless}& & & \\
			&Metro-Haul   & & & & & & \cite{ayoub2018filterless} & & \cite{metro-haulproject} \cite{dochhan2020metro} \\
			\hline
	\end{tabular}}
\end{table}
%(??? Identificar claramente o grupo Híbrido e Puro. Veja como exemplo o Vc fez na tabela seguinte.): OK

Table \ref{tab:papersByArchitecture} shows the main highlighted works that mention the presented architectures. Note that the same work in the literature can reference yet another architecture. This is because such architectures were used as an object of comparison with proposed solutions in the respective works, as occurs with the legacy architectures Pure OTN and OTN/DWDM. The UDWSN and MEON architecture has often been more investigated, not least because both are  DEON inspired by EON~\cite{shen2018ultra}, which is the transmission system that has proved to be ideal for the segment of metropolitan optical networks of the future \cite{7847391}. Regarding UDWSN, research has mainly evaluated cost and performance in the use of resources \cite{shen2018ultra, 8025162OTN, Zhang2018ExploitingEO}. 
As for MEON, these studies have focused specifically on the construction of new optical devices \cite{moreolo2018modular}, resource optimization and traffic engineering \cite{yan2018tidal,yan2020area, 8853968} . In addition, for at least seven of these architectures, a unique work was identified, demonstrating the need for more research in the field of metropolitan optical network architectures.

\subsection{Features Comparison}

Table \ref{Tab:ComparingArchitectures} presents a feature comparison between the main MON architectures highlighted above. The aspects considered in the comparison reflect an overview of the technological diversity present in the recent architectural proposals. The most general classification separates the architectures into two major groups: multi-layer architectures of electro-optical scope, and single-layer.
Fourteen architectures were surveyed, seven multi-layer and seven single-layer. Considering that, among multi-layer architectures, two of them are legacy (Pure OTN and OTN/DWDM) and one of them is logical architecture with legacy optical layer (HnS / WDM), while the rest are recently proposed architectures, which are, UDWSN, ARMONIA, DISCUS and MOMC. Single-layer architectures are all recent solutions, being still in the planning and testing phase.

This movement allows us to infer that all segments of the metropolitan network will be based on fully optical infrastructure, and as legacy networks will continue to coexist, multi-layer infrastructures have been optimized to achieve maximum performance. In addition, these multi-layer networks can also have their optical layer updated to achieve greater flexibility and dynamics, as has been shown through the ARMONIA and MOMC architectures, for example, even though the electronic layer cannot keep up with this movement in the same measure.

Other classifications presented are related to the presence or absence of a filter in the optical layer. In this sense, the With Filter (WF), Filterless (FL), and Semi-Filterless (sFL) classifications, as previously presented. These classifications give a general idea about issues of structural size and operational cost. While FL networks are less costly to set up and operate, WF networks are more complex and more expensive due to the widespread use of ROADMs, one of the highest value optical network elements \cite{shen2018ultra}.
The low port density means that more hardware elements like these are required. However, in FL networks the optical paths for transmission require good detection at the destination since the signals are mixed in the spectrum. Likewise, FL networks do not allow the reuse of the optical spectrum. These problems are not felt in WF networks because the filtering elements act to let only the signals that have been established pass through, making the optical channels available as long as it is unoccupied. This problem can be mitigated with the use of coherent receivers, but this solution, in turn, although has less impact on costs~\cite{boffi2020multi}, incurs an increase in signal processing time \cite{sambo2017sliceable}.
The vast majority of featured architectures are fully WF. FL architectures may be deployed in metropolitan networks in the future, in some segment of the network, and this does not necessarily require the removal of existing filters in networks already in operation. This idea is perceptible through experiments with sFL networks, which point towards a mixed environment, with islands of unfiltered nodes in the middle of the significantly present filter nodes. These experiments, even in symbolic scenarios with testbeds, already indicate the emergence of new demands, such as, for example, identifying the candidate nodes for replacement or upgrade operations, to continue guaranteeing performance and functionality in the face of the growing demands on the part of traffic.

The classification of architectures according to the network segment they were designed, also reinforces the idea of a mixed environment. While some of these architectures focus only on the MC segment or the MA segment, other architectures contemplate both segments in an end-to-end view from the transmission point of view or even extend beyond the metropolitan network environment. This categorization was illustrated in Figure \ref{fig:mapeamento}. 
%(??? Tem que incluir na coluna "Segment" a indicação das arquiteturas que consideram o acesso (A)): OK - Feito Tab.4
The first architectures mentioned, first with Pure OTN and then with OTN / DWDM, concentrated on the core segment of the metropolitan network. When they were planned and implemented, traffic volumes were much lower and the bandwidth resources were sufficient to serve services in the access segment. The changes between these generations have obtained achievements such as reducing latency and bandwidth multiplexing, but with the recent emergence of disruptive technologies demonstrating new business potential, the UDWSN, ARMONIA, MOMC, and MEON architectures have been advocated as more promising solutions. These modern architectures rely on flexible transport mechanisms based on coherent systems, greater spectral efficiency with greater range, as well as varying hierarchical levels to obtain varying degrees of data aggregation. This group of architectures is feasible to serve the 5G network and its services because it allows IT resources to be easily moved from the core to the edge of the network. In turn, the FMN, DuFiNet, and DnW architectures, which face problems with the limitation of the transmission range and quality, represent solutions with low cost and efficiency to be implemented along with the others.

Other attributes that can be compared between metropolitan optical network architectures are transmission band, topologies, the most prevalent type of switch and transponder, utilized modulation formats, frequency spectrum spacing of the optical spectrum and the number of transmission channels. %(??? Tabela trocar "Ch" por "Channels"): OK

Most architectures for metropolitan networks, especially those that are intended to solve problems with a focus on the MC segment, give preference to the C band due to its low transmission loss property. As a way to increase capacity by offering a greater number of channels, it is considered to expand the exploration of the optical spectrum to also include the L band, as proposed in the FMN architecture~\cite{paolucci2020disaggregated}.
Transmission band expansion projects generally require the purchase of extra equipment, such as amplifiers, couplers, and specific splitters by certain bands, for example, which can also represent an increase in energy expenditure with the expansion of the infrastructure. In order to try to keep the cost of updating the networks low in this sense, new low-cost and complex equipment can be implemented, but the performance of the network can be impacted due to the lower range and higher noise figure. As these issues are not so pronounced in optical access networks due to the shorter distances of the segment, it is probably for this reason that architectures with a focus on the MA segment and in access networks, such as SIMON, TDGMON, and FMN, are the ones that most bet on this approach of expanding the transmission band. The interconnection of the metro segments and access in the DISCUS architecture is done through LR-PON and has chains of amplifiers to guarantee the long reach, and for that reason, reach and scalability can be compromised by the high cost and energy consumption required by the amplifiers implanted.
It is possible to conjecture that the architectures that consider expansion of the transmission band beyond the C band, are also those that foresee the highest granularities of channels, and these in smaller numbers, when compared with the other architectures. As the main objective to be achieved with this increase in the number of bands used is to increase the availability of resources for transmission, this issue opens space for more questions when it is verified that a large number of channels can also be obtained in the same band C with the use of flexible grid transmission technologies. However, it is necessary to jointly investigate other topics related to CAPEX/OPEX and maturity of the technology used, since increasing the number of channels with less granularity also implies the implementation of optical elements with finer adjustments in the bandwidth tuning, for which there is a technological limitation.     

% \usepackage{multirow}
% \usepackage[normalem]{ulem}
\begin{landscape}
\begin{table}[]
    \tiny
        \caption{General aspects regarding MON architectures. \scriptsize{(NA and NM stand for Not Applicable and Not Mentioned, respectively)}}
                \label{Tab:ComparingArchitectures}
\begin{adjustbox}{width=1.4\textwidth}
\begin{tabular}{|l|l|l|l|l|l|l|l|l|l|l|l|l|l|l|}
\hline
\multirow{2}{*}{} & \multicolumn{1}{c|}{\multirow{2}{*}{\textbf{Architecture}}} & \multicolumn{1}{c|}{\multirow{2}{*}{\textbf{WF}}} & \multicolumn{1}{c|}{\multirow{2}{*}{\textbf{FL}}} & \multicolumn{1}{c|}{\multirow{2}{*}{\textbf{sFL}}} & \multicolumn{1}{c|}{\multirow{2}{*}{\textbf{Segment}}} & \multicolumn{1}{c|}{\multirow{2}{*}{\textbf{Band}}} & \multicolumn{1}{c|}{\multirow{2}{*}{\textbf{Topology}}} & \multicolumn{1}{c|}{\multirow{2}{*}{\textbf{\begin{tabular}[c]{@{}c@{}}Transmition\\ System\end{tabular}}}} & \multicolumn{1}{c|}{\multirow{2}{*}{\textbf{SDN}}} & \multicolumn{1}{c|}{\multirow{2}{*}{\textbf{OXC}}} & \multicolumn{1}{c|}{\multirow{2}{*}{\textbf{Transponder}}} & \multicolumn{1}{c|}{\multirow{2}{*}{\textbf{Modulation}}} & \multicolumn{1}{c|}{\multirow{2}{*}{\textbf{Spacing}}} & \multicolumn{1}{c|}{\multirow{2}{*}{\textbf{Channels}}} \\
 & \multicolumn{1}{c|}{} & \multicolumn{1}{c|}{} & \multicolumn{1}{c|}{} & \multicolumn{1}{c|}{} & \multicolumn{1}{c|}{} & \multicolumn{1}{c|}{} & \multicolumn{1}{c|}{} & \multicolumn{1}{c|}{} & \multicolumn{1}{c|}{} & \multicolumn{1}{c|}{} & \multicolumn{1}{c|}{} & \multicolumn{1}{c|}{} & \multicolumn{1}{c|}{} & \multicolumn{1}{c|}{} \\ \hline

\multirow{16}{*}{\rotatebox{90}{Multi-layer}} & \multirow{2}{*}{Pure OTN\cite{infinera2020cost}} & \multirow{2}{*}{X} & \multirow{2}{*}{} & \multirow{2}{*}{} & \multirow{2}{*}{MC} & \multirow{2}{*}{C} & \multirow{2}{*}{Ring} & \multirow{2}{*}{WDM} & \multirow{2}{*}{No} & \multirow{2}{*}{AWG} & \multirow{2}{*}{Line Card} & \multirow{2}{*}{NA} & \multirow{2}{*}{50GHz} & \multirow{2}{*}{96} \\
 &  &  &  &  &  &  &  &  &  &  &  &  &  &  \\ \cline{2-15} 
 & OTN/DWDM\cite{DBLP:journals/corr/abs-1901-04301} & X &  &  & MC & C & Mesh & WDM & No & WSS-ROADM & IMDD & OOK & 50GHz & 96 \\ \cline{2-15} 
 & \multirow{3}{*}{UDWSN\cite{zhang2016ultra}} & \multirow{3}{*}{X} & \multirow{3}{*}{} & \multirow{3}{*}{} & \multirow{3}{*}{\begin{tabular}[c]{@{}l@{}}MC\\ MA\end{tabular}} & \multirow{3}{*}{C} & \multirow{3}{*}{\begin{tabular}[c]{@{}l@{}}Mesh\\ Chain\\ Tree\end{tabular}} & \multirow{3}{*}{\begin{tabular}[c]{@{}l@{}}DEON\\ EON\end{tabular}} & \multirow{3}{*}{No} & \multirow{3}{*}{\begin{tabular}[c]{@{}l@{}}WSS-ROADM\\ AWG-OADM\end{tabular}} & \multirow{3}{*}{\begin{tabular}[c]{@{}l@{}}Co-BVT\\ IMDD\end{tabular}} & \multirow{3}{*}{\begin{tabular}[c]{@{}l@{}}QPSK\\ OOK\\ PAM-4\end{tabular}} & \multirow{3}{*}{\begin{tabular}[c]{@{}l@{}}5GHz\\ 6.25GHz\\ 12.5GHz\end{tabular}} & \multirow{3}{*}{\begin{tabular}[c]{@{}l@{}}800\\ 640\\ 320\end{tabular}} \\
 &  &  &  &  &  &  &  &  &  &  &  &  &  &  \\
 &  &  &  &  &  &  &  &  &  &  &  &  &  &  \\ \cline{2-15} 
 & \multirow{2}{*}{HnS/WDM\cite{she2017metro}} & \multirow{2}{*}{X} & \multirow{2}{*}{} & \multirow{2}{*}{} & \multirow{2}{*}{\begin{tabular}[c]{@{}l@{}}MC\\ MA\end{tabular}} & \multirow{2}{*}{C} & \multirow{2}{*}{HnS in Mesh} & \multirow{2}{*}{WDM} & \multirow{2}{*}{Yes} & \multirow{2}{*}{ROADM (multi degree)} & \multirow{2}{*}{NM} & \multirow{2}{*}{NM} & \multirow{2}{*}{50GHz} & \multirow{2}{*}{88} \\
 &  &  &  &  &  &  &  &  &  &  &  &  &  &  \\ \cline{2-15} 
 & \multirow{3}{*}{ARMONIA\cite{kretsis2020armonia}} & \multirow{3}{*}{X} & \multirow{3}{*}{} & \multirow{3}{*}{} & \multirow{3}{*}{\begin{tabular}[c]{@{}l@{}}MC\\ MA\\ A\end{tabular}} & \multirow{3}{*}{C} & \multirow{3}{*}{\begin{tabular}[c]{@{}l@{}}Mesh\\ \\ Ring\end{tabular}} & \multirow{3}{*}{Nyquist WDM} & \multirow{3}{*}{Yes} & \multirow{3}{*}{ROADM} & \multirow{3}{*}{\begin{tabular}[c]{@{}l@{}}Co-Receiver\\ BVT\end{tabular}} & \multirow{3}{*}{\begin{tabular}[c]{@{}l@{}}DP-BPSK\\ DP-QPSK\end{tabular}} & \multirow{3}{*}{12.5GHz} & \multirow{3}{*}{320} \\
 &  &  &  &  &  &  &  &  &  &  &  &  &  &  \\
 &  &  &  &  &  &  &  &  &  &  &  &  &  &  \\ \cline{2-15} 
 & \multirow{3}{*}{DISCUS\cite{ruffini2017access}} & \multirow{3}{*}{X} & \multirow{3}{*}{} & \multirow{3}{*}{} & \multirow{3}{*}{\begin{tabular}[c]{@{}l@{}}MC\\ MA\\ A\end{tabular}} & \multirow{3}{*}{C} & \multirow{3}{*}{\begin{tabular}[c]{@{}l@{}}Mesh Flated\\ Horseshoe\end{tabular}} & \multirow{3}{*}{DWDM} & \multirow{3}{*}{Yes} & \multirow{3}{*}{WSS-ROADM} & \multirow{3}{*}{Commercial 100G} & \multirow{3}{*}{DP-QPSK} & \multirow{3}{*}{50GHz} & \multirow{3}{*}{NM} \\
 &  &  &  &  &  &  &  &  &  &  &  &  &  &  \\
 &  &  &  &  &  &  &  &  &  &  &  &  &  &  \\ \cline{2-15} 
 & \multirow{2}{*}{MOMC\cite{larrabeiti2019all}} & \multirow{2}{*}{X} & \multirow{2}{*}{} & \multirow{2}{*}{} & \multirow{2}{*}{\begin{tabular}[c]{@{}l@{}}MC\\ MA\\ A\end{tabular}} & \multirow{2}{*}{C} & \multirow{2}{*}{\begin{tabular}[c]{@{}l@{}}Mesh\\ Ring\end{tabular}} & \multirow{2}{*}{WDM} & \multirow{2}{*}{Yes} & \multirow{2}{*}{CDC/SDM-ROADM} & \multirow{2}{*}{WDM Transponder} & \multirow{2}{*}{NRZ-OOK} & \multirow{2}{*}{50GHz} & \multirow{2}{*}{80} \\
 &  &  &  &  &  &  &  &  &  &  &  &  &  &  \\ \hline
\multirow{8}{*}{\rotatebox{90}{Single-layer}} & \multirow{2}{*}{MEON\cite{rottondi2013routing}} & \multirow{2}{*}{X} & \multirow{2}{*}{} & \multirow{2}{*}{} & \multirow{2}{*}{MC} & \multirow{2}{*}{C} & \multirow{2}{*}{Ring} & \multirow{2}{*}{\begin{tabular}[c]{@{}l@{}}DEON\\ EON\end{tabular}} & \multirow{2}{*}{No} & \multirow{2}{*}{BV-OXC} & \multirow{2}{*}{Coherent BVT} & \multirow{2}{*}{x-QAM} & \multirow{2}{*}{\begin{tabular}[c]{@{}l@{}}5GHz\\ 10GHz\end{tabular}} & \multirow{2}{*}{\begin{tabular}[c]{@{}l@{}}200\\ 100\end{tabular}} \\
 &  &  &  &  &  &  &  &  &  &  &  &  &  &  \\ \cline{2-15} 
 & SIMON\cite{muciaccia2019proposal} & X &  &  & \begin{tabular}[c]{@{}l@{}}MC\\ MA\\ A\end{tabular} & \begin{tabular}[c]{@{}l@{}}C\\ S\\ O\end{tabular} & Star-in-ring & WDM & Yes & \begin{tabular}[c]{@{}l@{}}3D-ROADM\\ 4D-ROADM\end{tabular} & \begin{tabular}[c]{@{}l@{}}WDM-Transponder\\ IMDD\end{tabular} & OOK & \begin{tabular}[c]{@{}l@{}}50GHz\\ 100GHz\end{tabular} & 100 \\ \cline{2-15} 
 & TDGMON\cite{lin2019three} & X &  &  & \begin{tabular}[c]{@{}l@{}}MA\\ A\end{tabular} & NM & Ring and Grid & WDM & No & AWG-OADM & CWDM SFP & DPSK NRZ & \begin{tabular}[c]{@{}l@{}}193.1 THz\\ 193.2 THz\\ 193.3 THz\\ 193.4 THz\end{tabular} & 4 \\ \cline{2-15} 
 & FMN\cite{paolucci2020disaggregated} &  & X &  & \begin{tabular}[c]{@{}l@{}}MC\\ MA\end{tabular} & \begin{tabular}[c]{@{}l@{}}C\\ L\end{tabular} & Horseshoe & EON & Yes & \begin{tabular}[c]{@{}l@{}}2D-ROADM\\  (extremities)\end{tabular} & Coherent BVT & \begin{tabular}[c]{@{}l@{}}QPSK\\ 16-QAM\end{tabular} & 37.5GHz & \begin{tabular}[c]{@{}l@{}}79(C)\\ 109(L)\end{tabular} \\ \cline{2-15} 
 & DuFiNet\cite{uzunidis2018dufinet} &  & X &  & \begin{tabular}[c]{@{}l@{}}MC\\ MA\end{tabular} & C & Ring & WDM & No & \begin{tabular}[c]{@{}l@{}}WSS-ROADM\\ (extremities)\end{tabular} & \begin{tabular}[c]{@{}l@{}}Coherent BVT\\ IMDD\end{tabular} & \begin{tabular}[c]{@{}l@{}}PM-QPSK\\ PM-16-QAM\\ NRZ\\ 4-PAM\end{tabular} & \begin{tabular}[c]{@{}l@{}}25 GHz\\ 37.5 GHz\\ 50 GHz\end{tabular} & \begin{tabular}[c]{@{}l@{}}133\\ 40\\ 20\end{tabular} \\ \cline{2-15} 
 & DnW\cite{8346157SemiFilterless} &  &  & X & \begin{tabular}[c]{@{}l@{}}MC\\ MA\end{tabular} & C & Horseshoe & WDM & No & 2D-ROADM & Coherent Detection & \begin{tabular}[c]{@{}l@{}}QPSK\\ OOK\end{tabular} & 50GHz & 80 \\ \cline{2-15} 
 & Metro-Haul\cite{ayoub2018filterless} &  &  & X & \begin{tabular}[c]{@{}l@{}}MC\\ MA\end{tabular} & C & Mesh & EON & Yes & WSS-ROADM & \begin{tabular}[c]{@{}l@{}}Coherent S-BVT\\ BVT\end{tabular} & \begin{tabular}[c]{@{}l@{}}DP-QPSK\\ DP-16QAM\end{tabular} & 12.5GHz & 200 \\ \hline
\end{tabular}
\end{adjustbox}
\end{table}
\end{landscape}

As for the spacing of the frequency grid and the number of channels available, there is a wide variety with a higher prevalence of the already standardized $50$ $GHz$ and $12.5$ $GHz$, as well as new ultra-narrow possibilities such as $5$ $GHz$ and $6.25$ $GHz$. Larger spacing, such as those that are already used today, suffer from the inefficiency of use because the data rates practiced in the metro and access segments are relatively small. In this sense, to make better use of the available resource, traffic aggregation techniques are used, which is commonly done in the electronic layer. The problem is that, at the same time, time-sensitive applications can be impacted with the use of this strategy, leading to an increase in the number of segmented levels in the metropolitan network, with specific layers for the task of aggregating and distributing traffic, as occurs in ARMONIA and MOMC.
With 5G networks, which will require very low latency, 5G New Radio interfaces for the fronthaul \cite{larrabeiti2020upcoming} are under development to define frequency spacing of a maximum of $6$ $GHz$ to connect the Cloud Radio Access Network (C-RAN) with an optical network. UDWSN and MEON, with a DEON approach, are architectures that promise equipment adjusted to work with a flexible grid of small granularity and can be the solution to meet the new service demands of mobile networks while ensuring efficient use of the spectrum.

\subsection{Research Themes by Architecture}

Table \ref{tab:FocusNarchitecture} shows a classification of the main research focuses in the most recent literature identified for each type of architecture highlighted. The identification of these themes shows the main gaps that need further studies, as well as showing the points in common with proposals that can be applied to more than one architecture in question. The research focus range from proposing new physical elements of the hardware infrastructure, network performance, planning, control and management, which is quite common when new network architectures are proposed. 
In addition, another aspect mentioned concerns traffic engineering, based mainly on new traffic patterns identified in metropolitan networks. It is worth mentioning that some problems investigated are architectural agnostics, but characteristic for a given transmission system. 
Thus, most works consider metropolitan networks with an EON and DEON transmission system, that is, the MEON architecture. However, architectures with WDM transmission will continue to be a reality and research suggests proposals for performance optimization. %A rede de backbone aparece na tabela para reunir trabalhos que são representativos para as próprias redes metropolitanas em cada coluna. Isso ocorre porque muitas das soluções apontadas começaram a ser estudadas exatamente na redes núcleo. 

%Roteamento em WDM em nive de fibra \cite{shiraki2020highly}

\begin{table}
	\caption{Main topics covered in the literature for each architecture.}
	\label{tab:FocusNarchitecture}
	\resizebox{\columnwidth}{!}{%
		\begin{tabular}{|l|l|l|l|l|l|l|l|} 		
			\hline
			\multicolumn{2}{|c|}{\multirow{2}{*}{\textbf{Architectures}}} & \multicolumn{6}{c|}{\textbf{Focus}} \\ \cline{3-8} 
		\multicolumn{2}{|c|}{} & \multicolumn{1}{c|}{\textbf{\begin{tabular}[c]{@{}c@{}}Hardware \\ Infrastructure\end{tabular}}} & \multicolumn{1}{c|}{\textbf{\begin{tabular}[c]{@{}c@{}}Traffic \\ Engineering\end{tabular}}} & \multicolumn{1}{c|}{\textbf{Costs}} & \multicolumn{1}{c|}{\textbf{Performance}} & \multicolumn{1}{c|}{\textbf{Protection}} & \multicolumn{1}{c|}{\textbf{\begin{tabular}[c]{@{}c@{}}Control and \\ Managemet\end{tabular}}} \\ \hline
			
			\multirow{7}{*}{\rotatebox{90}{Multi-layer}}& Pure OTN & & & \cite{infinera2020cost} &  &  & %\\\hline
			\\%\cline{2-8}
			
			& OTN/DWDM & \cite{li2017digital} &  & \cite{infinera2020cost} & \cite{kosmatos2019building} \cite{shiraki2020highly} \cite{MONIZ2019105608} & \cite{9153741} & \cite{vilalta2020experimental} \cite{da2019otn} %\\\hline
			\\%\cline{2-8}
			
			& UDWSN           & \cite{fabrega2020cost}  &  &     \cite{Zhang2018ExploitingEO}  & \cite{Zhang2018ExploitingEO} \cite{8025162OTN}  &      &  %\\\hline    
			\\ %\cline{2-8}
			
			& HnS/WDM         & \cite{li2017digital}   & \cite{TROIA2020100551} \cite{troia2019dynamic} \cite{zhu2019machine} &   & \cite{kosmatos2019building} \cite{filer2019low}\cite{garrich2020joint} & \cite{9153741} &%\\\hline 
			\\ %\cline{2-8}
			
			& ARMONIA     &   &   &   & \cite{kretsis2020armonia} &  & %\\\hline
			\\ %\cline{2-8}
			
			& DISCUS     &   &   &   & \cite{ruffini2017access} &  & %\\\hline
			\\ %\cline{2-8}
			
			& MOMC            &\cite{larrabeiti2019all} \cite{moreolo2018modular}\cite{nadal2019sdn}  & & &\cite{comellas2020spatial} &  &  \cite{9132992}            %\\\cline{2-8}
			\\\hline
			
			\multirow{7}{*}{\rotatebox{90}{Single-layer}}& MEON &  \cite{fabrega2020cost}\cite{moreolo2018modular} & \cite{yan2018tidal} \cite{8734478} \cite{yan2020area} \cite{kokkinos2019pattern} & \cite{8853968}          & \cite{liu2018joint} \cite{sambo2017sliceable} \cite{comellas2020spatial} \cite{beletsioti2020learning} \cite{ayoub2019routing}\cite{din2019rbcmlsa} &  & \cite{casellas2018enabling} \\ %\cline{2-8}
			
			& SIMON           &  &  &  & & &  \\%\cline{2-8}
			& TDGMON          &  &  &  & & &   \\%\cline{2-8}
			& FMN             &  &  &  & \cite{ruiz2019smart} & \cite{shariati2020real} &    \\	%\cline{2-8}
			& DuFiNet         &  &  &  & &\cite{shariati2020real}  &     \\	%\cline{2-8}
			& DnW             & \cite{cugini2018flexibleSemiFL} & & & & &  \\%\cline{2-8}
			& Metro-Haul      & \cite{9041836}  & \cite{dochhan2020metro} \cite{TROIA2020100551} \cite{troia2019dynamic} \cite{zhu2019machine} & \cite{9042241}&  & \cite{9153741} &  \\ 	\hline
	\end{tabular}}
\end{table}
%(??? Vc achou melhor não separar nesta tabela entre Híbrido e Puro?): OK

Many of the research insights regarding Metro-Haul architecture (architecture of the Metro-Haul project) can be directed to FL and sFL architectures. This is possible because Metro-Haul covers such subtypes, as mentioned in Subsection \ref{subsubsec:metroSemiFL}. 
As an example of this, in \cite{9042241} an economic feasibility study is carried out in FL networks to decide on the implementation of a control plan that guarantees dynamic operations or on the implementation of new filter nodes. In this sense, experiments are carried out with various configurations of scenarios that run through these types of architecture, and which consider the number of different equipment deployed, as well as traffic fluctuations in the metro network.

As the Pure OTN architecture has been used as a comparative scenario in some works, then recent works restricted to this network are not identified. For the case of OTN/DWDM, due to its proximity to HnS/WDM, many identified proposals can be applied to both, and in terms of hardware infrastructure, the idea is mainly to use transponders with the IMDD system \cite{li2017digital} in the metro-core to achieve $100$ $Gb/s$ rates with spectral efficiency, due to the cost-benefit. 
What is striking is that the same type of equipment is also suggested to maintain a low CAPEX in the UDWSN architecture, according to the scenario shown in~\cite{Zhang2018ExploitingEO}, but restricted to the ends of the metropolitan network, that is, only in the metro-access segment.
UDWSN focuses on leveraging part of the legacy infrastructure, and as a result, partially upgrading a DWDM MON can incur the long-term waste of resources, even if the future goal is to drastically change architecture to a flexible approach. As can be seen in the fourth column of the table, few studies have been analysed cost. This demonstrates a large gap for future research since most of these architectural proposals are very recent. The ARMONIA and DISCUS architectures, in their only outstanding works, perform a conceptual demonstration, present the respective architectures, as well as discuss an initial performance analysis.

Segments not fully explored are also identified. The description of the necessary and sufficient infrastructure for the architectures in the metro-access segment, that is, the part of the metro with the shortest distances, such as those of type FL and, also SIMON and TDGMON, are only presented in their respective seminal papers, referenced in the Table \ref{Tab:ComparingArchitectures}. 
For these ecosystems, some generic approaches to analyzing the optical signal \cite{shariati2020real} can be relevant both to improve transmission performance where there is no optical filter, identifying, for example, cases of signal overlap and as a metric for implementing path protection mechanisms. However, it is also necessary to deepen research in this field, in a more targeted way for each architecture individually. 

Finally, it's noted that, as there are more new proposals for single-layer networks, most of the approaches presented are based on testbed, so the next studies are likely to expand the test ecosystems for much larger topologies. The lesson that such works leave is that the multilayer approach is useful to make the most efficient use of the optical resources since transporting the aggregated traffic can lead to a greater chance of completely combining the data rate with the bandwidth of the available channels.

\subsection{Future Trends}
In view of the issues presented in this section, it was possible to understand the various proposals of the scientific literature for the MON of the future. This Subsection aims to highlight the identified trends that will possibly drive new business and service models. 

As with core networks, metropolitan networks \cite{8999006} trend will be to provide solutions to adapt the various technologies present in its infrastructure, due to the existence of several islands of equipment of different generations. New technologies, which until recently were investigated for deployment in long-distance network infrastructures, are now being explored for metropolitan networks. 
This is a more general inference that can be made based on the various studies presented that investigate new, lower-cost solutions while continuing to leverage legacy infrastructure. However, the cost is still a prohibitive issue. Consequently, the development of this theme in the light of the results obtained so far presents a series of questions that are still unresolved. As these are very recent architectures, several questions can be raised as an object of research. 

Considering the urgency of these issues and observing the main similarities/pattern of unanswered questions among all architectures, it is possible to conclude with such analysis the next research trends, which will offer solutions to the demands of the providers. The following can be highlighted as likely trends.

\subsubsection*{\textbf{Elastic Transmission System}}
MONs in operation today are a mix of generations of technologies, operating in a static manner with channels established to operate for months. The WDM system has undergone updates, and even a flexible version has already been thought of as an alternative deployment with reasonable complexity and low cost~\cite{li2017digital}. However, all the elastic infrastructure already installed or ready for deployment in the field is completely static and it is not yet possible to explore all the potential of the elastic functionalities. EON is the ideal solution to reduce the use of amplifiers in the metro, as it provides several possibilities of rates with high spectral efficiency, ideal to take advantage of, given that the transmission range is not such a relevant factor in small and medium distances typically found in this ecosystem. It is then necessary that a new intelligent control plan be proposed and developed so that all possibilities of EON are widely explored \cite{larrabeiti2020upcoming}. The elastic transmission system of the future should provide dynamic, and programmable control, as well as facilitated management to effectively manipulate the infrastructure deployed in the field.

\subsubsection*{\textbf{Mixed Ecosystem}}
The metropolitan network is expected to be customizable through the implementation of several nodes with heterogeneous architectures forming an environment with several generations of technologies, especially separated by network segment. As was shown with ARMONIA and MOMC, the increase in the number of levels/hierarchies is a possibility to manage traffic bottlenecks where it is not possible to implement modern and expensive equipment. Although this solution is costly from the point of view of latency requirements. There will be islands of nodes without a filter, mainly in the MA segment \cite{shen2018ultra}, which will have greater capillarity and, for this reason, will need to be low cost. In the MC segment, part of the nodes will be able to maintain electronic switches, in particular OTN switches, to perform traffic aggregation and better use the resources of optical networks. As this equipment has existed and has been in operation for some time, many will be removed from points where speed is an important factor but will continue to operate in specific locations defined according to the network design and the respective topology. The main reason behind this adaptation is the need to reduce energy consumption. Also, the coexistence of different transmission systems will be a reality in the long term, until there is a total replacement of fixed grid technologies by the flexible grid, which is inevitable considering the strict requirements of the most modern applications.
Also, the deployment of IT resources scattered throughout the metropolitan network will promote convergence of technologies for the introduction of edge computing, fog-computing, among others, with the implementation of mDCs and regional DCs \cite{moreolo2018modular}.

 \subsubsection*{\textbf{Convergence of services/products between network operators and content providers}}
 While network operators will face the challenge of keeping telecommunications networks functioning optimally, with infrastructure expansions planned at least $15$ years to deal with new content demands, content providers will increasingly scale their infrastructures including new servers in a distributed and balanced way, which can take place every two or three years, to keep up with traffic growth~\cite{7847391}. The convergence between both businesses may occur through the virtualization of infrastructure and resources, which will unite both providers around a single objective, creating new market segments with the offer of new chains of aggregated services as disruptive innovation.

%Esse cenário abre espaço para pesquisa que trate da definição sobre quantos e quais nós continuarão/passarão a implementar OTN. É necessário evidências sobre os melhores posicionamentos/localizações desses nós. Além disso, considerando apenas os nós da rede que combinam OTN e ROADMs, outro problema que necessita de investigação é o momento mais adequado para a interrupção do caminho óptico para a realização de uma agregação de dados, para, dessa maneira, aproveitar a melhor a capacidade dos canais disponíveis \cite{shen2018ultra}. Essa questão deve sua complexidade aos variados padrões de tráfego presentes nas redes metropolitanas, suas inerentes restrições.

\subsubsection*{\textbf{Network Slicing (NS)}}
Not only will the physical infrastructure be redesigned, but also its resources in the form of software. Flexible networks allow greater freedom for the virtualization of optical resources~\cite{7847391}, and virtualization should be the main law to benefit the creation of new vertical services in a mixed physical environment. Through this process, the same equipment in the network may be the structure for planning and implementing several logical networks, totally isolated and independent from each other, established according to the profile of each business, and mainly vacating dedicated resources when not needed \cite{hernandez2020techno}. 
Network slicing will benefit a variety of communication businesses because it will integrate a service umbrella that can be built on demand for each type of service. However, the main challenges related to these trends are the management of collective resources and the integration between different architectures that requires several layers of software, from resource planning and optimization tools, including dynamic resource allocation algorithms \cite{vilalta2020experimental}. 
This concept of ensuring that multiple customers exploit the same physical resources simultaneously can guarantee not only the generation of revenue, but also a considerable reduction in CAPEX/OPEX. MEON is the main architecture in which the potential for NS has been investigated \cite{troia2019dynamic}. 

\subsubsection*{\textbf{Software-Defined Metro Network (SDMN) or Intelligent Control Plane}}
This metropolitan network concept proposes transparency for network switching technologies. Optical, packet or burst switching is combined from infrastructure with white boxes to provide a platform capable of providing distributed network functions for inter-DC virtualization services \cite{vilalta2020experimental}. This idea of disruptive and virtualized connectivity will favor the trends of edge computing and network slicing, instantiating resources and services for multiple customers in an individualized way. SDMN is controlled through a WAN infrastructure manager (WIN), such as OpenDaylight, or ONOS, an open-source operating system with convergent topology abstraction, in which several network layers and technologies are presented as a single logical graph.

\subsubsection*{\textbf{Disaggregated networks}}
The concept of unbundled networks breaks with the scenario of closed optical network infrastructure with hardware and control from the same owner and allows the adoption of different elements from the most diverse manufacturers in the composition of the network architecture with a customizable control interface~\cite{hernandez2020techno, casellas2018enabling, boffi2020multi, nadal2019sdn, 7847391}. This is possible thanks to new data models developed to handle a wide variety of optical network elements that enable control and management through SDN, with common management through ONOS (open network operating system). The main advantage of this approach is the reduction in the CAPEX and OPEX of metropolitan networks based on the cheaper hardware that is used, both with the reduction of restrictions and the acquisition of closed components from the same manufacturer, as well as the opening for use and integration of legacy hardware. Then there is the possibility of offering many other services, such as unbundled components, such as, for example, vBBU (virtual Broadband Base Unit) and vRouter (virtual Router), which consequently result in lower costs in other layers of the network, in addition to allow greater programmability.
In optical networks, one of the main initiatives in this regard is the Open ROADM Multi-Source Agreement (MSA) Project~\cite{openROADM}, citing previously, that has proposed an optical switch, transponders, and pluggable optics, in the form of white boxes with the open management interface and YANG data model. Equipment like this has been evaluated in the project ADRENALINE testbed~\cite{9132992, moreolo2018modular,nadal2019sdn}, an optical transport network with fixed/flexible grid WDM technology, SDN/NFV, and edge and core cloud platform to deliver end-to-end $5G$ and IoT services across all network segments.

\subsubsection*{\textbf{Artificial Intelligence (AI) for optical data transport}}
Agile and programmable technological solutions for elastic networks will require the development of automated features that can be quickly employed. The level of automation provides a favorable environment for the exploration of data analytics, which has been shown as an AI research field for optical networks, developing both in the sense of network engineering and in the sense of traffic engineering. In terms of network operation, Learning Automata (LA) has been proposed to decide the allocation of resources that results in less blockage and greater energy efficiency \cite{beletsioti2020learning}. 
Also in the Metro-Haul project, machine-learning (ML) has been used in the allocation resources that serve different services with dynamic traffic~\cite{garrich2020joint}. On the other hand, with respect to traffic engineering solutions, recurrent neural network (RNN) have been used to forecast traffic in different areas of the metropolitan network \cite{paul2019traffic}, while backpropagation techniques have been used to predict trends in changes in traffic load \cite{zhu2019machine}.
In this way, new opportunities to explore AI solutions arise in the field of unbundled networks and communication networks for $5G$, taking advantage of centralized control and optimizing operations on these networks~\cite{casellas2018enabling}. 
Thus, solutions are expected to facilitate the control, management and monitoring of various network configuration parameters, which contribute to rapid decision making so that the performance of the entire system is improved. However, it is necessary to take into account that the elaboration of AI models to solve problems in a dynamic traffic environment incurs enormous energy expenditures and requires a lot of dedicated computational power.

\subsubsection*{\textbf{New Business Demands}}
With the implementation of flexible grid transport technologies, and their consequent low-granular optical spectrum grid, the number of channels in the optical spectrum increases considerably. As a result, spectrum rent is a market trend that can become popular, amidst a scenario of edge and fog computing deployment, geo-distributed in the metropolitan network~\cite{shen2018ultra, 8596108}. 
The optical spectrum-as-a-service (OSaaS)~\cite{shen2018ultra} rental service is similar to the black fiber rental service offered by several operators today, but at a lower cost, since the customer does not necessarily need to maintain and operate the equipment needed to handle fiber resources. The manipulation of optical fiber alone represents an extra complexity due to the greater responsibility for maintaining and monitoring another layer of technology. Instead, the customer acquires the rights to use a fixed portion of the optical spectrum, representing a few channels according to their demand, or to acquire a dynamic portion in the frequency space and time (different time slots) that may be shared with other customers who are also users of the same service.
Transport-as-a-Service (TaaS)~\cite{7847391} will be created from the need to increase the number of optical layers at hierarchical levels for network sharing. Currently, the networks are designed with a minimum of three layers, but the ARMONIA and MOMC architectures offer an opportunity to explore this new type of business.
Optimization-as-a-Service (OaaS) for disaggregated network architectures will be able to provide solutions on demand for resource allocation, traffic engineering, implementation of new interfaces, and even monitoring of the system as a whole. Furthermore, breaking paradigms through the use of closed and proprietary systems will allow the customer of these services to implement their software solutions for greater personalized control \cite{garrich2020joint}.

%%%%%%%%%%%%%%%%%%%%%%%%%%%%%%%%%%%%%%%%%%%%%%%%%%%%%%%%%%%%%%
%%%%%%%%%%%%%%%%%%%%%%%%%%%%%%%%%%%%%%%%%%%%%%%%%%%%%%%%%%%%%%
%%%%%%%%%%%%%%%%%%%%%%%%%%%%%%%%%%%%%%%%%%%%%%%%%%%%%%%%%%%%%%
\section{Literature Gaps}\label{sec:Trends}
Faced with an evolutionary scenario with so many alternatives of MON architectures, to be designed and implemented gradually, and through the main and irrefutable trends for metro business, there is a real need to investigate and learn more about all available solutions, since they are very recent. As expected, metro networks will be the point of convergence among several technologies such as $5G$ networks, home optical fiber, industry $4.0$, IoT, as well as different levels of distributed processing and storage points, such as cutting edge data centers \cite{7847391}.
As an initial direction, the main research opportunities which demand a greater need for further development are discussed in the following sections, organized by architecture.
\subsection*{OTN/DWDM}
As the new OXC proposed in \cite{9041836} is almost immune to traffic conditions at any point in the network, it is necessary to test the impact of the deployment of this equipment on different network architectures. It is also necessary to investigate other low-cost equipment to perform infrastructure combined updates in the metro varied segments.
\subsection*{UDWSN}
The intensity modulation transmission solutions \cite{fabrega2020cost} using coherent detection, combined with direct/external modulation can be experienced/applied at specific points into the metro optical network architecture to decrease costs and improve performance. This issue requires further investigation, especially for a potential upgrade of the transmitters used in the UDWSN, to analyze the cost-performance. Further testing with this transmission solution can also be investigated within the scope of other metro architectures to achieve high capacity and cost-effectiveness since most of these architectures implement low capacity transmitters.
\subsection*{HnS/WDM}
The BPANN service reconfiguration strategy based on traffic-aware prediction model with machine learning proposed in \cite{zhu2019machine} can be applied to design other logical topologies without increasing CapEx/OpEx in order to exploit legacy infrastructure, while solving the unbalanced network resource in metropolitan networks.

Beyond the proposed model in \cite{TROIA2020100551}	to dynamically provision network slices considering power consumption, latency and capacity based on forecast of mobile traffic network,  it is necessary to include other traffic forecast models considering others traffic patterns and profiles in metropolitan networks. Additional investigations in this field can extend the life of the architecture before the next infrastructure update.

The capacity planning tool, Net2Plan~\cite{garrich2020joint}, recently proposed and tested in the contest of Metro-Haul project, needs to be applied to other network architectures in order to potentially reduce the need to deploy new equipment, in the upgrade process, to deal with future demands.
\subsection*{MOMC}
The structures of nodes and transponders with a granularity of $25$ $GHz$, proposed in \cite{boffi2020multi}, need to be further investigated to identify the ideal locations for your installations. It is especially important for multi-segment metropolitan network architectures, which require a combination of service rate granularity and equipment capacity. Issues related to the costs of implementing these new solutions require further investigation. Besides, finer spectrum granularity can be explored to match cross-hierarchical requirements and capacity adaptation for smaller demands \cite{boffi2020multi}.

The challenges for MONs in terms of elastic $Tb/s$ services from $5G$ network and edge-computing, highlighted in \cite{larrabeiti2020upcoming} for the next decade, requires further investigation for cost-efficient reliable deployment of mDC in COs in the MAN and for match low fronthaul rates and OFDM symbol burst size. The situation is urgent because currently many existing architectures are unable to cope with future demand patterns.

The transmission quality is investigated in for a modular SDM system proposed in \cite{calabretta2019photonic}. Moreover, it is expected further investigation about costs, capacity and scalability issues.

The study in \cite{hernandez2020techno} demonstrates that the disaggregation paradigm is not feasible for small networks with few nodes in terms of costs. However, there are no answers for cases where specific environments are considered, for example, some segment of the metropolitan network (metro-core, metro-aggregation, metro-access), and which ones can benefit most from total or partial disaggregation.

A testbed with programmable (SDN-enabled) sliceable bandwidth/bitrate variable transceivers (S-BVTs) in a disaggregated network is presented in \cite{nadal2019sdn}. Additionally, investigations about infrastructure scalability issues and related impacts in terms of costs and power consumption are also required.% \cite{nadal2019sdn}.
\subsection*{FMN}
Regarding strategies for updating filterless networks to include other transmission bands \cite{paolucci2020disaggregated}, there is space to investigate CAPEX and OPEX in real-world scenarios of traffic and capacity requirements, since the focus of this work is on the quality of multiband transmission.
\subsection*{TDGMON}
The three-dimensional architecture presented in \cite{lin2019three} for metro-access network is analyzed from the point of view of the scalability of number of ONUs and quality of transmission. Further investigations on the issues of cost and higher transmission rates, in order to meet the requirements of technologies such as edge computing and $5G$ are required.
\subsection*{Metro-Haul}
In addition of use-case demonstration of resource allocation for video applications with high performance requirements \cite{dochhan2020metro}, further investigations are needed to explore other types of traffic services, taking into account tidal-traffic patterns for more accurate solutions.

\subsection*{MEON}
On the other hand, due to the greater attention that MEON architecture has received since the introduction of the approach of the flexible network to core networks, the roadmap for the next research fronts concerning this technology is better defined. The main open issues identified are the ones presented as follows.
\begin{itemize}
\item Dynamic allocation with adaptive rates in the MEON architecture improves the use of the spectrum \cite{8734478}. However, further investigation is required considering a greater variation of types and profile parameters of traffic to minimize service latency while increasing throughput.
\item The model for shaping traffic in order to improve resource allocation and reduce energy consumption considers physical-layer interactions to minimizes the amount of total time-averaged transponder power consumption in~\cite{8853968}. Additionally, also other elements (OXC, amplifiers, etc) of optical network  can be included in the analysis to further research.
\item Besides the RSA solution for allocating resources with a model of traffic variation which is area-aware based on mobile network traffic \cite{yan2020area}, there is space to expand the solution to include RMLSA algorithms and explore other AI-based traffic forecasting methods.
\end{itemize}
These issues are of great complexity mainly because they involve different segments, with multi-layered configurations, comprising several application domains, in addition to representing an important investment cost. Network operators will need to scale the segments of the metropolitan network quickly, as services begin to be deployed, and choose different criteria for decision making in this regard, both on which requirements are a priority for business performance and on periods and places of greatest need. Even the edge of the network in the MA segment will have to expand to support the flows of these services, and therefore it will need to be thought of together with the underlying segments.

\subsection*{MON in General}
There is a large research gap on traffic patterns analysis in metropolitan networks. As has been seen, the emergence of new services combined with the population dynamics of cities and changes in the distribution of traffic, there can often be an imbalance in the use of resources available on the network, since this ecosystem is based on elements with great performance differences. With this, there is the possibility of new redesigns and traffic distribution beyond the phenomenon of tidal traffic. At the same pace, online solutions are expected to react in the direction of time and adapt to network conditions. Consider as an atypical example the enormous impact that the COVID-$19$ Pandemic had on the traffic profile of large cities. 

The deployment of $5G$ mobile networks will depend on the features that optical network architectures can offer. After all, fiber is still the fastest vehicle for data transport. $5G$ service providers and operators will decide on the appropriate costs and strategies to take advantage and optimize the existing infrastructure, or, on the other hand, invest in the implementation of a new environment with easy integration and simplified operations.
This is not just an industry issue, as the trend towards infrastructure convergence and resource sharing, as in backbone networks, will lead to the debate on security, survival, and the mixing of technologies from several different generations, among other issues. For example, the first-generation fibers will not be removed from the infrastructure ducts but will work in conjunction with new versions such as multinucleated fibers. It is also important to mention that there is no ideal architecture, as the choice for a solution depends heavily on the providers' applications and business segments.

%%%%%%%%%%%%%%%%%%%%%%%%%%%%%%%%%%%%%%%%%%%%%%%%%%%%%%%%%%%%%%
%%%%%%%%%%%%%%%%%%%%%%%%%%%%%%%%%%%%%%%%%%%%%%%%%%%%%%%%%%%%%%
%%%%%%%%%%%%%%%%%%%%%%%%%%%%%%%%%%%%%%%%%%%%%%%%%%%%%%%%%%%%%%
\section{Final Considerations} \label{sec:conclusao}
The emergence of new computing paradigms, such as edge computing and IoT, and the proliferation of new access technologies, such as PON and $5G$, together have been shifting the traffic load closer to the cities, thus putting pressure on the metropolitan optical networks infrastructure.

The  recent  literature  surveys  proposes  various  solutions traffic engineering for optical networks,  but  without  taking into  account  the  characteristics  of  the  underlying network architecture.

On  the  other  hand,  projects  for  the  implementation  of  new  edge  nodes are flourishing and, with this wide range of possibilities, many metropolitan network architectures have been proposed, especially since $2017$.

This  work  focuses  on  bringing  together  these  architectural  proposals to discuss its main characteristics, advantages and disadvantages. Architectures like UDWSN \cite{zhang2016ultra}, HnS/WDM \cite{she2017metro}, ARMONIA~\cite{kretsis2020armonia}, DISCUS~\cite{ruffini2017access}, MOMC~\cite{larrabeiti2019all}, MEON \cite{rottondi2013routing}, SIMON \cite{muciaccia2019proposal}, TDGMON~\cite{lin2019three}, FMN \cite{paolucci2020disaggregated}, DuFiNet \cite{uzunidis2018dufinet}  and DnW \cite{8346157SemiFilterless}, for example, were designed with different objectives and concepts, and because they are very recent, they are not yet consolidated.

A comparative analysis among architectures identify the featured research topics and its future trends, finally the literature gaps are highlighted. In summary, the exciting future of MON research and operation lies into the use of elastic transmission technologies in open hybrid tailored architectures, all controlled by autonomic intelligence control planes.

\begin{center}
    {\scriptsize
    \begin{longtable}{|l|p{0.7\textwidth}|}
        \caption{List of Acronyms.} \label{tab:Acronym} \\

        \hline
        \multicolumn{1}{|c|}{\textbf{Acronym}} & \multicolumn{1}{c|}{\textbf{Description}} \\
        \hline 
        \endfirsthead

        \multicolumn{2}{c}%
        {{\bfseries \tablename\ \thetable{} -- continued from previous page}} \\
        \hline
        \multicolumn{1}{|c|}{\textbf{Acronym}} & \multicolumn{1}{c|}{\textbf{Description}} \\
        \hline 
        \endhead

        \hline
        % \multicolumn{2}{|r|}{{Continued on next page...}} \\
        % \hline
        \endfoot

		AAU&Active Antenna Units \\
		ADM&Add/Drop Multiplexer\\
		AI & Artificial Inteligence \\
		AMEN & Access Metro Edge Node \\
		AN&Access Nodes\\
		ARN & Active Remote Node \\
		AWG & Arrayed Waveguide Gratings \\
		A-WDM&Analog-Wavelenght Division Multiplexing\\
		BER & Bit Error Rate \\
		BHP & Burst Header Packets\\
		BnS &Broadcast-and-Select\\
		BP-ANN &Back Propagation based Artificial Neural Network \\
		BVT & Bandwidth Variable Transponder \\
		C&Optic Couplers\\
		CAPEX & Capital Expenditure  \\
		CD&Colorless and Directionless\\
		CD-ROADM&Colorless and Directionless Reconfigurable Optical Add Drop Multiplexers\\
		CDC & Colorless, Directionless and Contetionless \\
		CDC-ROADMs&Colorless, Directionless and Contentionless Reconfigurable Optical Add Drop Multiplexers\\
		CO & Central Office \\
		Co&Coherent\\
		CT&Coherent Transponder\\
		CORD & Central Office Re-architected as a Datacenter \\
		CPRI&Common Public Radio Interface\\
		CU&Centralized Unit\\
		CWDM&Coarse Wavelength Division Multiplexing\\
		DC & Data Center \\
		DCF&Dispersion Compensating Fibers\\
		DCM & Dispersion Compensation Methods \\
		DnW & Drop-and-Waste \\
		DP&DualPolarization\\
		DS&Downstream\\
		DSLAM&Digital Subscriber Line Access Multiplexer\\
		DSP & Digital Signal Processing \\
		D-WDM&Digital to WDM\\
		DU&Distributed Unit\\
		DWDM & Dense Wavelenght Division Multiplexing \\
		DuFiNet & Dual Fibre Network \\
		E2E & End-to-End \\
		EDFA & Erbium-Doped Fiber Amplifier  \\
		EON & Elastic Optical Network  \\
		FCN&Flexible Control Nodes \\
		FL & Filterless  \\
		Flex-OCSM&Centralized Flexible Optical Carrier Source Module\\
		FMF&Few-Mode Fiber\\
		FON&Filterless Optical Networks\\
		FSU & Frequency Slot Units\\
		HnS&Hub-and-Spoke\\
		ILP&Integer Linear Programming\\
		IMDD & Intensity-Modulation Direct-Detection \\ 
		IMT&International Mobile Telecommunications Group\\
		ISP&Internet Service Provider\\
		IP&Internet Protocol\\
		ISP & Internet Service Provider \\ 
		ITU & International Telecommunication Union   \\ 
		LCOS&Liquid Crystal on Silicon\\
		LR-PON&Long Reach Passive Optical Networks\\
		LSTM&Long Short-Term Memory\\
		MA &Metro-Acess \\
		MC & Metro-Core\\
		MCENs&MetroCore Edge Nodes\\
		MCF&Multi-Core Fiber\\
		MCS&Multicast Switch\\
		mDC &micro Data-Center\\
		MD-ROADM&Multi-degree Reconfigurable Optical Add and Drop Multiplexer\\
		MEM&Micro-Electrical-Mirror\\
		Metro &Metropolitan\\
		MEON&Metro Elastic Optical Network\\
		ML-SBVT&Multi-Laser Sliceable Bandwidth Variable Transceivers\\
		MIMO&Multiple-Input Multiple-Output\\
		MMF&Multi-Mode Fiber\\
		MN&Metro Nodes\\
		MNOs&Mobile Network Operators\\
		MOMC&Modular Optical Metro-Core\\
		MON&Metropolitan Optical Network\\
		M-OTN&Mobile-Optimized Optical Transport Network\\
		MP&Muxponders\\
		MW-SBVT&Multi-Wavelength Sliceable Bandwidth Variable Transceivers\\
		NRZ&Non return-to-zero\\
		MWDM&Metro  Wavelength-Division Multiplexing \\
		NWDM&Nyquist Wavelength-Division Multiplexing\\
		OADM&Optical Add and Drop Multiplexer\\
		OaaS&optimization-as-a-service (OaaS)\\
		OCh&Optical Channel\\
		ODU&Optical Data Unit\\
		OEO&Optical-Eletrical-Optical\\
		OFDM&Orthogonal Frequency Division Multiplexing \\
		OLS&Optical Line Systems\\
		ONU&Optical Network Unit\\
		OOFDM&Optical Orthogonal Frequency Division Multiplexing\\
		OPEX&Operational Expenditure\\
		OSNR&Optical Signal to Noise\\
		OSS&open source software\\
		OTDM&Orthogonal time division multiplexing\\
		OTN&Optical Transport Network\\
		OTU&Optical Transport Unit\\
		OXC&Optical Crossconnect\\
		P2MP&Point-to-Multipoint\\
		P2P&Point-to-Point\\
		PAM&Pulse Amplitude Modulation\\
		PASSION & ProgrAmmable transmission and switching modular systems based on Scalable Spectrum/space aggregation for future agIle high capacity metrO Networks\\
		PE&Passive Element\\
		POP&Points of Presence\\
		PSM&Photonic Space Switch Matrix\\
		qNWDM&quasi-Nyquist Wavelength-Division Multiplexed\\
		QoE&Quality of Experience\\
		QoS&Quality of Service\\
		RnC&Route-and-Couple\\
		ROADM&Reconfigurable Optical Add and Drop Multiplexer\\
		RRU&Remote Radio Units\\
		RSA&Routing and Spectrum Assignment\\
		RSSA&Routing, Spatial channel and Spectrum Assignment\\
		RWA&Routing and Wavelength Assignment\\
		S&Splitters\\
		SBVT&Sliceable Bandwidth Variable Transceivers\\
		SDN&Software Defined Networking\\
		sFL&Semi-Filterless\\
		SMF&Single-Mode Fiber\\
		SOA&Semiconductor Optical Amplifier\\
		%SOI OF&&&\\
		SPs&Switchponders\\
		SSS&Spectrum Selective Switch\\
		SWDMM&Shortwave Wavelength Division Multiplexing\\
		US&Upstream\\
		TE&Traffic Engineering\\
		TFP&Time-Frequency Packing\\
		ToS&Type of Service\\
		TPs&Transponders\\
		UDWSN&Ultra Dense-Wavelength Switched Network\\
		VCSEL&Vertical Cavity Surface Emitting Laser\\
		VNF&Virtual Network Functions \\
		WAN&Wide Area Network\\
		WDM&Wavelength Division Multiplexing\\
		WSS&Wavelength Selective Switch			 \\
    \end{longtable}
    }
\end{center}

\bibliographystyle{IEEEtran}
%\bibliography{survey.bbl}

\end{document}